\documentclass[a4paper,12pt]{article}
\usepackage[english]{babel}
\usepackage{jheppub2}
\usepackage{subfig}
\usepackage[T1]{fontenc} 
\usepackage{graphicx}
\usepackage{epsfig}
\usepackage{amsmath}
\usepackage{amssymb}
\usepackage{float}
\usepackage{placeins}
\usepackage{braket}
\usepackage{enumitem}
\allowdisplaybreaks[4]
\makeatletter
\renewcommand\paragraph{\@startsection{paragraph}{4}{\z@}%
            {-2.5ex\@plus -1ex \@minus -.25ex}%
            {1.25ex \@plus .25ex}%
            {\normalfont\normalsize\bfseries}}
\makeatother
\title{Low-energy theorem revisited and OPE in massless QCD}

\author[a]{Marco Bochicchio}
\author[b,c]{Elisabetta Pallante}

\affiliation[a]{Physics Department, INFN Roma1, Piazzale Aldo Moro 2, 00185, Rome, Italy}
\affiliation[b]{Van Swinderen Institute for Particle Physics and Gravity, University of Groningen, 9747 AG, The Netherlands}
\affiliation[c]{Nikhef, Science Park, Amsterdam, The Netherlands}

\emailAdd{marco.bochicchio@roma1.infn.it}
\emailAdd{e.pallante@rug.nl}
\abstract{We revisit a low-energy theorem (LET) of NSVZ type in SU($N$) QCD with $N_f$ massless quarks derived in \cite{MBR} by implementing it in dimensional regularization. The LET relates $n$-point correlators in the lhs to $n+1$-point correlators with the extra insertion of $\Tr F^2$ at zero momentum in the rhs.
First, we demonstrate that, for $2$-point correlators of an operator $O$ in the lhs, the LET implies that, in general, the integrated $3$-point correlator in the rhs needs in perturbation theory an infinite additive renormalization in addition to the multiplicative one.
Second, we relate the above counterterm -- that is completely fixed by the LET -- to a corresponding  divergent contact term in a certain coefficient of the OPE of $\Tr F^2$ with $O$ in the momentum representation, thus extending by means of the LET to any operator $O$ an independent argument that first appeared for $O=\Tr F^2$ in \cite{Z3}.
Third, we verify by direct computation that the latter divergent contact term first computed in \cite{Z1} to order $g^4$ in perturbation theory and to all orders in \cite{Z3} actually agrees with the one implied by the LET.
Fourth, we evaluate the divergent contact terms for the above OPE coefficient both in the coordinate and momentum representation and discuss their relation. 
Fifth, we demonstrate that in the asymptotically free phase of QCD the aforementioned  counterterm in the LET -- though divergent order by order in perturbation theory -- is actually finite
nonperturbatively after resummation to all perturbative orders.
Finally, we briefly recall the implications of the LET in the gauge-invariant framework of dimensional regularization for the perturbative and nonperturbative renormalization in large-$N$ QCD.
The implications of the LET inside and above the conformal window of SU($N$) QCD with $N_f$ massless quarks will appear in a forthcoming paper.}

\DeclareMathOperator{\Tr}{Tr}
\newcommand{\be}{\begin{equation}}
\newcommand{\ee}{\end{equation}}
\newcommand{\nn}{\nonumber}
\newcommand{\bea}{\begin{eqnarray}}
\newcommand{\eea}{\end{eqnarray}}
\newcommand{\bfig}{\begin{figure}}
\newcommand{\efig}{\end{figure}}
\newcommand{\bc}{\begin{center}}
\newcommand{\ec}{\end{center}}

\newcommand{\f}[2]{\frac{#1}{#2}}

\newcommand{\eps}{{\epsilon}}
\newcommand{\cO}{{\mathcal O}}
\newcommand{\cF}{{\mathcal F}}
\newcommand{\td}{{\tilde{d}}}
\newcommand{\tg}{{\tilde{g}}}
\newcommand{\tD}{{\tilde{\Delta}}}
\newcommand{\dr}{\text{dim.\,reg.}}
\newcommand{\Lrgi}{\Lambda_{\scriptscriptstyle{UV}}}

\begin{document}
\maketitle
\flushbottom

\section{Introduction, physics motivations and conclusions} \label{1}

In the present paper we work out in dimensional regularization several equivalent versions of a low-energy theorem (LET) of NSVZ type in SU($N$) QCD with $N_f$ massless quarks derived in \cite{MBR}. The LET relates $n$-point correlators in the lhs to $n+1$-point correlators with the extra insertion of $\Tr F^2$  integrated over space-time -- i.e. at zero momentum --  in the rhs.\par
Apart from its intrinsic interest, the above study has several applications that we divide in two installments.\par
The first one -- in the present paper -- concerns the relation of the LET with a certain OPE coefficient of $\Tr F^2$ with an operator $O$ defined by the solution of the Callan-Symanzik equation in the momentum representation, $C^{(F^2,O)}_1(p)$, its perturbative and nonperturbative renormalization in the asymptotically free (AF) phase of QCD, and the nonperturbative renormalization in large-$N$ QCD.\par
The second one -- in a forthcoming paper \cite{BP2} -- concerns the LET in the phases inside and above the conformal window of QCD, and at the Wilson-Fisher conformal fixed point in $4 - 2\epsilon$ dimensions. In fact, the second installment depends crucially on some delicate results in the first one that we explain as follows. \par
First, one of our key results on the LET applied to $2$-point correlators $C^{(O,O)'}_0(z)$
$=\langle O(z)O(0)\rangle'$ at distinct points $z \neq 0$ of a multiplicatively renormalizable operator $O$ in the lhs is that the integrated $3$-point correlator in the rhs needs in general an infinite additive renormalization as $\epsilon \rightarrow 0$ -- completely controlled by the LET -- in addition to the multiplicative one order by order in perturbation theory. The corresponding counterterm consists
in the subtraction of an integrated -- i.e. at zero momentum -- divergent contact term proportional to a $\delta^{(4)}$ multiplying the $2$-point correlator of $O$. \par
Second, we demonstrate that the above divergent contact term coincides with the corresponding countertem for $C^{(F^2,O)}_1(p)$, which for the special case $C^{(F^2,F^2)}_1(p)$ agrees with a previous result  \cite{Z3} obtained by means of an independent argument.\par 
Third, we find out by direct computation that the aforementioned counterterm for $C^{(F^2,F^2)}_1(p)$ does not
totally arise in dimensional regularization from the Fourier transform of $C^{(F^2,O)'}_1(x)$ restricted at distinct points. As a consequence, an intrinsically divergent contact term -- that we refer to as a divergent  proper contact term -- must be added to $C^{(F^2,O)'}_1(x)$ in order to reproduce the correct infinite additive renormalization of $C^{(F^2,F^2)}_1(p)$ and, more generally, of $C^{(F^2,O)}_1(p)$.
This observation will play a key role in our second installment, where we apply the LET to QCD inside and above the conformal window.\par
Fourth, we demonstrate that in the AF phase of the theory the additive renormalization in the rhs of the LET turns out to be finite after the nonperturbative resummation to all perturbative orders.\par
Finally, we employ the LET in dimensional regularization to verify in a manifestly gauge-invariant framework some computations \cite{MBL} involving a hard-cutoff regularization of the bare LET. 
We briefly recall the latter results and their connection to the present paper.
 The LET has two versions \cite{MBR}, one involving in the lhs the logarithmic derivative with respect to the gauge coupling, and one involving the logarithmic derivative with respect to the RG-invariant scale, $\Lambda_{\scriptscriptstyle{UV}}$, in YM theories AF in the UV.
Originally, the LET has been employed \cite{MBR} to study the nonperturbative renormalization of large-$N$ confining massless QCD-like theories and, in particular, massless QCD \cite{H,V,Migdal,W}, since the second version of the LET and the nonperturbative renormalization \cite{MBR} of $\Lambda_{\scriptscriptstyle{QCD}}$ in the large-$N$ 't Hooft \cite{H} and Veneziano \cite{V} expansions control \cite{MBR} the structure of the nonperturbative counterterms for the YM action. In fact, the second version of the LET provides an obstruction \cite{MBL} to the existence of canonical string models implementing the open/closed string duality that would realize nonperturbatively 
the large-$N$ 't Hooft expansion of QCD and massive $\mathcal{N}=1$ SUSY QCD in the confining phase.
Besides, the first version of the LET demonstrates how the open/closed string duality may be implemented in canonical string models \cite{MBL} perturbatively realizing massless QCD-like theories to order $g^2$.
In this context an essential tool to actually compute the possible divergences in the rhs of the LET has been the OPE of $O$ with $\Tr F^2$, both in perturbation theory and in its RG-improved form by a hard-cutoff regularization of the space-time integral in the rhs  \cite{MBL,BB}. The logic above can be inverted and the relevant contribution to order $g^2$ in the rhs of the LET for the OPE of $O$ with $\Tr F^2$ can be recovered \cite{BB} from the anomalous dimension of $O$ in the lhs  by assuming the LET for the $2$-point correlator of $O$.
We should also mention the approach to the OPE in \cite{F} that is closely related to the LET, but actually independent of the present paper that employs dimensional regularization. \par

\section{Plan of the paper} \label{2}

In section \ref{3} we work out several equivalent versions of the LET in dimensional regularization and establish their main implications with respect to the aforementioned additive renormalization.\par
In section \ref{4} we investigate the relation between the LET and the OPE coefficient $C^{(F^2,O)}_1(x)$ including contact terms. Moreover, we specifically compute the finite and divergent proper contact terms in $C^{(F^2,F^2)}_1(x)$. \par
In section \ref{5} we compare the perturbative version of the LET with its nonperturbative resummation to all perturbative orders. \par
In section \ref{6} we solve the LET in perturbation theory to order $g^2$, where the theory is exactly conformal. \par
In section \ref{60} we demonstrate by direct computation that divergent proper contact terms occur in $C^{(F^2,F^2)}_0(z)$ to order $g^2$ as in $C^{(F^2,F^2)}_1(x)$. \par
In section \ref{7}  we employ one version of the LET in dimensional regularization to verify the implications of the corresponding version in hard-cutoff regularization, both perturbatively and nonperturbatively.  \par
In the appendices \ref{C}-\ref{F} we report several ancillary computations.

\section{LET} \label{3}
 The LET applies to YM theories in $d=4$ dimensions. 

\subsection{LET and Wilsonian normalization of the action}
\label{4.1}

 Starting from the Euclidean functional integral of SU($N$) YM theories in $d=4$ dimensions,
the LET has been derived for bare correlators with the Wilsonian normalization of the action \cite{MBR}:
\be
\label{eq:PI}
\braket{ \cO_1\cdots\cO_n}_0=\f{\int\,\cO_1\cdots\cO_n \,e^{-\f{N}{2g_0^2}\int \,\Tr\cF^2d^4x+\cdots} }{\int\,e^{-\f{N}{2g_0^2}\int \,\Tr\cF^2d^4x+\cdots}}
\ee 
with $g_0$ the bare 't Hooft coupling $g_0^2=g^2_{0 YM} N$, $\Tr$ the trace in the fundamental representation,
$\cO_i$ local bare operators independent of $g_0$, $\Tr \mathcal{F}^2 \equiv \Tr (\mathcal{F}_{\mu\nu}\mathcal{F}_{\mu\nu})$, $\mathcal{F}_{\mu\nu} = \partial_{\mu}A_{\nu} - \partial_{\nu} A_{\mu} + i [A_{\mu},A_{\nu}]$, and the sum over repeated indices understood. 
We explicitly write in eq. \eqref{eq:PI} only the term in the action that depends on $g_0$ and, therefore, enters the derivation of the LET. We immediately arrive at the LET by deriving eq. \eqref{eq:PI} with respect to $-1/g_0^2$:
\be
\label{eq:LETWcoord}
\f{\partial}{\partial\log g_0} \braket{\cO_1\cdots\cO_n}_0 = \f{N}{2g_0^2}\,\int\, \braket{\cO_1\cdots\cO_n \cF^2(x)}_0 - \braket{\cO_1\cdots\cO_n}_0\braket{\cF^2(x)}_0\, d^4x
\ee
with $\cF^2\equiv 2\Tr\cF^2$.

\subsection{LET and canonical normalization of the action} \label{4.2}

 We rescale the gauge fields in eq. \eqref{eq:PI} by the factor $\frac{g_0}{\sqrt N}$ in order to rewrite the LET for the canonically normalized YM action \cite{BB}.
Defining $\frac{g_0^2}{N} \Tr F^2= \Tr \mathcal{F}^2$ and $(\frac{g_0}{\sqrt N})^{c_{O_k}} O_k= \mathcal{O}_k$ for some $c_{O_k}$ -- for example, $c_{F^2}=2$ --  after the rescaling, with $O_k$ now dependent on $g_0$ but canonically normalized and $F_{\mu\nu} = \partial_{\mu}A_{\nu} - \partial_{\nu} A_{\mu} + i \f{g_0}{\sqrt N}[A_{\mu},A_{\nu}]$, we get the identity \cite{BB}:
\bea 
\label{res}
\braket{ \mathcal{O}_1 \cdots \mathcal{O}_n}_0 = \prod^{k=n}_{k=1}(\frac{g_0}{\sqrt N})^{c_{O_k}}   \braket{ O_1\cdots O_n }_{0}
\eea
with the expectation values in the lhs and rhs defined by the Wilsonian and canonical normalization, respectively.
The LET in eq. \eqref{eq:LETWcoord}
 is rewritten in terms of canonically normalized bare local operators \cite{BB} by employing eq. \eqref{res}:
\bea
\label{eq:LETCcoord_sec4}
&&\sum^{k=n}_{k=1} c_{O_k} \braket{O_1\cdots O_n}_0 +
\f{\partial}{\partial\log g_0} \braket{O_1\cdots O_n}_0 \nn\\
&&\hspace{0.5truecm}= \f{1}{2}\,\int\, \braket{O_1\cdots O_n F^2(x)}_0 
- \braket{O_1\cdots O_n}_0\braket{F^2(x)}_0\, d^4x
\eea
with $F^2\equiv 2\Tr F^2$.
In fact, we only consider the applications of the LET for $n=2$ and $O_1=O_2=O$:
\bea \label{LET2}
&&2 c_{O} \braket{O(z)  O(0)}_0 +
\f{\partial}{\partial\log g_0} \braket{O(z) O(0)}_0 \nn\\
&&= \f{1}{2}\,\int\, \braket{O(z) O(0) F^2(x)}_0 
- \braket{O(z) O(0)}_0\braket{F^2(x)}_0\, d^4x
\eea

\subsection{Dimensional regularization} 

 In general, the bare correlators are divergent and need regularization and, eventually, renormalization.
Therefore, the first issue to apply the LET to YM theories is to find a regularization of the bare correlators and, correspondingly, of the LET. Dimensional regularization preserves gauge invariance to all orders of perturbation theory and it is our natural choice. It may be performed both in the coordinate and momentum representation for the correlators of the fundamental fields and the corresponding composite operators.

\subsection{Dimensional regularization of YM theories}
\label{5.1}

In dimensional regularization the space-time dimension 
is shifted:
\be
d\to \td=d-2\eps
\ee
with $\eps$ positive and small and, in the present paper, $d=4$. The canonical dimension of the bare operators in the action follows by dimensional analysis, for the action to be dimensionless in $\td$ dimensions. 
In particular, the canonically normalized YM action formally reads in $\td$ dimensions:
\be\label{eq:caction}
S= \f{1}{2} \int \Tr F^2_0\,d^\td x
\ee
with $\Tr F^2_0=\Tr(F_{\mu\nu}F_{\mu\nu})$ the bare operator. Given $[d^\td x] = -\td$, with $[.]$ the mass dimension of its argument, $[S]=0$ implies:
\be
[\Tr F^2_0]\equiv\tD_{F^2_0}=\td
\ee
As $F_{\mu\nu}=\partial_\mu A_{\nu} -\partial_\nu A_{\mu} +i\f{g_0}{\sqrt{N}}[A_{\mu}, A_{\nu}]$, the canonical dimension of the bare gauge field $A_{\mu}$ follows:
\bea \label{F0}
[A_{\mu}]&=&\f{\tD_{F^2_0}}{2}-1\nn\\\
&=&\f{\td}{2}-1\nn\\
&\overset{d=4}{=}&1-\eps
\eea
Besides, since $[\partial_\mu]=1$, eq. \eqref{F0} implies for the bare gauge coupling:
\bea\label{eq:g0DR}
[g_0]&=&-2[A_\mu ]+\f{\td}{2} \nn\\
 &=&2-\f{\td}{2}\nn\\
 &\overset{d=4}{=}&\eps
\eea
We recall some relations needed to rewrite the LET in terms of the renormalized coupling and correlators in dimensionally regularized YM theories. 
 According to eq. \eqref{eq:g0DR} we get:
\bea\label{eq:gg0}
g_0=Z_g(g,\eps) \mu^\eps g
\eea
where $g$ is the dimensionless renormalized gauge coupling and $Z_g$ the renormalization factor that in $\overline{MS}$-like renormalization schemes is a series of pure poles in $\eps$ -- as it is the renormalization factor $Z_{O}$ for a multiplicatively renormalizable operator $O$. 
 Moreover, the beta function reads in $\td=4-2\eps$ dimensions:
\be\label{eq:betaeps}
\beta(g,\epsilon )=-\epsilon g +\beta(g)
\ee
with:
\bea
\beta (g,\epsilon )=\f{dg}{d\log\mu} 
\eea
and:
\bea
\beta(g) &=& -g\,\f{d\log Z_g}{d\log\mu}\nn\\
&=&\f{dg}{d\log\mu}\bigg|_{\eps=0}
\eea
the beta function in $d=4$ dimensions. We obtain from eq. \eqref{ZF2}:
 \bea\label{eq:gg0rel0}
  \f{\partial\log g}{\partial\log g_0} &=& \left ( 1+ \f{\partial\log Z_g}{\partial\log g}\right )^{-1}\nn\\
  &=&1- \f{\beta(g)}{\eps g}\nn\\
  &=&Z_{F^2}^{-1}(g,\eps)
  \eea
 that follows from the logarithmic derivative of eq. \eqref{eq:gg0} by equating eq. \eqref{eq:betaeps} to:
  \bea
  \beta(g,\eps)&=& -\eps g \left ( 1+ \f{\partial\log Z_g}{\partial\log g}\right )^{-1}
  \eea
Finally, from the anomalous dimension $\gamma_O$ of a multiplicatively renormalizable operator $O=Z_O O_0$:
  \be\label{gamma}
  \gamma_O(g) = - \f{d\log Z_O}{d\log\mu}
  \ee
we obtain: 
  \bea\label{eq:Zrel0}
  \f{\partial\log Z_O^{-2}}{\partial\log g_0}&=&-2\f{\partial\log Z_O}{\partial\log g_0}\nn\\
  &=&-2\f{d\log Z_O}{d\log\mu}\f{d\log\mu}{d\log g}\f{\partial\log g}{\partial\log g_0}\nn\\
  &=&2\gamma_O(g)\left(\f{1}{g}\f{d g}{d\log\mu} \right)^{-1}\left ( 1+ \f{\partial\log Z_g}{\partial\log g}\right )^{-1}\nn\\
  &=&2\gamma_O(g)\left(\f{\beta(g,\eps)}{g} \right)^{-1}\left ( 1- \f{\beta(g)}{\eps g}\right )\nn\\
  &=&-2\f{\gamma_O(g)}{\eps}\left ( 1- \f{\beta(g)}{\eps g}\right )^{-1}\left ( 1- \f{\beta(g)}{\eps g}\right )\nn\\
   &=&-2\f{\gamma_O(g)}{\eps}
  \eea
employing eqs. \eqref{gamma} and \eqref{eq:gg0rel0}.

\subsection{LET in dimensional regularization} \label{5.2}

 The LET in $\td$ dimensions  
with the Wilsonian normalization of bare operators 
may be written in terms of the dimensionless bare coupling $\bar{g}_0=g_0\mu^{\td/2-2}$ according to eq. \eqref{eq:g0DR}: 
\bea
\label{eq:LETWcoordd}
&&\f{\partial}{\partial\log \bar{g}_0} \braket{\cO_1\cdots\cO_n}_0 = 
\f{N}{2\bar{g}_0^2}\mu^{2-\f{\td}{2}}\,\int\, \braket{\cO_1\cdots\cO_n \cF^2(x)}_0 \nn\\
&&- \braket{\cO_1\cdots\cO_n}_0\braket{\cF^2(x)}_0\, d^\td x 
\eea
where the $\mu$ dependence in the rhs is due to $[\cF^2_0]=4$.  
Rewriting eq. \eqref{eq:LETWcoordd} 
in terms of canonically normalized bare operators
we get in $\td$ dimensions:
\bea
\label{eq:LETCcoordd}
&&\sum^{k=n}_{k=1} c_{O_k} \braket{O_1\cdots O_n}_0 +
\f{\partial}{\partial\log {g}_0} \braket{O_1\cdots O_n}_0 \nn\\
&&\hspace{0.5truecm}= \f{1}{2}\,\int\, \braket{O_1\cdots O_n F^2(x)}_0 
- \braket{O_1\cdots O_n}_0\braket{F^2(x)}_0\, d^\td x 
\eea
where the integral over space-time in the rhs is defined by analytic continuation and, since $[F^2_0]=\td$, no dependence on the scale $\mu$ appears in the rhs. As for the notation, every time that in an equation the integral over space-time is denoted by:
\bea
\int  d^\td x 
\eea
dimensional regularization of the correlators is understood as well.
For $n=2$ with $O_1=O_2=O$ we get:
\bea
\label{eq:LETCuv_n2}
&&2c_{O} \braket{O(z) O(0)}_0 +
\f{\partial}{\partial\log {g}_0} \braket{O(z) O(0)}_0  \nonumber \\
&&= \f{1}{2} \int\, \braket{O(z) O(0) F^2(x)}_0 - \braket{O(z)O(0)}_0\braket{F^2(x)}_0\, d^\td x
\eea

\subsection{Bare LET}

We limit ourselves to correlators of  gauge-invariant scalar operators $O(z)$ that are multiplicatively renormalizable up to the mixing with operators that are BRST variations and operators that vanish by the equations of motion (EOM) \cite{EOM,EOM1,EOM2}.
We neglect the mixing with BRST operators, since their insertion in gauge-invariant correlators vanishes.
The mixing with EOM operators produces at most contact terms. 
Therefore, we set $z \neq 0$ once for all, so that we may ignore the above mixing both in the lhs and rhs of the LET and express the LET in terms of the multiplicatively renormalized operator $O(z)$.
Then, the LET reads in terms of the bare operator $F_0^2(x)$ and the multiplicatively renormalized one $O(z)$:
 \bea \label{bareLET22}
&&\braket{O(z)O(0)}\left( 2c_O - \f{2\gamma_O(g)}{\eps}+ Z_{F^2}^{-1}\f{\partial\log \braket{O(z)O(0)}}{\partial\log g}\right) \nn\\
&&=\f{1}{2}\int \braket{O(z) O(0) F_0^2(x)} - \braket{O(z)O(0)}\braket{F_0^2(x)} d^\td x 
\eea
with:
\bea
Z_{F^2}^{-1}= 1-\f{\beta(g)}{\eps g}
\eea
Eq. \eqref{bareLET22} is obtained from eq. \eqref{eq:LETCuv_n2} as follows.
 The renormalized operator $O$ is related to the bare one $O_0$ by a multiplicative renormalization, hence:
\bea \label{eq:RvsB0}
O_0(z)&=&Z_O^{-1}(g,\eps) O(z)
\eea
The lhs of eq. \eqref{bareLET22} is then obtained by multiplying by the factor $Z_O^{2}$ the lhs of eq. \eqref{eq:LETCuv_n2} that reads:
 \bea
 \label{eq:lhs_gen0}
&& 2c_O\, Z_O^{-2}\braket{O(z)O(0)} +
g_0\f{\partial}{\partial {g}_0} Z_O^{-2}\braket{O(z)O(0)}
+Z_O^{-2}g_0\f{\partial}{\partial {g}_0} \braket{O(z)O(0)} \nn\\
&&= Z_O^{-2}\braket{O(z)O(0)}\left( 2c_O +\f{\partial\log Z_O^{-2}}{\partial\log g_0}
+\f{\partial\log \braket{O(z)O(0)}}{\partial\log g_0}
\right) \nn\\
&&= Z_O^{-2}\braket{O(z)O(0)}\left( 2c_O - \f{2\gamma_O(g)}{\eps}+\left ( 1-\f{\beta(g)}{\eps g}\right ) \f{\partial\log \braket{O(z)O(0)}}{\partial\log g}\right)
\eea
where in the last line we have employed:
\bea
\frac{\partial}{\partial\log g_0} =\frac{\partial \log g}{\partial\log g_0} \frac{\partial}{\partial\log g}
\eea
and eqs. \eqref{eq:gg0rel0} and \eqref{eq:Zrel0}.

\subsection{Renormalized LET} \label{3.7}

In general, both sides of the LET in eq. \eqref{bareLET22} diverge as $\epsilon \rightarrow 0$ order by order in perturbation theory. Hence, we derive from eq. \eqref{bareLET22} two renormalized versions of the LET requiring that the lhs -- and therefore also the rhs -- is finite as $\epsilon \rightarrow 0$.\par
They differ by a finite term in the renormalized object in the lhs and we find it convenient to present each of them, (I) and (II), in three -- though totally equivalent -- ways as follows.\par
Version (IA):
 \bea \label{LET1}
\f{\partial \braket{O(z)O(0)}}{\partial\log g} 
&=&\f{1}{2}Z_{F^2} \int \braket{O(z) O(0) F_0^2(x)} - \braket{O(z)O(0)}\braket{F_0^2(x)} d^{\td}x \nn\\
&&-Z_{F^2}\left( 2c_O - \f{2\gamma_O(g)}{\eps}\right)\braket{O(z)O(0)}
\eea
(IA) shows that, since the lhs above is by construction finite as $\epsilon \rightarrow 0$ because it is expressed only in terms of a renormalized object, also the rhs must be finite. Moreover, it shows that the integrated correlator in the rhs needs an additive renormalization in addition to the multiplicative one. \par
Version (IB):
\bea \label{LET1A}
&&\f{\partial \braket{O(z)O(0)}}{\partial\log g} \nn\\
&&=\f{1}{2}Z_{F^2} \int \braket{O(z) O(0) F_0^2(x)} - \left(2c_O - \f{2\gamma_O(g)}{\eps}\right)\big(\delta^{(\td)}(x-z)+\delta^{(\td)}(x)\big)\braket{O(z)O(0)} \nn\\
&&\,\,\,\,\,\,- \braket{O(z)O(0)}\braket{F_0^2(x)} d^{\td}x 
\eea
where $\delta^{(\td)}(x)$ is defined by analytic continuation by means of the equation \cite{Collins}:
\bea
\int \delta^{(\td)}(x)\, d^{\td}x =1
\eea
and it is a well-defined distribution in integer dimensions.
(IB) shows that the aforementioned additive renormalization is equivalent to the subtraction of finite and divergent contact terms as $\epsilon \rightarrow 0$ from the bare $3$-point correlator in the rhs. By locality and dimensional analysis, in $d=4$ dimensions this counterterm is proportional to the product of the $2$-point correlator $\braket{O(z)O(0)}$ and the sum of contact terms $\delta^{(4)}(x)+\delta^{(4)}(x-z)$. 
Only after the subtraction involving the above mentioned finite and divergent contact terms, the integrated bare $3$-point correlator becomes multiplicatively renormalizable. \par
Version (IC):
\bea \label{LET1C}
&&\f{\partial\braket{O(z)O(0)}}{\partial\log g} \nn\\
&&=\f{1}{2}Z_{F^2} \int \braket{O(z) O(0) F_0^2(x)} - 2 c_O\big(\delta^{(\td)}(x-z)+\delta^{(\td)}(x)\big)\braket{O(z)O(0)} \nn\\
&&\,\,\,\,\,\,- \braket{O(z)O(0)}\braket{F_0^2(x)} d^{\td}x+Z_{F^2} \f{2\gamma_O(g)}{\eps} \braket{O(z)O(0)}
\eea
(IC) shows that subtracting only the finite contact terms in (IB) from the bare correlator in the rhs changes the divergent contact terms to be subtracted from the multiplicatively renormalized one, in order to obtain the very same finite fully renormalized object defined in the lhs.\par
In fact, there is a finite ambiguity also in the definition of the fully renormalized object in the lhs that affects the infinite additive renormalization of the integrated bare $3$-point correlator in the rhs.\par
Version (IIA):
 \bea \label{LET2A}
&&\f{\partial \braket{O(z)O(0)}}{\partial\log g} + 2c_O \braket{O(z)O(0)}\nn\\
&&=\f{1}{2}Z_{F^2} \int \braket{O(z) O(0) F_0^2(x)} - \braket{O(z)O(0)}\braket{F_0^2(x)} d^{\td}x \nn\\
&&\,\,\,\,\,\,+Z_{F^2} \f{2\gamma_O(g)-2c_O \f{\beta(g)}{g}}{\eps} \braket{O(z)O(0)}
\eea
where we have employed (IA) and the identity:
\bea
-\frac{2c_O}{1-\f{\beta(g)}{\eps g}}+2c_O=-\frac{2c_O\frac{\beta(g)}{g}}{\epsilon \left(1-\f{\beta(g)}{\eps g}\right)}
\eea
As anticipated, the finite change in (IIA) with respect to (IA) in the definition of the fully renormalized object in the lhs of the LET produces an infinite change in the needed additive renormalization of the bare $3$-point correlator in the rhs.\par
Version (IIB):
\bea \label{LET2B}
&&\f{\partial\braket{O(z)O(0)}}{\partial\log g}+ 2c_O \braket{O(z)O(0)} \nn\\
&&=\f{1}{2}Z_{F^2} \int \braket{O(z) O(0) F_0^2(x)} + \f{2\gamma_O(g)-2c_O \frac{\beta(g)}{g}}{\eps} \big(\delta^{(\td)}(x-z)+\delta^{(\td)}(x)\big)\braket{O(z)O(0)} \nn\\
&&\,\,\,\,\,\,- \braket{O(z)O(0)}\braket{F_0^2(x)} d^{\td}x 
\eea
that is the analogue of (IB).  \par
Version (IIC):
\bea \label{LET2C}
&&\f{\partial\braket{O(z)O(0)}}{\partial\log g}+ 2c_O \braket{O(z)O(0)} \nn\\
&&=\f{1}{2}Z_{F^2} \int \braket{O(z) O(0) F_0^2(x)} - \f{2c_O \frac{\beta(g)}{g}}{\eps} \big(\delta^{(\td)}(x-z)+\delta^{(\td)}(x)\big)\braket{O(z)O(0)}\nn\\
&&\,\,\,\,\,\,- \braket{O(z)O(0)}\braket{F_0^2(x)} d^{\td}x+Z_{F^2} \f{2\gamma_O(g)}{\eps} \braket{O(z)O(0)}
\eea
that is the analogue of (IC).

\subsection{Renormalized LET and operator mixing of $F^2$ in massless QCD}

The various forms of the LET in terms of the bare operator $F^2_0$ apply to any YM theory.
However, it is convenient to express the LET in terms of the renormalized operator $F^2$.
The renormalization of $F^2$ depends on the specific choice of the YM theory.\par
In massless QCD $F^2$ mixes \cite{Spiridonov} with the quark Lagrangian density $\bar\psi \gamma_{\mu}D_{\mu} \psi$ that vanishes by the EOM, whose insertion in the correlators produces at most contact terms. 
Yet, contact terms cannot be ignored in the rhs of the LET because of the space-time integration.\par
Nevertheless, even if $F^2_0$ mixes with $\bar\psi \gamma_{\mu}D_{\mu} \psi$, (IIA) shows that the integrated correlator in the rhs is made finite by the multiplicative renormalization $Z_{F^2}$ in addition to the additive one, despite the possible nontrivial mixing of $F^2$. Hence, as far as the LET is 
concerned, we may define $F^2 \equiv Z_{F^2} F_0^2$ and ignore the mixing.

\subsection{RG-invariant form of the LET}

It is worth writing the LET in a form that involves the RG-invariant operator (appendix {\ref C}):
\bea
- \epsilon F^2_0 &=&- \epsilon Z^{-1}_{F^2}(g,\epsilon) Z_{F^2}(g,\epsilon) F^2_0 \nonumber \\
 &=& - \epsilon Z^{-1}_{F^2}(g,\epsilon) F^2 \nonumber \\
 &=& - \epsilon \left ( 1-\f{\beta(g)}{\epsilon g}\right )  F^2 \nonumber \\
 &=& \f{\beta(g,\epsilon)}{g} F^2 \nonumber \\
 \eea
Multiplying eq. \eqref{bareLET22} by $- \epsilon$ and employing eq. \eqref{eq:gg0rel0} we obtain:
\bea \label{bareLET220}
&&\braket{O(z)O(0)}\left( 2 \gamma_O(g)  - 2c_O \epsilon+ \f{\beta(g,\epsilon)}{g} \f{\partial\log \braket{O(z)O(0)}}{\partial\log g}\right) \nn\\
&&=\f{1}{2} \f{\beta(g,\epsilon)}{g} \int \braket{O(z) O(0) F^2(x)} - \braket{O(z)O(0)}\braket{F^2(x)} d^{\td}x 
\eea
Interestingly, in this form of the LET no infinite additive renormalization arises. \par
Finally, setting $O_0=F^2_0$ and multiplying eq. \eqref {eq:LETCuv_n2} by $(-\epsilon)^3$ we get: \\
\bea
\label{eq:LETCuv_n20}
&&-2c_{O} \epsilon   \braket{ \f{\beta(g,\epsilon)}{g} F^2(z)  \f{\beta(g,\epsilon)}{g} F^2(0)} +
 \f{\beta(g,\epsilon)}{g} \f{\partial}{\partial\log {g}} \braket{ \f{\beta(g,\epsilon)}{g} F^2(z)  \f{\beta(g,\epsilon)}{g} F^2(0)} \nonumber \\
&&= \f{1}{2}\int\, \braket{ \f{\beta(g,\epsilon)}{g} F^2(z)  \f{\beta(g,\epsilon)}{g} F^2(0)  \f{\beta(g,\epsilon)}{g} F^2(x)} \nn\\
&& \,\,\,\,\,\,- \braket{ \f{\beta(g,\epsilon)}{g} F^2(z) \f{\beta(g,\epsilon)}{g} F^2(0)}\braket{ \f{\beta(g,\epsilon)}{g} F^2(x)}\, d^{\td} x 
\eea

\subsection{LET involving $\Lambda_{\scriptscriptstyle{UV}}$}

In an AF YM theory the LET can be rewritten in terms of the RG-invariant scale $\Lambda_{\scriptscriptstyle{UV}}$ \cite{MBR}.   
We obtain in dimensional regularization:
\bea \label{bareLETLQCD0}
&&\braket{O(z)O(0)}\left( 2 \gamma_O(g)  - 2c_O \epsilon- \f{\partial\log \braket{O(z)O(0)}}{\partial\log \Lambda_{\scriptscriptstyle{UV}} }\right) \nn\\
&&=\f{1}{2} \f{\beta(g,\epsilon)}{g} \int \braket{O(z) O(0) F^2(x)} - \braket{O(z)O(0)}\braket{F^2(x)} d^{\td}x 
\eea
employing the chain rule $\f{\partial}{\partial\log g}=\f{\partial \Lambda_{\scriptscriptstyle{UV}}}{\partial\log g}\f{\partial}{\partial\Lambda_{\scriptscriptstyle{UV}}}$, the defining relation of the RG invariance of $\Lambda_{\scriptscriptstyle{UV}}$ (appendix \ref{D.1}):
\bea
\bigg(\mu\f{\partial}{\partial \mu}+\beta(g,\epsilon)\f{\partial}{\partial g}\bigg)\Lambda_{\scriptscriptstyle{UV}}=0
\eea
and the identity:
\bea
\f{\partial \Lambda_{\scriptscriptstyle{UV}}}{\partial\log \mu}=\Lambda_{\scriptscriptstyle{UV}}
\eea
that follows from $\Lambda_{\scriptscriptstyle{UV}}=\mu f(g,\eps)=e^{\log\mu}f(g,\eps)$ for a certain function $f(g,\eps)$ with  $g=g(\mu)$. 
The lhs of the LET has a finite limit for $\eps\to 0$. Therefore, also the rhs must be finite as $\eps\to 0$. 
Hence, in the AF phase of the theory this in turn implies the finiteness of the space-time integral itself in the rhs as $\eps\to 0$, so that eq. \eqref{bareLETLQCD0} reads in the limit $\eps\to 0$:
\bea \label{bareLETLQCDd4}
&&\braket{O(z)O(0)}\left( 2 \gamma_O(g)  - \f{\partial\log \braket{O(z)O(0)}}{\partial\log \Lambda_{\scriptscriptstyle{UV}} }\right) \nn\\
&&=\f{1}{2} \f{\beta(g)}{g} \int \braket{O(z) O(0) F^2(x)} - \braket{O(z)O(0)}\braket{F^2(x)} d^4x 
\eea
where the integral in the rhs and all the correlators are in $d=4$ dimensions.

\section{LET and OPE} \label{4}

We show in the following that 
the infinite additive renormalization of the integrated $3$-point correlator in the rhs of the LET
is related to the corresponding divergence
of the multiplicatively renormalized OPE coefficient $Z_{F^2}C_{1}^{(F_0^2,O)}(p)$ in the momentum representation:
\be\label{eq:ct00}
\frac{1}{2} [Z_{F^2} \int \braket{O(z)O(0)F_0^2(x)} d^{\td}x]_{\textrm{div}}= [Z_{F^2} C_{1}^{(F_0^2,O)}(p)]_{\textrm{div}} \braket{O(z)O(0)}
\ee
The divergence is represented by a contact term in the momentum representation, i.e. a polynomial in momentum, and it is just a constant in our specific case. The OPE coefficient in the coordinate representation reads:
\bea
Z_{F^2} F_0^2(x)O(0) =  \cdots +Z_{F^2} C_{1}^{(F_0^2,O)}(x)O(0) + \cdots
\eea
Actually, we define $C_{1}^{(F_0^2,O)}(x)$ as a certain extension, including a distribution supported at coinciding points $x=0$, of the OPE coefficient ${C_{1}^{(F_0^2,O)}}'(x)$
at distinct points $x\neq 0$. The above extension is naturally defined as the inverse Fourier transform of the OPE coefficient in the momentum representation $Z_{F^2} C_{1}^{(F_0^2,O)}(p)$
in the sense of distributions, both in dimensional regularization and $d=4$ by taking the limit $\epsilon \rightarrow 0$:
\bea \label{FText}
Z_{F^2} C_{1}^{(F_0^2,O)}(x)= \frac{1}{(2 \pi)^{4-2\epsilon}}  \int Z_{F^2} C_{1}^{(F_0^2,O)}(p)  e^{ip\cdot x} d^{4-2\epsilon}p
\eea
that is inverted as:
\bea \label{FText1}
Z_{F^2} C_{1}^{(F_0^2,O)}(p)= \int Z_{F^2} C_{1}^{(F_0^2,O)}(x)  e^{-ip\cdot x} d^{4-2\epsilon}x
\eea
 This definition involves implicitly an interplay between the solutions of the CS equation in the momentum and coordinate representation and the corresponding contact terms.

\subsection{OPE and contact terms from the CS equation in the momentum representation}\label{sec4.1}

We start by recalling the CS equation in the coordinate representation for $2$-point correlators and OPE coefficients of multiplicatively renormalizable operators at distinct points. 
The renormalized $2$-point correlator at distinct points  -- denoted by a prime -- in $d=4$ dimensions satisfies the CS equation (appendix \ref{CS0}):
\bea
\left(z\cdot\f{\partial}{\partial z} + \beta(g)\f{\partial}{\partial g} + 2(\Delta_{O_0}+\gamma_{O}(g)) \right) \braket{O(z)O(0)}' \bigg|_{d=4}= 0
\eea
and analogously for the OPE coefficient at distinct points:
\bea
\left(x\cdot\f{\partial}{\partial x} + \beta(g)\f{\partial}{\partial g} +4+\gamma_{F^2}(g) \right) {C_{1}^{(F^2,O)}}'(x) \bigg|_{d=4}= 0
\eea
Similarly, the dimensionally regularized correlator satisfies (appendix \ref{CS1}):
\bea
\left(z\cdot\f{\partial}{\partial z} + \beta(g,\epsilon)\f{\partial}{\partial g} + 2(\tilde \Delta_{O_0}+\gamma_{O}(g)) \right) \braket{O(z)O(0)}'\bigg|_{\tilde d=4-2\epsilon}= 0
\eea
and the OPE coefficient:
\bea\label{CSepsC1x}
\left(x\cdot\f{\partial}{\partial x} + \beta(g,\epsilon)\f{\partial}{\partial g} + 4-2 \epsilon+\gamma_{F^2}(g) \right) {C_{1}^{(F^2,O)}}'(x) 
\bigg|_{\tilde d=4-2\epsilon}= 0
\eea
where $\Delta_{O_0}$, $\tilde\Delta_{O_0}=\Delta_{O_0}-\delta_O\eps$ are the canonical dimensions of $O$ in $d=4$ and $\tilde d=4-2\eps$ dimensions, $\Delta_{O}=\Delta_{O_0}+\gamma_O$, $\tilde\Delta_{O}=\tilde\Delta_{O_0}+\gamma_O$ the scaling dimensions of $O$, and $\gamma_O$ its anomalous dimension, respectively.\par
We stress that all the above equations only capture the multiplicative renormalization of the correlators and OPE coefficients at distinct points, so that their solutions cannot be extended at coinciding points, even in the dimensionally regularized case.
Indeed, to take into account contact terms, they need to be modified to include additive renormalizations, as we demonstrate momentarily.\par
The general solution 
for the $2$-point correlator at distinct points in $d=4$ dimensions is (appendix \ref{CS0}): 
\bea \label{co}
 \braket{O(z)O(0)}_{d=4}&=&
\f{\mathcal{G}^{(O)}_2(g(z))}{|z|^{2 \Delta_{O_0}}} Z^{(O)2}(g(z),g(\mu))
\eea
with $|z|=\sqrt {z^2}$, and
for the OPE coefficient:
\bea \label{co0}
{C_{1}^{(F^2,O)}}'(x)_{d=4}&=&
\f{\mathcal{G}^{(F^2,O)}(g(x))}{|x|^{4}} Z^{(F^2)}(g(x),g(\mu))\eea
Moreover, in $\tilde d =4-2\epsilon$ dimensions, we get for the $2$-point correlator (appendix \ref{CS1}):
\bea
\label{CSStilde00}
 \braket{O(z)O(0)}_{\tilde d=4-2\epsilon}&=&\f{\mathcal{ G}^{(O)}_2(\tilde g(z))}{|z|^{2 \tilde \Delta_{O_0}}}  Z^{(O)2}(\tilde g(z),g(\mu))
 \eea 
and OPE coefficient:
\bea
\label{cooreg}
{C_{1}^{(F^2,O)}}'(x)_{\tilde d=4-2\epsilon}&=&\f{\mathcal{ G}^{(F^2,O)}(\tilde g(x))}{|x|^{ \tilde \Delta_{F^2_0}}}  Z^{(F^2)}(\tilde g(x),g(\mu))
 \eea
The above solutions may be extended at coinciding points as distributions to include contact terms.
In fact, the -- possibly divergent -- contact terms are most conveniently obtained in the momentum representation as $\epsilon \rightarrow 0$, where the operators are not multiplicatively renormalizable in general.\par
To clarify this issue, we discuss the CS equations for the 2-point correlator and OPE coefficient in the momentum representation that include the aforementioned additive renormalization (appendix \ref{F}). \par
We denote the fully renormalized $2$-point correlator in the momentum representation in $\td=4-2\eps$ dimensions as the OPE coefficient of the identity $C_{0}^{(O,O)}(p)$:
\bea\label{C0OO_app0}
C_{0}^{(O,O)}(p)=Z_{O}^2 C_0^{(O_0,O_0)}(p)+ p^{2\Delta_{O_0}-4}\mu^{2(1-\delta_O)\eps}Z_{0 \textrm{c.t.}} 
\eea
It decomposes into the sum of the multiplicatively renormalized contribution $Z_{O}^2$ $C_0^{(O_0,O_0)}(p)$ and the -- in general divergent as $\epsilon \rightarrow 0$ -- additive renormalization proportional to $Z_{0 \textrm{c.t.}}$.
The corresponding CS equation in dimensional regularization reads (appendix \ref{F}):
\be \label{CSC0momreg_mu0}
\bigg(\mu \frac{\partial}{\partial \mu} +\beta(g,\epsilon) \f{\partial}{\partial g}+ 2\gamma_{O}(g)
\bigg) C_0^{(O,O)}(p,\mu,g(\mu)) = p^{2\Delta_{O_0}-4}\mu^{2(1-\delta_O)\eps}\gamma_{0 \textrm{c.t.}}(g) 
\ee
where:
\bea\label{gamma0ct0}
\gamma_{0 \textrm{c.t.}}(g) &=&\mu\f{dZ_{0 \textrm{c.t.}}}{d\mu} +2\gamma_{O}Z_{0 \textrm{c.t.}}+2(1-\delta_O)\eps Z_{0 \textrm{c.t.}} \nn\\
&=&\beta(g,\epsilon)\f{\partial Z_{0 \textrm{c.t.}}}{\partial g}+2\gamma_{O}Z_{0 \textrm{c.t.}}+2(1-\delta_O)\eps Z_{0 \textrm{c.t.}} 
\eea
Hence, the CS equation in $d=4$ dimensions reads: 
\be \label{CSC0momreg_p_4}
\bigg(\mu \frac{\partial}{\partial \mu} +\beta(g) \f{\partial}{\partial g}+ 2\gamma_{O}(g)\bigg) C_0^{(O,O)}(p,\mu,g(\mu)) = p^{2\Delta_{O_0}-4}\gamma_{0 \textrm{c.t.}}(g) 
\ee
Perturbatively, in dimensional regularization as $\epsilon \rightarrow 0$, the fully renormalized OPE coefficient $C_{1}^{(F^2,O)}(p)$ similarly decomposes into the sum of a nontrivial function of $\frac{p^2}{\mu^2}$ -- $Z_{F^2} C_1^{(F_0^2,O)}(p)$ -- that is obtained by multiplicative renormalization of the bare OPE coefficient $C_1^{(F_0^2,O)}(p)$ and a -- in general divergent as $\epsilon \rightarrow 0$ -- constant term $Z_{1 \textrm{c.t.}}$:
\bea \label{29}
C_{1}^{(F^2,O)}(p)=Z_{F^2} C_1^{(F_0^2,O)}(p)+Z_{1 \textrm{c.t.}} 
\eea
The corresponding CS equation in dimensional regularization 
 reads (appendix \ref{F}): 
\bea \label{momreg}
\bigg(\mu \frac{\partial}{\partial \mu} +\beta(g,\epsilon) \f{\partial}{\partial g}+\gamma_{F^2}(g)\bigg) C_1^{(F^2,O)}(p,\mu,g(\mu)) = \gamma_{1 \textrm{c.t.}}(g) 
\eea
with:
\bea
\gamma_{1 \textrm{c.t.}}(g) &=&\mu\f{dZ_{1 \textrm{c.t.}}}{d\mu} +\gamma_{F^2}(g)Z_{1 \textrm{c.t.}}\nn\\
&=&\beta(g,\epsilon)\f{\partial Z_{1 \textrm{c.t.}}}{\partial g}+\gamma_{F^2}(g)Z_{1 \textrm{c.t.}}
\eea
Hence, the CS equation in $d=4$ dimensions reads:
\bea
\bigg(\mu \frac{\partial}{\partial \mu} +\beta(g) \f{\partial}{\partial g}+ \gamma_{F^2}(g)\bigg) C_1^{(F^2,O)}(p,\mu,g(\mu)) =  \gamma_{1 \textrm{c.t.}}(g) 
\eea
We demonstrate by explicit computation for $O=F^2$ (section \ref{sec4.2}) that 
 -- contrary to eqs. \eqref{FText} and \eqref{FText1}, which by definition are the inverse of each other in dimensional regularization at least in the limit $\epsilon \rightarrow 0$ -- in general:
\bea \label{ineq}
Z_{F^2} C_{1}^{(F_0^2,O)}(p)  \neq \int Z_{F^2} {C_{1}^{(F_0^2,O)}}'(x) e^{-ip\cdot x} d^{4-2\epsilon}x 
\eea
as $\epsilon \rightarrow 0$, with the objects in the coordinate and momentum representation defined by means of the corresponding CS equations above. 
The interpretation of the inequality is that in general not all the divergent contact terms that occur in the momentum representation as $\epsilon \rightarrow 0$ arise from the divergence of the Fourier transform of the OPE coefficient at distinct points as $\epsilon \rightarrow 0$. \par 
In general, in order for the Fourier transform of $Z_{F^2} {C_{1}^{(F_0^2,O)}}(x)$ to reproduce $Z_{F^2} {C_{1}^{(F_0^2,O)}}(p)$, $Z_{F^2} {C_{1}^{(F_0^2,O)}}(x)$ must contain in the sense of distributions some extra divergent contact terms as $\epsilon \rightarrow 0$, $[Z_{F^2} {C_{1}^{(F_0^2,O)}}(x)-Z_{F^2} {C_{1}^{(F_0^2,O)}}'(x)]_{\textrm{div}}$, that add to the ones that arise from the Fourier transform of  $Z_{F^2} {C_{1}^{(F_0^2,O)}}'(x)$ at distinct points. 
We refer to (minus) 
the divergent contact terms of the multiplicatively renormalized OPE coefficient in the momentum representation as additive counterterms,
while we refer to the divergent contact terms that are extensions of the OPE coefficient in the coordinate representation at coinciding points $[Z_{F^2} {C_{1}^{(F_0^2,O)}}(x)-Z_{F^2} {C_{1}^{(F_0^2,O)}}'(x)]_{\textrm{div}}$ as  divergent proper contact terms.\par
To summarize, according to eq. \eqref{ineq}, in the above language the divergent proper contact terms do not vanish in general, as we demonstrate below.

\subsection{A paradigmatic example: the OPE coefficient for $O=F^2$}
\label{sec4.2}
In massless QCD the occurrence of the additive renormalization in the OPE coefficient $C_1^{(F^2,F^2)}(p)$ of $F^2$ with itself -- i.e. in the special case $O=F^2$ --
in the momentum representation has been discovered to order $g^4$ \cite{Z1} and $g^6$ \cite{Z2} in perturbation theory, and computed in a closed form in terms of the beta function and its first derivative \cite{Z3} (appendix \ref{A}), so that the  fully renormalized object reads in our notation:
\begin{equation}\label{Cf}
C_{1}^{(F^2,F^2)}(p)=Z_{F^2} C_1^{(F^2_0,F^2)}(p)+Z_{F^2} \f{2g\f{\partial}{\partial g}\left (\f{\beta(g)}{g}\right )-4 \frac{\beta(g)}{g}}{\eps} 
\end{equation}
Interestingly, the above additive counterterm is scheme independent to order $g^4$ \cite{Z1}:
\bea
Z_{F^2} \f{2g\f{\partial}{\partial g}\left (\f{\beta(g)}{g}\right )-4 \frac{\beta(g)}{g}}{\eps}= -\f{4\beta_1g^4}{\eps}+ \cdots
\eea
and must cancel a corresponding scheme-independent divergence in $Z_{F^2} C_1^{(F^2_0,F^2)}(p)$ to order $g^4$ that we report in our notation in the $\overline{MS}$ scheme \cite{Z1}:
\bea\label{C1p4}
Z_{F^2} C_1^{(F_0^2,F^2)}(p)&=&4-4 B_{1,1}g^2-4 B_{1,2}g^4+\f{4\beta_1g^4}{\eps} \nn\\
&&\hspace{-1.truecm}-4\beta_0g^2\log\f{p^2}{\mu^2}
+4\beta_0^2g^4\log^2\f{p^2}{\mu^2}+4B_{1,3}g^4\log\f{p^2}{\mu^2}
 +\cdots
\eea
with the beta-function universal coefficients:
\bea
\beta_0&=&\f{1}{(4\pi)^2}\left( \f{11}{3}-\f{2}{3}\f{N_f}{N}\right)\nn\\
\beta_1&=&\f{1}{(4\pi)^4}\left( \f{34}{3}-\f{13}{3}\f{N_f}{N}+\f{N_f}{N^3}\right)
\eea
and the scheme-dependent coefficients:
\bea\label{relB}
B_{1,1}&=&\f{1}{(4\pi)^2}\left( -\f{49}{9}+\f{10}{9}\f{N_f}{N}\right)\nn\\
B_{1,2}&=& \f{1}{(4\pi)^4}\left( -\f{11509}{81}+66\zeta_3 +(13-12\zeta_3) \f{N_f}{N^2}\f{N^2-1}{N}+\left(\f{3095}{81}+12\zeta_3\right)\f{N_f}{N}\right.
\nn\\
&&\left.-\f{100}{81}\f{N_f^2}{N^2}\right)\nn\\
B_{1,3}&=& -4\beta_1+2\beta_0B_{1,1}
\eea
where we have conveniently rewritten $B_{1,3}$ \cite{Z1} in terms of $\beta_0, \beta_1$ and $B_{1,1}$.\par
Hence, as pointed out in \cite{Z1}, it follows from eq. \eqref{C1p4} that the bare OPE coefficient $ C_1^{(F_0^2,F^2)}(p)$ in the momentum representation is not multiplicatively renormalizable because a divergent contact term $+\f{4\beta_1g^4}{\eps}$ arises in $Z_{F^2} C_1^{(F^2_0,F^2)}(p)$ to order $g^4$ that is cancelled by the additive renormalization in eq. \eqref{Cf} to order $g^4$.\par
Our aim now is to reconstruct the OPE coefficient $Z_{F^2} C_1^{(F^2_0,F^2)}(x)$ in the coordinate representation to order $g^4$
from the OPE coefficient $Z_{F^2} C_1^{(F^2_0,F^2)}(p)$ in the momentum representation according to eqs. \eqref{FText} and \eqref{FText1} as $\epsilon \rightarrow 0$. 
In doing so we verify the inequality in eq. \eqref{ineq} up to order $g^4$, thus demonstrating the occurrence of both additive counterterms and divergent proper contact terms, according to the terminology introduced above.\par
To find the OPE coefficient in the coordinate representation whose Fourier transform is $Z_{F^2} C_1^{(F_0^2,F^2)}(p)$ in eq. \eqref{C1p4} it is mandatory to start with an ansatz for the multiplicatively renormalized coefficient in the coordinate representation at distinct points:
\bea \label{C1xeps_gen}
 C_1^{(F^2,F^2)'}(x) 
=  \frac{Z^{(F^2)}(\tilde g(x), g(\mu))}{x^{4-2\eps}}  (a \tilde {g}^2(x)+ b \tilde {g}^4(x) + \cdots)
\eea 
that is solution of the CS equation in $\td=4-2\eps$ dimensions in eq. \eqref{CSepsC1x} for $O=F^2$, where to the relevant order (appendix \ref{CS1}):
\bea\label{grun}
\f{\tilde g^2(x)}{g^2(\mu)}&=&\f{|x\mu|^{2\eps}}{1-\beta_0 g^2(\mu)\f{|x\mu|^{2\eps}-1}{\eps}}
\eea
and:
\bea\label{Zrun}
Z^{(F^2)}(\tilde g(x), g(\mu))&=&\f{\tilde g^2(x)}{g^2(\mu)} |x\mu|^{-2\eps}
\eea
By inserting eqs. \eqref{grun} and \eqref{Zrun} in eq. \eqref{C1xeps_gen} and setting $g(\mu)=g$, the matching of the logarithmic terms to order $g^2$ in eq. \eqref{C1p4} fixes $a=4\beta_0/\pi^2$ and $b=-4\tilde{B}_{1,3}/\pi^2$ with $\tilde{B}_{1,3}$ to be determined a posteriori. Hence, we get:
\bea \label{C1xeps_exp}
&&C_1^{(F^2,F^2)'}(x) 
=  \f{4\beta_0}{\pi^2}\f{\mu^{2\eps}g^2}{x^{4-4\eps}} \bigg(1-\beta_0 g^2\f{|x\mu|^{2\eps}-1}{\eps}\bigg)^{-2}
\nn\\&&\hspace{2.5truecm}
-\f{4\tilde{B}_{1,3}}{\pi^2}\f{\mu^{4\eps}g^4}{x^{4-6\eps}} \bigg(1-\beta_0 g^2\f{|x\mu|^{2\eps}-1}{\eps}\bigg)^{-3}+\cdots\nn\\
&&\hspace{1.0truecm}
=\f{4\beta_0}{\pi^2}
\f{\mu^{2\eps}g^2}{x^{4-4\eps}}\bigg(1+2\beta_0 g^2 \f{|x\mu|^{2\eps}-1}{\eps}\bigg) 
-\f{4\tilde{B}_{1,3}}{\pi^2}
\f{\mu^{4\eps}g^4}{x^{4-6\eps}} +\cdots\nn\\
&&\hspace{1.0truecm}
=\f{4\beta_0}{\pi^2}
\f{\mu^{2\eps}g^2}{x^{4-4\eps}}\bigg(1-\f{2\beta_0 g^2}{\eps}\bigg) 
+\f{8\beta_0^2}{ \pi^2\,\eps}\f{\mu^{4\eps}g^4}{ x^{4-6\eps}}
-\f{4\tilde{B}_{1,3}}{\pi^2}
\f{\mu^{4\eps}g^4}{x^{4-6\eps}} +\cdots
\eea 
where the second equality is the perturbative expansion for small $g$ with $\eps$ fixed and the dots stand for order $g^6$ contributions. Its Fourier transform (FT) is: 
\bea\label{FTC1xeps}
&&{\mbox{FT}}\left[  C_1^{(F^2,F^2)'}(x) \right]= 
\f{4\beta_0g^2}{\pi^2}
\bigg( 1-\f{2\beta_0 g^2}{\eps}\bigg) 
{\mbox{FT}}\left[\f{\mu^{2\eps}}{x^{4-4\eps}} \right]
\nn\\&&\hspace{3.5truecm}
+\f{4g^4}{\pi^2}\bigg(
\f{2\beta_0^2}{\eps}-\tilde{B}_{1,3}\bigg)
{\mbox{FT}}\left[\f{\mu^{4\eps}}{x^{4-6\eps}} \right] + \cdots\nn\\
&&=
4\beta_0g^2\bigg(\f{1}{\eps}-\log\f{p^2}{\mu^2}+2+3\Gamma'(1)-\log\f{\pi}{4}\bigg)
+8\beta_0^2g^4\bigg\{
-\f{1}{2\eps^2}+\f{1}{2}\log^2\f{p^2}{\mu^2}\nn\\
&&~~-\Big(\f{1}{\eps}-\log\f{p^2}{\mu^2}\Big)\Big(2+3\Gamma'(1)-\log\f{\pi}{4}\Big)
+\Big(\f{1}{\eps}-2\log\f{p^2}{\mu^2}\Big)\Big(\f{3}{2}+\f{5}{2}\Gamma'(1)-\log\f{\sqrt{\pi}}{4}\Big)\nn\\
&&~~-\f{1}{2}\Big(2+3\Gamma'(1)-\log\f{\pi}{4}\Big)^2
+\Big(\f{3}{2}+\f{5}{2}\Gamma'(1)-\log\f{\sqrt{\pi}}{4} \Big)^2
+\f{1}{4}-\f{1}{4}\Gamma^{'2}(1)+\f{1}{4}\Gamma''(1)
\bigg\}\nn\\
&&~~-4\tilde B_{1,3} g^4\bigg(\f{1}{2\eps}-\log\f{p^2}{\mu^2}
+\f{3}{2}+\f{5}{2}\Gamma'(1)-\log\f{\sqrt{\pi}}{4}\bigg)
+\cdots
\eea
where in the second equality we have employed the FT in $d$ dimensions:
\bea
{\mbox{FT}}\left[\f{\mu^{d-2\Delta}}{x^{2\Delta}} \right]&\equiv& 
\int d^{d}x\, \f{\mu^{d-2\Delta}}{x^{2\Delta}}e^{-ipx} = 
\pi^{{d}/{2}}\f{\Gamma({d}/{2}-\Delta)}{\Gamma(\Delta)}\left(\f{p^2}{4\mu^2}\right)^{\Delta-{d}/{2}}
\eea
that for $d =4-2\eps$ and perturbatively in $\eps$ gives us for  $\Delta = 2-2\eps$:
\bea\label{FT4}
&&{\mbox{FT}}\left[\f{\mu^{2\eps}}{x^{4-4\eps}} \right]= \pi^2\bigg(\f{\pi p^2}{4\mu^2}\bigg)^{-\eps}\f{\Gamma(\eps)}{\Gamma(2-2\eps)}\nn\\
&&= \pi^2\bigg(\f{\pi p^2}{4\mu^2}\bigg)^{-\eps}\bigg(\f{1}{\eps}+2+3\Gamma'(1)
+\eps \Big(4+6\Gamma'(1)(1+\Gamma'(1))-\f{3}{2}\Gamma''(1)\Big)+\cdots\bigg)\nn\\
&&=\pi^2\bigg(\f{1}{\eps}-\log\f{\pi p^2}{4\mu^2}+2+3\Gamma'(1)
+\eps \bigg(\f{1}{2}\Big(\log\f{\pi p^2}{4\mu^2}-2-3\Gamma'(1)\Big)^2\nn\\
&&~~+2+\f{3}{2}\Gamma^{'2}(1)-\f{3}{2}\Gamma''(1)\bigg)+\cdots\bigg)
\eea
and for $\Delta = 2-3\eps$:
\bea\label{FT6}
&&{\mbox{FT}}\left[\f{\mu^{4\eps}}{x^{4-6\eps}} \right]= 
\pi^2\bigg(\f{\sqrt{\pi} p^2}{4\mu^2}\bigg)^{-2\eps}\f{\Gamma(2\eps)}{\Gamma(2-3\eps)}\nn\\
&&=\pi^2\bigg(\f{\sqrt{\pi} p^2}{4\mu^2}\bigg)^{-2\eps}
\bigg( \f{1}{2\eps}+\f{3}{2}+\f{5}{2}\Gamma'(1)
+\f{\eps}{2}\Big(9+15\Gamma'(1)(1+\Gamma'(1))-\f{5}{2}\Gamma''(1)\Big)+\cdots\bigg)
\nn\\
&&=\pi^2\bigg( \f{1}{2\eps}-\log\f{\sqrt{\pi} p^2}{4\mu^2}
+\f{3}{2}+\f{5}{2}\Gamma'(1)
+\eps\bigg(  \Big(\log\f{\sqrt{\pi} p^2}{4\mu^2} -\f{3}{2}-\f{5}{2}\Gamma'(1)\Big)^2\nn\\
&&~~+\f{9}{4}+\f{5}{4}\Gamma^{'2}(1)-\f{5}{4}\Gamma''(1)
\bigg)+\cdots
\bigg)
\eea 
Setting in eq. \eqref{FTC1xeps}:
\be
\tilde B_{1,3}=B_{1,3} -2\beta_0^2\Big(2+3\Gamma'(1)-\log\f{\pi}{4}\Big)
+4\beta_0^2\Big(\f{3}{2}+\f{5}{2}\Gamma'(1)-\log\f{\sqrt{\pi}}{4}\Big)
\ee
we obtain:
\bea\label{FTprime}
&&{\mbox{FT}}\left[  C_1^{(F^2,F^2)'}(x) \right]=
 4\beta_0{g}^2\left( \f{1}{\eps}-\log\f{ p^2}{\mu^2}+2+3\Gamma'(1)-\log\f{\pi}{4}\right)
\nn\\
&&~~~~+8\beta_0^2{g}^4
\bigg\{ -\f{1}{2\eps^2}+\f{1}{2}\log^2\f{p^2}{\mu^2}
-\f{1}{2\eps}\Big(2+3\Gamma'(1)-\log\f{\pi}{4}\Big)\nn\\
&&~~~~-\f{1}{2}\Big(2+3\Gamma'(1)-\log\f{\pi}{4} \Big)^2 
-\Big(\f{3}{2}+\f{5}{2}\Gamma'(1)-\log\f{\sqrt{\pi}}{4} \Big)^2\nn\\
&&~~~~+\Big(2+3\Gamma'(1)-\log\f{\pi}{4} \Big)\Big(\f{3}{2}+\f{5}{2}\Gamma'(1)-\log\f{\sqrt{\pi}}{4} \Big)
+\f{1}{4}-\f{1}{4}\Gamma^{'2}(1)+\f{1}{4}\Gamma''(1)\bigg\}
\nn\\
&&~~~~-4B_{1,3} g^4 \bigg(\f{1}{2\eps} - \log\f{ p^2}{\mu^2}+\f{3}{2}+\f{5}{2}\Gamma'(1)-\log\f{\sqrt{\pi}}{4} \bigg)
+\cdots\nn\\
&&=4\beta_0{g}^2\left( \f{1}{\eps}-\log\f{ p^2}{\mu^2}\right)+8\beta_0^2{g}^4
\bigg( -\f{1}{2\eps^2}+\f{1}{2}\log^2\f{p^2}{\mu^2}\bigg)+4B_{1,3} g^4\log\f{ p^2}{\mu^2}+\f{8\beta_1g^4}{\eps}\nn\\
&&~~~~-\f{4\beta_0g^4}{\eps}\bigg(B_{1,1}+\beta_0\Big(2+3\Gamma'(1)-\log\f{\pi}{4} \Big)\bigg)+\textrm{finite contact terms}+\cdots
\eea
where in the last equality we have employed $B_{1,3}=-4\beta_1+2\beta_0 B_{1,1}$ in eq. \eqref{relB} to rewrite the order $g^4/\eps$ terms, and:
\bea\label{finitenolog}
&&\textrm{finite contact terms}=
 4\beta_0{g}^2\left( 2+3\Gamma'(1)-\log\f{\pi}{4}\right)
\nn\\
&&~~~~+8\beta_0^2{g}^4
\bigg\{ 
-\f{1}{2}\Big(2+3\Gamma'(1)-\log\f{\pi}{4} \Big)^2 
-\Big(\f{3}{2}+\f{5}{2}\Gamma'(1)-\log\f{\sqrt{\pi}}{4} \Big)^2\nn\\
&&~~~~+\Big(2+3\Gamma'(1)-\log\f{\pi}{4} \Big)\Big(\f{3}{2}+\f{5}{2}\Gamma'(1)-\log\f{\sqrt{\pi}}{4} \Big)
+\f{1}{4}-\f{1}{4}\Gamma^{'2}(1)+\f{1}{4}\Gamma''(1)\bigg\}
\nn\\
&&~~~~-4B_{1,3} g^4 \bigg(\f{3}{2}+\f{5}{2}\Gamma'(1)-\log\f{\sqrt{\pi}}{4} \bigg)
\eea
From eq. \eqref{FTprime} it is clear that the Fourier transform of the OPE coefficient at distinct points does not reproduce the object in eq. \eqref{C1p4}, thus proving the inequality in eq. \eqref{ineq} up to order $g^4$. Specifically, it differs from it by finite and divergent proper contact terms.\par
To further proceed we define an extension of the OPE coefficient including the finite and divergent proper contact terms according to the ansatz: 
\bea\label{tottildeFT10}
C_1^{(F^2,F^2)}(x)  &=& C_1^{(F^2,F^2)'}(x) 
\nn\\&&
+\delta^{(\td)}(x)\bigg[Z_{F^2}\bigg(4 -4\tilde B_{1,1} {g}^2-4\tilde B_{1,2} {g}^4
+\cdots\bigg)\bigg]_{\textrm{up to order } g^4} 
\eea
The corresponding Fourier transform reads:
\bea\label{tottildeFT1}
{\mbox{FT}}\left[  C_1^{(F^2,F^2)}(x) \right] &=& {\mbox{FT}}\left[ C_1^{(F^2,F^2)'}(x) \right] 
\nn\\&&
+\bigg[Z_{F^2}\bigg(4 -4\tilde B_{1,1}{g}^2-4\tilde B_{1,2}{g}^4
+\cdots\bigg)\bigg]_{\textrm{up to order } g^4} 
\eea
where:
\be\label{B11til}
\tilde B_{1,1}=B_{1,1}+\beta_0\Big(2+3\Gamma'(1)-\log\f{\pi}{4} \Big)
\ee
$\tilde B_{1,2}$ is determined a posteriori to match the finite contributions to order $g^4$ in eq. \eqref{C1p4}, and (appendix \ref{C}):
\bea\label{Zp0}
Z_{F^2}(g,\epsilon)&=& 1-\f{\beta_0g^2}{\epsilon}-
\f{\beta_1g^4}{\epsilon} + \f{\beta_0^2 g^4}{\epsilon^2} +\cdots
\eea
The ansatz above is obtained multiplying by $Z_{F^2}$ the finite proper contact terms up to order $g^4$, thus obtaining:
\bea\label{Zctans}
&&\bigg[Z_{F^2}\bigg(4 -4\tilde B_{1,1}{g}^2-4\tilde B_{1,2}{g}^4
+\cdots\bigg)
\bigg]_{\textrm{up to order } g^4} =\nn\\
&&\bigg[\bigg(1-\f{\beta_0 g^2}{\epsilon}-
\f{\beta_1 g^4}{\epsilon} + \f{\beta_0^2 g^4}{\epsilon^2} +\cdots\bigg) 
\bigg(4 -4\tilde B_{1,1}{g}^2-4\tilde B_{1,2}{g}^4+\cdots\bigg)
\bigg]_{\textrm{up to order } g^4} \nn\\
&&=4-4\tilde B_{1,1}{g}^2-4\tilde B_{1,2}{g}^4
-4\f{\beta_0g^2}{\eps}+ 4\f{\beta_0^2 g^4}{\epsilon^2}
-4\f{\beta_1 g^4}{\epsilon} 
+4\tilde B_{1,1}\f{\beta_0 g^4}{\epsilon}
 +\cdots
\eea
Hence, by means of  eqs. \eqref{FTprime}, \eqref{B11til} and \eqref{Zctans},
eq. \eqref{tottildeFT1} reads:
\bea\label{FTtotalx}
&&{\mbox{FT}}\left[ C_1^{(F^2,F^2)}(x) \right]=
4\beta_0{g}^2\left( \f{1}{\eps}-\log\f{ p^2}{\mu^2}\right)+8\beta_0^2{g}^4
\bigg( -\f{1}{2\eps^2}+\f{1}{2}\log^2\f{p^2}{\mu^2}\bigg)\nn\\
&&~~~~+4B_{1,3} g^4\log\f{ p^2}{\mu^2}+\f{8\beta_1g^4}{\eps}
-4\tilde B_{1,1} \f{\beta_0g^4}{\eps}
+\textrm{finite contact terms}\nn\\
&&~~~~
+4-4\tilde B_{1,1}{g}^2-4\tilde B_{1,2}{g}^4
-4\f{\beta_0g^2}{\eps}+ 4\f{\beta_0^2 g^4}{\epsilon^2}
-4\f{\beta_1 g^4}{\epsilon} 
+4\tilde B_{1,1}\f{\beta_0 g^4}{\epsilon}
 +\cdots\nn\\
&&=4-4\tilde B_{1,1}{g}^2-4 \tilde B_{1,2}{g}^4+\textrm{finite contact terms}\nn\\
&&~~~~-4\beta_0{g}^2\log\f{ p^2}{\mu^2}
+4\beta_0^2{g}^4\log^2\f{p^2}{\mu^2} +4B_{1,3} g^4 \log\f{p^2}{\mu^2} 
+\f{4\beta_1 g^4}{\eps}+\cdots\nn\\
&&=4-4B_{1,1}{g}^2-4 B_{1,2}{g}^4+\f{4\beta_1 g^4}{\eps}\nn\\
&&~~~~-4\beta_0{g}^2\log\f{ p^2}{\mu^2}
+4\beta_0^2{g}^4\log^2\f{p^2}{\mu^2} +4B_{1,3} g^4 \log\f{p^2}{\mu^2} 
+\cdots
\eea
where in the last equality we have employed eq. \eqref{B11til} and:
\bea
\tilde B_{1,2}&=&B_{1,2} -B_{1,3}\Big(\f{3}{2}+\f{5}{2}\Gamma'(1)-\log\f{\sqrt{\pi}}{4} \Big)
+2\beta_0^2\bigg\{
-\f{1}{2}\Big(2+3\Gamma'(1)-\log\f{\pi}{4} \Big)^2 \nn\\
&&-\Big(\f{3}{2}+\f{5}{2}\Gamma'(1)-\log\f{\sqrt{\pi}}{4} \Big)^2
+\Big(2+3\Gamma'(1)-\log\f{\pi}{4} \Big)\Big(\f{3}{2}+\f{5}{2}\Gamma'(1)-\log\f{\sqrt{\pi}}{4} \Big)\nn\\
&&+\f{1}{4}-\f{1}{4}\Gamma^{'2}(1)+\f{1}{4}\Gamma''(1)
\bigg\}
\eea
The result in eq. \eqref{FTtotalx} reproduces eq. \eqref{C1p4} to order $g^4$.
We have thus verified the inequality in eq. \eqref{ineq} for $O=F^2$ up to order $g^4$ in perturbation theory. Indeed, 
the object in the coordinate representation whose Fourier transform reproduces the multiplicatively renormalized coefficient $Z_{F^2} C_1^{(F_0^2,F^2)}(p)$ in eq. \eqref{C1p4} is provided by eq. \eqref{tottildeFT10}
as $\eps\to 0$, with $C_1^{(F^2,F^2)'}(x)$ in eq. \eqref{C1xeps_exp}. 
It differs from the multiplicatively renormalized $ C_1^{(F^2,F^2)'}(x)$ at distinct points by divergent and finite proper contact terms. 
Specifically, to order $g^2$ the multiplicatively renormalized $C^{(F^2,F^2)}_1(p)$ is finite, so that no additive counterterm occurs, while the Fourier transform of the multiplicatively renormalized $C^{(F^2,F^2)'}_1(x)$ at distinct points is divergent, thus implying the occurrence of divergent proper contact terms in $C^{(F^2,F^2)}_1(x)$ to cancel the latter divergence. 
To order $g^4$ the multiplicatively renormalized $C^{(F^2,F^2)}_1(p)$ is divergent and the additive renormalization occurs, but the divergence -- as to order $g^2$ -- does not entirely arise from the Fourier transform of $C^{(F^2,F^2)'}_1(x)$.

\subsection{Contact terms in the OPE from the LET}

 Having clarified the nature of contact terms in the Fourier transform of the OPE coefficient of $Z_{F^2}F^2_0(x)$ with $O(0)$, we demonstrate the relation between its divergences and the ones of the integrated $3$-point correlator in the rhs of the LET according to eq. \eqref{eq:ct00}.
Obviously, it suffices to consider the OPE of $F^2(x)$ with $O(0)$, the OPE with $O(z)$ being entirely analogous:
\be\label{eq:OPEb01}
Z_{F^2} F_0^2(x) O(0) =   \cdots + Z_{F^2} C_{1}^{(F^2_0,O)}(x)O(0) + \cdots
\ee
In general, for dimensional reasons the above OPE coefficient is the only one that may contribute in the integrated $3$-point correlator the product of a dimension $4$ -- possibly divergent -- contact term times the $2$-point correlator at distinct points $\langle O(z) O(0) \rangle$ that has to be cancelled by the counterterm with the same structure in the rhs of the LET, both in its perturbative and RG-improved version. The argument is as follows. \par
 It is convenient to consider at the same time the theory in $d=4$ dimensions and its dimensionally regularized version in $\tilde d=4-2 \epsilon$ dimensions.
The theory in $d=4$ dimensions is exactly conformal perturbatively to order $g^2$ in a conformal scheme \cite{ConfQCD}, which may differ by a finite renormalization from a generic scheme, because the beta function $\beta(g)=-\beta_0 g^3 +\cdots$ affects the solution of the CS equation only starting from order $g^4$. 
The theory in $\tilde d=4-2 \epsilon$ dimensions is exactly conformal perturbatively to order $g^0$, because the beta function $\beta(g,\epsilon)=- \epsilon g +\cdots$ affects the solution of the CS equation only starting from order $g^2$. \par
 It follows that up to the perturbative order where the theory is conformal, under the standard assumption that $O$ belongs to a basis of mutually orthogonal conformal primary operators, along with the term in the OPE in eq. \eqref{eq:OPEb01}
only the terms in the OPE of $Z_{F^2}F_0^2(x)$ with the descendants of $O(0)$ may a priori also contribute to the rhs of the LET, once inserted in the v.e.v. with $O(z)$. Indeed, all the $2$-point correlators of $O(z)$ with different primary operators and their descendants vanish identically, given the orthogonality of the basis.
Such a basis of Hermitian primary operators certainly exists, since the $2$-point correlators of primary operators with different conformal dimensions vanish by the conformal symmetry, and the $2$-point correlators of operators with the same conformal dimension are a real symmetric matrix that can always be diagonalized -- times a universal conformal structure. \par
In fact, the OPE of $Z_{F^2}F_0^2(x)$ with the descendants of $O(0)$ cannot contribute to the divergence of the integrated $3$-point correlator: The corresponding OPE coefficients have conformal dimensions lower than $4$ -- their integrals being therefore UV finite -- and the insertion of the descendants gives rise to derivatives of the $2$-point correlator, instead of the correlator itself. \par
A similar argument holds to higher orders in perturbation theory and nonperturbatively, both in $d=4$ and $\tilde d=4-2 \epsilon$ dimensions, in AF massless QCD.
In both cases, for dimensional reasons, only operators $O'$ with the same canonical dimension of $O$ may a priori contribute to the relevant OPE. Yet, their insertion gives rise to $2$-point correlators $\braket{O(z)O'(0)}$ that -- by the above conformal argument, because of the orthogonality of $O$ and $O'$ in the conformal limit -- are necessarily suppressed by powers of the coupling  $g^{2n}(\mu)$ perturbatively, with $n$ a positive integer.
To actually rule out their contribution to the divergences in the rhs of the LET in the AF case, we show that $\braket{O(z)O(0)}$ is in fact linearly independent of $\braket{O(z)O'(0)}$ with coefficients valued in the renormalized coupling $g(\mu)$, so that no linear combination of $\braket{O(z)O'(0)}$ may occur in the divergent part of the integrated $3$-point correlator in the rhs of the LET. To this aim, it suffices to observe that in $d=4$ dimensions in general
the asymptotics of $\braket{O(z)O'(0)}$ as $z \rightarrow 0$:
\bea \label{co'}
 \braket{O(z)O'(0)}_{d=4}&=&
\f{\mathcal{G}^{(O,O')}_2(g(z))}{|z|^{2\Delta_{O_0}}} Z^{(O)}(g(z),g(\mu))Z^{(O')}(g(z),g(\mu))\nn\\
&\sim& \f{g^{2n}(z)}{|z|^{2\Delta_{O_0}}} g^{\f{\gamma_0^{(O)}+\gamma_0^{(O')}}{\beta_0}}(z)
\eea
is linearly independent of the one of $\braket{O(z)O(0)}_{d=4}$:
\bea \label{co1}
 \braket{O(z)O(0)}_{d=4}&=&
\f{\mathcal{G}^{(O)}_2(g(z))}{|z|^{2\Delta_{O_0}}} Z^{(O)2}(g(z),g(\mu))\nn\\
&\sim&\f{1}{|z|^{2\Delta_{O_0}}} g^{\f{2\gamma_0^{(O)}}{\beta_0}}(z)
\eea
 and analogously in $\tilde d=4-2 \epsilon$ dimensions (appendix \ref{CS1}).
The only exception may occur for $\gamma_0^{(O')}+2n\beta_0=\gamma_0^{(O)}$ that, in the terminology of \cite{MB00,BB00}, is the resonant condition for the operator mixing of $O$ with $O'$, necessary for the nonexistence of a renormalization scheme where $O,O'$ are multiplicatively renormalizable.\par
We conclude that, if the nonresonant condition $\gamma_0^{(O')}+2n\beta_0 \neq \gamma_0^{(O)}$ is satisfied, the OPE  coefficient of $Z_{F^2}F_0^2$ with $O'$, the latter having the same canonical dimension of $O$, does not contribute  to the divergent part of the integrated $3$-point correlator in the rhs of the LET for a given $2$-point correlator of $O$ in the lhs. \par
However, we should stress that the nonresonant condition is sufficient but by no means necessary.
First, even if the UV asymptotics of $\braket{O(z)O(0)}$ and $\braket{O(z)O'(0)}$ coincide, the actual $2$-point correlators may still be functional independent to higher orders in the running coupling.
Second, even if the resonant condition is realized, the resulting correlators may still be functional independent for different reasons. For example, for $O=F^2$ in massless QCD, despite the resonant condition with $n=1$ is satisfied for $O'=\bar \psi \gamma_{\mu} D_{\mu} \psi$, the correlator $ \braket{F^2(z) \bar \psi \gamma_{\mu} D_{\mu} \psi(0)}$ at distinct points vanishes because of the EOM both in $d=4$ and $\tilde d=4-2 \epsilon$ dimensions.\par
Hence, under the above assumptions, we can relate the divergent contact terms that arise in the integrated 3-point correlator in the rhs of the LET as $\epsilon \rightarrow 0$ to those of the corresponding multiplicatively renormalized OPE coefficient in the momentum representation:
\be\label{eq:ct001}
\frac{1}{2} [Z_{F^2} \int \braket{O(z)O(0)F_0^2(x)} d^\td x]_{\textrm{div}}= [Z_{F^2} C_{1}^{(F_0^2,O)}(p)]_{\textrm{div}} \braket{O(z)O(0)}
\ee
where the Fourier transform in the rhs may be conveniently computed at nonvanishing momentum $p$ to avoid infrared divergences in perturbation theory, without affecting the divergent contact terms:
\bea \label{eq:ct0012}
[Z_{F^2} C_{1}^{(F_0^2,O)}(p)]_{\textrm{div}}= [Z_{F^2} C_{1}^{(F_0^2,O)}(0)]_{\textrm{div}}
 \eea
because they are the Fourier transform of a $\delta^{(4)}$ -- or a $\delta^{(\tilde d)}$ in $\tilde d=4-2 \epsilon$ dimensions (section \ref{3.7}) -- thus a momentum-independent constant. 
 We should notice that in perturbation theory there is a finite ambiguity in eq. \eqref{eq:ct001}, due to the fact that the lhs is computed in the coordinate representation and the rhs in the momentum representation. \par
 Finally, once the lowest order finite contact term $2 c_{O} \delta^{(4)}$ is determined consistently with perturbation theory according to (IIA),  under the above assumptions,
the LET unambiguously predicts the fully renormalized OPE coefficient in the momentum representation: 
\begin{equation} \label{CT}
C_{1}^{(F^2,O)}(p)=Z_{F^2} C_{1}^{(F_0^2,O)}(p)+Z_{F^2} \f{2\gamma_{O}(g)-2c_O \frac{\beta(g)}{g}}{\eps}
\end{equation}
in terms of the multiplicatively renormalized one and the  divergent contact term occurring in the rhs of the LET. Indeed, given that $\gamma_{F^2}(g)=g\f{\partial}{\partial g}\left (\f{\beta(g)}{g}\right)$ (appendix \ref{C}), the contact term above coincides for $O=F^2$ with the one computed in eq. \eqref{Cf} \cite{Z3}.

\section{Perturbative versus nonperturbative LET} \label{5}

\subsection{Perturbative LET}

 (IIA) shows that in general the $3$-point correlator with the insertion of $Z_{F^2}F_0^2(x)$ at zero momentum needs -- order by order in perturbation theory -- an infinite additive renormalization as $\epsilon \rightarrow 0$ to yield the finite fully renormalized object defined by the lhs:
 \bea\label{5.1b}
&&\f{\partial \braket{O(z)O(0)}}{\partial\log g} + 2c_O \braket{O(z)O(0)}\nn\\
&&=\f{1}{2}Z_{F^2} \int \braket{O(z) O(0) F_0^2(x)} - \braket{O(z)O(0)}\braket{F_0^2(x)} d^{\td}x \nn\\
&&\,\,\,\,\,\,+Z_{F^2} \f{2\gamma_O(g)-2c_O \f{\beta(g)}{g}}{\eps} \braket{O(z)O(0)}
\eea
Indeed, we have already compared the additive  counterterm in (IIA) to order $g^4$ in perturbation theory for $O=F^2$ with the corresponding  counterterm in the OPE coefficient $C_1^{(F^2,F^2)}(p)$ finding perfect agreement (section \ref{4}). Besides, in doing so we have discovered the existence of divergent proper contact terms in the coordinate representation. 

\subsection{Nonperturbative LET}

We take the limit $\epsilon \rightarrow 0$ of the LET in eq. \eqref{5.1b} in the AF phase:
 \bea \label{LET2A11}
&&\f{\partial \braket{O(z)O(0)}}{\partial\log g} + 2c_O \braket{O(z)O(0)}\nn\\
&&=\f{1}{2} \int \braket{O(z) O(0) F^2(x)} - \braket{O(z)O(0)}\braket{F^2(x)} d^4x \nn\\
&&\,\,\,\,\,\,- \f{2\gamma_O(g)-2c_O \f{\beta(g)}{g}}{\f{\beta(g)}{g}} \braket{O(z)O(0)}
\eea
where we have employed:
\bea
Z_{F^2} \f{2\gamma_O(g)-2c_O \f{\beta(g)}{g}}{\eps}=\f{2\gamma_O(g)-2c_O \f{\beta(g)}{g}}{\eps(1-\f{\beta(g)}{\eps g})}
\eea 
Hence, eq. \eqref{LET2A11} reads identically:
 \bea \label{LET2A12}
&&\f{\partial \braket{O(z)O(0)}}{\partial\log g} \nn\\
&&=\f{1}{2} \int \braket{O(z) O(0) F^2(x)} - \braket{O(z)O(0)}\braket{F^2(x)} d^4x \nn\\
&&\,\,\,\,\,\,- \f{2\gamma_O(g)}{\f{\beta(g)}{g}} \braket{O(z)O(0)}
\eea
Somehow unexpectedly, no infinite additive renormalization occurs nonperturbatively in $d=4$ dimensions in the AF phase, since in the last term no zero of the beta function arises  in the denominator but the one at $g=0$ that is cancelled by the zero of the same order due to the anomalous dimension $\gamma_O(g)$ in the numerator. \par
Another important difference with the perturbative case is that in the nonperturbative AF case an infinite additive renormalization cannot  either arise from the integrated OPE coefficient at distinct points, since the latter turns out to be UV finite because of the asymptotic freedom for small $x^2$ \cite{MBL}:
\bea
\int {C_1^{(F^2,O)}}'(x)  e^{-ip\cdot x} d^4x  \sim \int \frac{g^2(x)}{g^2(\mu)} \frac{g^2(x)}{x^4} e^{-ip\cdot x} d^4x < +\infty
\eea
where the first factor in the last integrand arises from the anomalous dimension of $F^2$ and the second one from the fact that the first contribution to the OPE coefficient at distinct points occurs to order $g^2$ in perturbation theory. As a consequence, no divergent proper contact term occurs 
nonperturbatively in the AF phase.

\section{Solving for the perturbative correlators to order $g^2$ by the LET} \label{6}

We find an a-priori solution of the LET for the correlators  up to order $g^2$ in perturbation theory, where a massless QCD-like theory is conformal invariant in $d=4$ dimensions.
Though LET (IC) and (IIB) are strictly equivalent, for technical reasons it is interesting to compute to order $g^2$ both versions: (IC) is further employed to verify (section \ref{7.1}) the corresponding version that involves a hard-cutoff regularization \cite{MBL,BB}, while (IIB) is suitable to match the finite and divergent proper contact terms in the OPE coefficient $C^{(F^2,O)}_1(x)$ up to order $g^2$ in perturbation theory.

\subsection{LET (IC) and (IIB) to order $g^2$}\label{sec:2IC_sdr}

We find the common solution of (IC): 
\bea \label{LET1CP}
&&\f{\partial\braket{O(z)O(0)}}{\partial\log g} \nn\\
&&=\f{1}{2}Z_{F^2} \int \braket{O(z) O(0) F_0^2(x)} - 2 c_O\big(\delta^{(\td)}(x-z)+\delta^{(\td)}(x)\big)\braket{O(z)O(0)} d^{\td}x \nn\\
&&\,\,\,\,\,\,+Z_{F^2} \f{2\gamma_O(g)}{\eps} \braket{O(z)O(0)}
\eea
and (IIB):
\bea \label{LET2BP}
&&\f{\partial\braket{O(z)O(0)}}{\partial\log g}+ 2c_O \braket{O(z)O(0)} \nn\\
&&=\f{1}{2} Z_{F^2}\int \braket{O(z) O(0) F_0^2(x)} \nn\\
&&\,\,\,\,\,\,+ Z_{F^2}\f{2\gamma_O(g)-2c_O \frac{\beta(g)}{g}}{\eps} \big(\delta^{(\td)}(x-z)+\delta^{(\td)}(x)\big)\braket{O(z)O(0)} d^{\td}x 
\eea
up to order $g^2$, where we have set $z \neq 0$ once for all, and $\braket{F_0^2(x)}=0$ perturbatively in dimensional regularization. \par
The lhs of the LET is computed  from the exact solution of the CS equation at distinct points in dimensional regularization by means of (appendix \ref{D.6}):
\bea
\f{\partial\braket{O(z)O(0)}_{\tilde d=4-2\epsilon}}{\partial\log g} = \braket{O(z)O(0)}_{\tilde d=4-2\epsilon} \f{\partial\log \braket{O(z)O(0)}_{\tilde d=4-2\epsilon}}{\partial\log g}
\eea
 expanded up to order $g^2$. We obtain:
\bea
 \f{\partial\log \braket{O(z)O(0)}_{\tilde d=4-2\epsilon}}{\partial\log g}\bigg|_{\textrm{up to order } g^2}\hspace{-0.3truecm}= \f{\partial\log N_2(g)}{\partial\log g}-2g\f{\partial\gamma_O}{\partial g} \log{|z\mu|} +\cdots    \bigg|_{\textrm{up to order } g^2}
 \eea
that coincides, up to terms that vanish as $\epsilon \rightarrow 0$, with the corresponding object computed by means of the conformal renormalized 2-point correlator in $d=4$ dimensions to order $g^2$:
\be\label{sec2:N2}
\braket{O(z)O(0)}_{d=4}=\f{N_2(g)\mu^{2\Delta_{O_0}-2\Delta_O}}{|z|^{2\Delta_O}}
\ee
Hence, (IC) reads: 
\bea \label{PLET1C}
&&\braket{O(z)O(0)}\left( 
 \f{\partial\log N_2(g)}{\partial\log g}-2g\f{\partial\gamma_O}{\partial g} \log{|z\mu|} +\cdots
\right)   \bigg|_{\textrm{up to order } g^2}
 \nn\\
&&=\f{1}{2}Z_{F^2} \int \braket{O(z) O(0) F_0^2(x)} - 2 c_O\big(\delta^{(\td)}(x-z)+\delta^{(\td)}(x)\big)\braket{O(z)O(0)} d^{\td}x \nn\\
&&\,\,\,\,\,\,+Z_{F^2} \f{2\gamma_O(g)}{\eps} \braket{O(z)O(0)}\,\bigg|_{\textrm{up to order } g^2}
\eea
and (IIB) reads: 
\bea \label{PLET2B}
&&\braket{O(z)O(0)}\left( 2c_O+
 \f{\partial\log N_2(g)}{\partial\log g}-2g\f{\partial\gamma_O}{\partial g} \log{|z\mu|} +\cdots
\right)   \bigg|_{\textrm{up to order } g^2}
 \nn\\
&&=\f{1}{2} Z_{F^2}\int \braket{O(z) O(0) F_0^2(x)} \nn\\
&&\,\,\,\,\,\, +Z_{F^2}\f{2\gamma_O(g)-2c_O \frac{\beta(g)}{g}}{\eps} \big(\delta^{(\td)}(x-z)+\delta^{(\td)}(x)\big)\braket{O(z)O(0)} d^{\td}x \,\bigg|_{\textrm{up to order } g^2}
\eea
Because of the structure of (IC) the following ansatz is natural for the bare $3$-point correlator  extended at coinciding points:
\bea\label{ICbarecorr_sdr}
\braket{O(z) O(0) F_0^2(x)}_{\tilde d=4-2\epsilon}\bigg|_{\textrm{up to order } g^2} \hspace{-0.5cm}&=&(2 c_O + \tilde B g^2)(\delta^{(\td)}(x-z)+\delta^{(\td)}(x)) \braket{O(z)O(0)}\nn\\
&\,\,\,\,\,\,\,&+Z^{-1}_{F^2} \braket{O(z) O(0) F^2(x)}'  \bigg|_{\textrm{up to order }  g^2}\nn\\
&=&(2 c_O + \tilde B g^2)(\delta^{(\td)}(x-z)+\delta^{(\td)}(x)) \braket{O(z)O(0)}\nn\\
&\,\,\,\,\,\,\,&+ \braket{O(z) O(0) F^2(x)}'  \bigg|_{\textrm{up to order }  g^2}
\eea
where the second equality follows from the fact that $\braket{O(z) O(0) F^2(x)}'$ is necessarily of order $g^2$ by the LET. Therefore, the bare correlator in eq. \eqref{ICbarecorr_sdr}
differs from the multiplicatively renormalized one at distinct points by finite proper contact terms and it is finite up to order $g^2$.
It follows from eq. \eqref{ICbarecorr_sdr} the multiplicatively renormalized correlator extended at coinciding points:
\bea\label{IIBbarecorr_sdr}
Z_{F^2}\braket{O(z) O(0) F_0^2(x)}\bigg|_{\textrm{up to order } g^2} 
\hspace{-0.6cm}&=&Z_{F^2}(2 c_O + \tilde B g^2)  (\delta^{(\td)}(x-z)+\delta^{(\td)}(x)) \braket{O(z)O(0)}\nn\\
&\,\,\,\,\,\,\,&+ \braket{O(z) O(0) F^2(x)}'  \bigg|_{\textrm{up to order }  g^2}
\eea
Hence,  inserting eq. \eqref{ICbarecorr_sdr} in the rhs of (IC), we conclude that 
to order $g^2$ the additive counterterm in the last line of the rhs of (IC) may only be compensated for by the divergence of the space-time integral of the bare 3-point correlator at distinct points:
\bea\label{int_mRG_sdr1}
&&\f{1}{2} \int \braket{O(z) O(0) F^2(x)}' d^\td x\bigg|_{\textrm{to order }  g^2}
\eea
 that to order $g^2$ coincides with the nultiplicatively renormalized one.\par
 Instead, the multiplicatively renormalized 3-point correlator extended at coinciding points in eq. \eqref{IIBbarecorr_sdr} that occurs in (IIB) differs from the correlator at distinct points by finite and divergent proper contact terms to order $g^2$, consistently with the explicit computation of the OPE coefficient for $O=F^2$.
Therefore, the finiteness of the rhs of (IIB) arises from a
more involved cancellation between the integral of the correlator at distinct points, the divergent proper contact terms in eq. \eqref{IIBbarecorr_sdr} and  the additive counterterm in the rhs of (IIB) to order $g^2$. \par
Yet, the two versions of the LET are equivalent, and we employ (IC) in the following.  
We need an ansatz for the $3$-point correlator at distinct points to order $g^2$ in $\td=4-2\epsilon$ dimensions: 
\be\label{3pconf_sdr}
\braket{O(z) O(0) F^2(x)}_{\tilde d=4-2\epsilon}'=\f{N_3(g)\mu^{2\tD_{O}+\Delta^*_{F^2}-2\tD_{O_0}-\tD_{F^2_0}}}{|z|^{2\tD_{O}-\Delta^*_{F^2}}|x|^{\Delta^*_{F^2}}|x-z|^{\Delta^*_{F^2}}}   \bigg|_{\textrm{to order }  g^2}
\ee
where $\Delta^*_{F^2}$ is uniquely determined by the two conditions:\par
(I) The correlator should be conformal in the limit $\epsilon \rightarrow 0$.\par
(II) The correlator should imply the correct OPE coefficient to order $g^2$ in $\td=$ 
\indent\indent $4-2\epsilon$ dimensions.\par
\noindent The ansatz in eq. \eqref{3pconf_sdr} fulfills property (I) provided that $\Delta^*_{F^2} \rightarrow 4$ as $\epsilon \rightarrow 0$, since $N_3(g)$ is of order $g^2$. \par
We point out that (II) is in general incompatible with an exact conformal symmetry in $\td=4-2\epsilon$ dimensions because of the soft breaking of the latter in the CS equation  to order $g^2$, due to the nonvanishing beta function $\beta(g,\eps)=-\eps g+\cdots$ induced by dimensional regularization.
We determine $\Delta^*_{F^2}$ from (II) by observing that up to order $g^2$:
\be\label{shortsing3OPE}
\braket{O(z) O(0) F^2(x)}' \sim C_1^{(F^2,O)'}(x)\braket{O(z) O(0)}   \bigg|_{\textrm{to order }  g^2}
\ee
should hold as $x^2$ approaches zero, with $C_1^{(F^2,O)'}(x)$ the OPE coefficient at distinct points.
$C_1^{(F^2,O)'}(x)$ satisfies eq. \eqref{CSepsC1x}, and the perturbative expansion to order $g^2$ of its solution in eq. \eqref{cooreg} reads: 
\bea\label{C1FOsepper}
{C_{1}^{(F^2,O)}}'(x)&=&\f{\mathcal{ G}^{(F^2,O)}(\tilde g(x))}{|x|^{ 4-2\eps}}  Z^{(F^2)}(\tilde g(x),g)\nn\\
&=&\f{(a\tilde g^2(x) +\cdots)}{|x|^{ 4-2\eps}}\f{\tilde g^2(x)}{g^2}|x\mu|^{-2\eps}\nn\\
&=& \f{ag^2 }{|x|^{ 4-2\eps}}\bigg(\f{|x\mu|^{2\eps}}{1-\beta_0 g^2\f{|x\mu|^{2\eps}-1}{\eps}}
\bigg)^2 |x\mu|^{-2\eps}+\cdots\nn\\
&=& \f{a\mu^{2\eps}g^2 }{|x|^{ 4-4\eps}}+\cdots   \bigg|_{\textrm{to order }  g^2}
 \eea
where we have employed $\tilde \Delta_{F^2_0}=4-2\eps$, eq. \eqref{grun} and eq. \eqref{Zrun}, with $g(\mu) =g$.
On the other hand, the short-distance singularity as $x^2\to 0$ of the $3$-point correlator in eq. \eqref{3pconf_sdr}  reads to order $g^2$: 
\bea\label{short3}
\braket{O(z) O(0) F^2(x)}'_{\tilde d=4-2\epsilon}&{\sim}&
\f{N_3(g)\mu^{\Delta^*_{F^2}-\tD_{F^2_0}}}{|x|^{\Delta^*_{F^2}}}
\f{\mu^{2\tD_{O}-2\tD_{O_0}}}{|z|^{2\tD_{O}}}   \bigg|_{\textrm{to order }  g^2} \nn\\
&\sim & \f{N_3(g)}{N_2(g)}\f{\mu^{\Delta^*_{F^2}-\tD_{F^2_0}}}{|x|^{\Delta^*_{F^2}}}\braket{O(z)O(0)}   \bigg|_{\textrm{to order }  g^2}
\eea
The comparison of eq. \eqref{C1FOsepper} with eq. \eqref{short3} and the asymptotics in eq. \eqref{shortsing3OPE} then imply that the corrected dimension due to the nonvanishing beta function in dimensional regularization is $\Delta^*_{F_0^2}=\tD_{F_0^2}-2\eps=4-4\eps$ in the denominator of eq. \eqref{3pconf_sdr}, while the powers of $\mu$ in the numerator compensate for the change of dimension. 
From eq. \eqref{shortsing3OPE} we also deduce the normalization of the OPE coefficient:
\be\label{normOPE}
ag^2=\f{N_3(g)}{N_2(g)}\bigg|_{O(g^2)}\equiv \f{N_3}{N_2}g^2
\ee
in terms of the normalization of the 3- and 2-point correlators. 
Therefore, to order $g^2$ we obtain:
\bea\label{int_mRG_sdr}
&&\f{1}{2} \int \braket{O(z) O(0) F^2(x)}' d^\td x \bigg|_{\textrm{up to order }  g^2}\nn\\
&&=
\f{1}{2}\int \f{\mu^{2\eps}N_3(g)}{|z|^{2\tD_{O_0}-\tD_{F_0^2}+2\eps}|x|^{\tD_{F_0^2}-2\eps}|x-z|^{\tD_{F_0^2}-2\eps}}\,d^\td x \bigg|_{\textrm{up to order }  g^2}
\nn\\
&&=\braket{O(z)O(0)}_{\td=4-2\eps}\f{\pi^2}{2}  \f{N_3(g)}{N_2(g)}
\bigg(\f{2}{\eps}+4\log|z\mu| -2\log\pi +2+2\Gamma'(1)\bigg)  \bigg|_{\textrm{up to order}\,  g^2}
\eea
where the above integral has been computed in appendix \ref{AppCC}.
Hence, the finiteness of the rhs of (IC) in eq. \eqref{PLET1C} implies:
\be\label{eq:eps_sdr}
-\f{\pi^2}{2}  \f{N_3(g)}{N_2(g)}  \bigg|_{\textrm{to order }  g^2}=\gamma_O(g)  \bigg|_{\textrm{to order }  g^2}
\ee
with $\gamma_O(g)=-\gamma_0^{(O)}g^2+\cdots$, that in turn implies for the OPE coefficient of a generic operator $O$ with $F^2$ at distinct points to order $g^2$:
\bea
{C_{1}^{(F^2,O)}}'(x)&=&
 \f{2\gamma_0^{(O)}}{\pi^2}\f{\mu^{2\eps}g^2 }{|x|^{ 4-4\eps}}+\cdots
\eea
 by means of eqs. \eqref{C1FOsepper} and \eqref{normOPE}. 
Besides, the matching to order $g^2$ of the logarithmic terms in the lhs and rhs of eq. \eqref{PLET1C}, which is independent of the regularization, leads to the constraint:
\be\label{eq:log0_sdr}
-2g\f{\partial\gamma_O}{\partial g} \log{|z\mu|} \bigg|_{\textrm{order } g^2}=4\f{\pi^2}{2}  \f{N_3(g)}{N_2(g)}\log|z\mu| \bigg|_{\textrm{order } g^2}
\ee
that implies:
\be\label{eq:log0_sim_sdr}
g\f{\partial\gamma_O}{\partial g} \bigg|_{\textrm{order } g^2}=-{\pi^2}  \f{N_3(g)}{N_2(g)} \bigg|_{\textrm{order } g^2}
\ee
Putting the above results together, we get:
\be\label{eq:div0_sdr}
g\f{\partial\gamma_O}{\partial g}\bigg|_{\textrm{order } g^2}=2\gamma_O(g)\bigg|_{\textrm{order } g^2}
\ee
whose only solution is that the anomalous dimension $\gamma_O(g)=-\gamma_0^{(O)}g^2$ is one-loop exact, which is obviously consistent with our perturbative assumption to order $g^2$.\par
Vice versa, since $\gamma_O(g)=-\gamma_0^{(O)}g^2$ to order $g^2$, no divergent contact term may arise in the bare correlator in eq. \eqref{ICbarecorr_sdr}, otherwise the LET would not be satisfied.
The LET further implies the matching of the finite terms to order $g^2$:
\bea
 \f{\partial \log N_2(g)}{\partial \log g} \bigg|_{\textrm{to order }  g^2}= \tilde B g^2 - \gamma_O(g) \Big(2+2\Gamma'(1)-2\log\pi \Big)  \bigg|_{\textrm{to order }  g^2}
\eea
that agrees with eq. \eqref{LETcons} for $O=F^2$.

\subsection{Verifying the LET for $O=F^2$}

It is interesting to further verify the predictions of the LET for $O=F^2$ by comparing with the perturbative computation \cite{Kataev,Z1,Z2,Z3,BB}.
First, to order $g^0$:
\bea\label{sec2OPEFF}
F^2(x) F^2(0) = 2c_{F^2} \, \delta^{(4)}(x) F^2(0) + \cdots
\eea
with $c_{F^2}=2$ \cite{Kataev,Z1,Z2,Z3,BB} according to the LET.\par
Second, eq. \eqref{eq:log0_sim_sdr}:
\bea
 -\f{\pi^2}{2}  \frac{4\beta_0}{\pi^2}g^2=-2\beta_0 g^2=\gamma_{F^2}(g)
\eea
is satisfied to order $g^2$, where $\gamma_{F^2}(g)=-2\beta_0 g^2 + \cdots$ (appendix \ref{AppCC}) and we read from the OPE of $F^2$ with itself \cite{Kataev,Z1,Z2,Z3,BB} that $N_2= \frac{48(N^2-1)}{\pi^4}$ and $N_3=N_2 \frac{4\beta_0}{\pi^2}g^2$ to their leading order, respectively.

\subsection{Conformal resummation to order $g^2$}

We find an a-priori solution of the resummed version of the LET in eq. \eqref{LET2A12} to order $g^2$ by employing the resummed conformal $3$-point correlator  in $d=4$ dimensions expanded to order $g^2$ after the space-time integration:
\bea \label{LET2Bpert}
&&\braket{O(z)O(0)}\left(\f{\partial\log N_2(g)}{\partial\log g}-2g\f{\partial\gamma_O}{\partial g} \log{|z\mu|} \right)   \bigg|_{\textrm{up to order } g^2}\nn\\
&&=\f{1}{2} \int Z_{F^2} \braket{O(z) O(0) F_0^2(x)} \nn\\
&&~~~~- \f{2\gamma_O(g)}{\frac{\beta(g)}{g}} \big(\delta^{(4)}(x-z)+\delta^{(4)}(x)\big)\braket{O(z)O(0)} d^4x \bigg|_{\textrm{up to order } g^2}
\eea
In the resummed conformal theory the multiplicatively renormalized $3$-point correlator in the coordinate representation must be finite, since no contact term may arise for $\gamma_{F^2} \neq 0$ for dimensional reasons:
\bea \label{ANS}
&&\braket{O(z) O(0) F_0^2(x)} \bigg|_{\textrm{up to order } g^2}=Z_{F^2}^{-1}\braket{O(z) O(0) F^2(x)}' \bigg|_{\textrm{up to order } g^2}
\eea
Moreover, in $d=4$ dimensions the space-time integral of the resummed conformal $3$-point correlator at distinct points: 
\bea\label{sec2_int}
&&\int \braket{O(z) O(0) F^2(x)}' d^4x\bigg|_{\textrm{up to order } g^2}\hspace{-0.3cm}=\int \f{N_3(g)\mu^{-2\gamma_O-\gamma_{F^2}}}{|z|^{2\Delta_O-\Delta_{F^2}}|x|^{\Delta_{F^2}}|x-z|^{\Delta_{F^2}}}d^4x\bigg|_{\textrm{up to order }  g^2}\nn\\
&&=\braket{O(z)O(0)} \pi^2\f{N_3(g)}{N_2(g)}|z\mu|^{-\gamma_{F^2}}
\f{\Gamma\big(2+\gamma_{F^2}\big)\Gamma\big(-\f{\gamma_{F^2}}{2}\big)^2}{\Gamma\big(2+\f{\gamma_{F^2}}{2}\big)^2 \Gamma\big(-\gamma_{F^2}\big)}
\bigg|_{\textrm{up to order }  g^2}
 \nn\\
&&=\braket{O(z)O(0)} \pi^2\f{N_3(g)}{N_2(g)}\left(\f{2}{\beta_0g^2}+4\log|z\mu|+\cdots\right)\bigg|_{\textrm{up to order } g^2}
\eea
is finite because $\Delta_{F^2}=4+\gamma_{F^2}$, with nonzero $\gamma_{F^2}=-2\beta_0g^2$ to order $g^2$  (appendix \ref{AppCC}).
Hence, the LET implies the following matching of the logarithmic terms in the lhs and rhs up to order $g^2$:
\bea 
&& -2g\f{\partial\gamma_O}{\partial g} \log{|z\mu|}    \bigg|_{\textrm{up to order } g^2}
 = 
\f{\pi^2}{2}\f{N_3(g)}{N_2(g)} 4\log|z\mu|
\bigg|_{\textrm{up to order } g^2}
\eea
that is regularization independent, and leads to the constraint:
 \be\label{eq:log0_cres}
 g\f{\partial\gamma_O}{\partial g}\bigg|_{\textrm{up to order } g^2}=-\pi^2\f{N_3(g)}{N_2(g)}\bigg|_{\textrm{up to order } g^2}
 \ee
The latter implies that $\f{N_3(g)}{N_2(g)}$ is of order $g^2$ and it reproduces the same constraint implied by (IC) and (IIB) in perturbation theory.
The finite contribution in the rhs of eq. \eqref{LET2Bpert} reads to order $g^0$ by means of eq. \eqref{sec2_int}:
 \bea
 0&=&
\f{\pi^2}{2}\f{N_3(g)}{N_2(g)}\f{2}{\beta_0g^2}-\f{2\gamma_O(g)}{\frac{\beta(g)}{g}} \bigg|_{\textrm{order } g^0}\nn\\
&=&-\f{1}{\beta_0g^2}g\f{\partial\gamma_O}{\partial g}-\f{2\gamma_O(g)}{\frac{\beta(g)}{g}} \bigg|_{\textrm{order } g^0}
\eea
that vanishes identically for $\gamma_O(g)=-\gamma_0^{(O)}g^2$ and $\frac{\beta(g)}{g}=-\beta_0 g^2$. The LET is then fulfilled because  
$\f{\partial\log N_2(g)}{\partial\log g}$ in the lhs vanishes identically to order $g^0$.
Yet, the finite terms cannot be determined to order $g^2$ consistently with the conformal symmetry, since:
  \bea\label{g2rhsconf}
&&\f{\pi^2}{2}\f{N_3(g)}{N_2(g)}\f{2}{\beta_0g^2}-\f{2\gamma_O(g)}{\frac{\beta(g)}{g}} \bigg|_{\textrm{order } g^2}\nn\\
&=&\f{\pi^2}{2}\f{N_3(g)}{N_2(g)}\bigg|_{\textrm{order } g^4}
\f{2}{\beta_0g^2}-\f{2\gamma_O(g)}{\frac{\beta(g)}{g}} \bigg|_{\textrm{order } g^2}
\eea
involve the evaluation of $\f{N_3(g)}{N_2(g)}\bigg|_{\textrm{order } g^4}$ and $\f{2\gamma_O(g)}{\frac{\beta(g)}{g}}\bigg|_{\textrm{order } g^2}$ that is not consistent with the conformal symmetry.

\section{Contact terms revisited}  \label{60}

The occurrence of divergent proper contact terms in $C^{(F^2,O)}_1(x)$ may seem surprising. In the following we establish whether they are an isolated phenomenon or a more general one, by investigating whether they exist in $C^{(F^2,F^2)}_0(x)$ as well. \par
Indeed, we demonstrate that to order $g^0$ the multiplicatively renormalized 
$Z_{F^2}$ $C^{(F_0^2,F_0^2)}_0(p)$ needs an infinite additive renormalization that fully originates from 
the Fourier transform of $C^{(F^2,F^2)'}_0(z)$ at distinct points, so that no divergent proper contact term occurs to this order. 
 Yet, we show that divergent proper contact terms do occur to order $g^2$ in $C^{(F^2,F^2)}_0(x)$ as in $C^{(F^2,F^2)}_1(x)$.

\subsection{$\braket{F^2(z)F^2(0)}$ to order $g^2$ in perturbation theory}

We expand the solution of the CS equation in $\td=4-2\eps$ dimensions at distinct points (appendix \ref{CS1}): 
\bea
 \braket{F^2(z)F^2(0)}'&=&\frac{1}{z^{8-4\eps}}\, \mathcal{G}_{2}^{(F^2)}(\tg(z))\, Z^{(F^2)2}(\tg(z), g(\mu)) 
 \eea
perturbatively up to order $g^2$. The RG-invariant function $\mathcal{G}_{2}^{(F^2)}$ reads:
 \be
  \mathcal{G}_{2}^{(F^2)}(\tg(z))=
 \mathcal{G}^{(F^2)}_2(0)(1+\delta_2^{(F^2)}\tg^2(z)+\cdots)
 \ee
 and, by means of eq. \eqref{ZF_IR_AF_eps}, we obtain: 
 \bea\label{FFpert_eps}
 \braket{F^2(z)F^2(0)}'&=&\frac{\mathcal{G}^{(F^2)}_2(0)}{z^{8-4\eps}}\, (1+\delta_2^{(F^2)}\tg^2(z)+\cdots)
 \, \f{\tg^4(z)}{g^4(\mu)}|z\mu|^{-4\eps} +\cdots
 \eea
By employing the perturbative expansion of the running coupling $\tg(z)$ in eq. \eqref{sol_eps_AF}:
 \bea
 \f{\tg^2(z)}{g^2(\mu)}&=&\f{|z\mu|^{2\eps}}{
1-\beta_0g^2(\mu)\f{|z\mu|^{2\eps}-1}{\eps} +\cdots}\nn\\
&=&|z\mu|^{2\eps}\left( 1+\beta_0g^2(\mu)\f{|z\mu|^{2\eps}-1}{\eps} +\cdots\right)
 \eea
 and setting $g(\mu)=g$, eq. \eqref{FFpert_eps} yields: 
 \bea\label{FFpert_eps2}
 \braket{F^2(z)F^2(0)}'&=&\frac{\mathcal{G}^{(F^2)}_2(0)}{z^{8-4\eps}}\, \bigg(1+\delta_2^{(F^2)}
 |z\mu|^{2\eps}g^2+\cdots\bigg)\nn\\
&&~~\bigg( 1+2\beta_0g^2\f{|z\mu|^{2\eps}-1}{\eps} +\cdots\bigg)+\cdots
\nn\\
&&\hspace{-3.cm}=\frac{\mathcal{G}^{(F^2)}_2(0)}{z^{8-4\eps}}\,\bigg(1+g^2\bigg(\delta_2^{(F^2)}
 |z\mu|^{2\eps}+2\beta_0\f{|z\mu|^{2\eps}-1}{\eps} 
 \bigg)+\cdots\bigg)\nn\\
 &&\hspace{-3.cm}=\mathcal{G}^{(F^2)}_2(0)\bigg\{\f{1}{z^{8-4\eps}}\left(1-\f{2\beta_0g^2}{\eps}\right)
 +\f{\mu^{2\eps}}{z^{8-6\eps}}g^2\left(\delta_2^{(F^2)}
+\f{2\beta_0}{\eps}\right)+\cdots
\bigg\} 
 \eea
 Its Fourier transform is:
\bea\label{FFFTeps}
\textrm{FT}\big[\braket{F^2(z)F^2(0)}'\big]&=&\int d^{\td}z\,e^{ipz}\braket{F^2(z)F^2(0)}'\nn\\
&=&\mathcal{G}^{(F^2)}_2(0)\bigg\{\textrm{FT}\left[\f{1}{z^{8-4\eps}}\right]\left(1-\f{2\beta_0g^2}{\eps}\right)
\nn\\&&~~
+\textrm{FT}\left[\f{\mu^{2\eps}}{z^{8-6\eps}}\right]g^2\left(\delta_2^{(F^2)}
+\f{2\beta_0}{\eps}\right)+\cdots
\bigg\}
 \eea
with $\td=4-2\eps$. It is convenient to first Fourier transform the correlator in eq. \eqref{FFpert_eps2} and then to expand in $\eps$. To this aim, we employ the general FT:
 \be
 \textrm{FT}\bigg[\f{1}{x^{2\Delta}}\bigg]=\pi^{\f{d}{2}}\f{\Gamma\big(\f{d}{2}-\Delta\big)}{\Gamma\big(\Delta\big)} \left(\f{p^2}{4}\right)^{\Delta - \f{d}{2}}
 \ee
that for $d=4-2\eps$ and $\Delta=4-2\eps$ yields to the relevant order in $\eps$:
 \bea\label{FT1eps}
 \textrm{FT}\bigg[\f{1}{x^{8-4\eps}}\bigg]&=& \left(\f{\pi p^2}{4}\right)^{2}
 \left(\f{\pi p^2}{4}\right)^{-\eps}\f{\Gamma(-2+\eps)}{\Gamma(4-2\eps)}\nn\\
 &=& \f{\pi^2p^4}{16} \left(\f{\pi p^2}{4}\right)^{-\eps}\f{1}{12\eps}
 \bigg(1+\eps\bigg(\f{31}{6}+3\Gamma'(1)\bigg)\nn\\
 &&\hspace{1.cm}+\eps^2\bigg(\f{601}{36}+\f{31}{2}\Gamma'(1)-\f{3}{2}\Gamma^{''}(1)+6\Gamma^{'2}(1)\bigg)+\cdots\bigg)
 \eea
 and, for $\Delta=4-3\eps$:
 \bea\label{FT2eps}
\textrm{FT}\bigg[\f{\mu^{2\eps}}{x^{8-6\eps}}\bigg]&=&\left(\f{\pi p^2}{4}\right)^{2}
\mu^{2\eps} \left(\f{\sqrt{\pi} p^2}{4}\right)^{-2\eps} 
\f{\Gamma(-2+2\eps)}{\Gamma(4-3\eps)}\nn\\
 &=& \f{\pi^2p^4}{16} \mu^{2\eps}\left(\f{\sqrt{\pi} p^2}{4}\right)^{-2\eps}\f{1}{24\eps}
 \bigg(1+\eps\bigg(\f{17}{2}+5\Gamma'(1)\bigg)\nn\\
&&\hspace{1.cm} +\eps^2\bigg(\f{179}{4}+\f{85}{2}\Gamma'(1)-\f{5}{2}\Gamma^{''}(1)+15\Gamma^{'2}(1)\bigg)+\cdots\bigg)
 \eea
 By means of eqs. \eqref{FT1eps} and \eqref{FT2eps}, eq. \eqref{FFFTeps} reads as $\eps\to 0$:
 \bea\label{FFFTeps-sep}
&&\textrm{FT}\big[\braket{F^2(z)F^2(0)}'\big]=\f{\pi^2\mathcal{G}^{(F^2)}_2(0)}{192}{p^4}\bigg\{
 \f{1}{\eps}-\log{p^2}  -\log\f{\pi}{4}+\f{31}{6}+3\Gamma'(1)\nn\\
&& -\f{g^2\beta_0}{\eps^2}
 +\f{g^2}{2\eps}\bigg(\delta_2^{(F^2)}-2\beta_0\Big(\f{11}{6}+\Gamma'(1)-\log \pi\Big)\bigg)
 +\f{g^2\beta_0}{\eps}\log\mu^2\nn\\
&&+ g^2\beta_0\log^2\f{p^2}{\mu^2} -\f{g^2\beta_0}{2}\log^2{\mu^2} \nn\\
&&-g^2
\bigg( \delta_2^{(F^2)}+2\beta_0\Big(\f{10}{3}+2\Gamma'(1)+\log{ 4}\Big)\bigg)\log\f{p^2}{\mu^2}\nn\\
&&-\f{g^2}{2}\bigg(\delta_2^{(F^2)}-2\beta_0\Big(\f{11}{6}+\Gamma'(1)-\log \pi\Big)\bigg)\log\mu^2
\nn\\
&&
-g^2\delta_2^{(F^2)}\bigg(\log\f{\sqrt{\pi}}{4}-\f{17}{4}-\f{5}{2}\Gamma'(1)\bigg)
+g^2\beta_0\bigg(\Big(\f{11}{6}+\Gamma'(1)\Big)\log\pi\nn\\
&& +4\Big(\f{5}{3}+\Gamma'(1)\Big)\log{4}+\f{409}{36}+\f{23}{2}\Gamma'(1)+\f{1}{2}\Gamma^{''}(1)+3\Gamma^{'2}(1)
\bigg)+\cdots
 \bigg\} 
 \eea
 We now compare the above result with the perturbative computation of $C_{0}^{(F^2,F^2)}(p)$  in the $\overline{MS}$ scheme \cite{Z1,Z3}:
  \bea\label{C0S_tot}
C_{0}^{(F^2,F^2)}(p)=Z_{F^2}^2 C_0^{(F^2_0,F^2_0)}(p)+ p^{4}Z_{0 \textrm{c.t.}} 
\eea
where in our notation the multiplicatively renormalized contribution to order $g^2$ reads \cite{Z1, Z3}:
\bea\label{C0S_mr}
 &&Z_{F^2}^2C_0^{(F_0^2,F_0^2)}(p)
 = \f{N^2-1}{4\pi^2}p^4\bigg\{1 +\f{1}{\eps}-\log\f{p^2}{\mu^2}
 +g^2\beta_0\log^2\f{p^2}{\mu^2}
 -\f{g^2\beta_0}{\eps^2}\nn\\
 &&~~
+2g^2(c_1-\beta_0)
\log\f{p^2}{\mu^2}
-\f{g^2c_1}{\eps}
+\f{g^2}{16\pi^2}\bigg(-12\zeta_3 +\f{485}{12}-\f{17}{2}\f{N_f}{N}\bigg)
+\cdots\bigg\}~~~~
\eea
and the additive counterterm up to order $g^2$ reads \cite{Z1, Z3}:
\bea\label{C0S_add}
p^4Z_{0 \textrm{c.t.}} &=& \f{N^2-1}{4\pi^2}p^4\bigg\{-\f{1}{\eps}
 +\f{g^2\beta_0}{\eps^2}+\f{g^2c_1}{\eps}
+\cdots\bigg\}
\eea
where, for later convenience, we have defined:
\be\label{c1CZ}
c_1=-\f{1}{16\pi^2}\bigg(\f{17}{2}-\f{5}{3}\f{N_f}{N}\bigg)
\ee
so that we obtain the coefficient of the subleading $\log\f{p^2}{\mu^2}$ to order $g^2$ 
in eq. \eqref{C0S_mr}:
\be\label{c1CZb}
2(c_1-\beta_0)=-\f{1}{16\pi^2}\bigg(\f{73}{3}-\f{14}{3}\f{N_f}{N}\bigg)
\ee
Eq. \eqref{FFFTeps-sep} reproduces the leading divergence as well as the leading $\log\f{p^2}{\mu^2}$ to order $g^0$ of the multiplicatively renormalized coefficient in eq. \eqref{C0S_mr}, once the overall normalization has been matched:
\bea\label{fixCZ2point_norm}
\mathcal{G}^{(F^2)}_2(0)&=&\f{48(N^2-1)}{\pi^4}
\eea
We conclude that no divergent proper contact term occurs to order $g^0$. \par
The matching of the subleading $\log\f{p^2}{\mu^2}$ to order $g^2$ in eqs. \eqref{FFFTeps-sep} and \eqref{C0S_mr} implies:
\bea\label{d2f2CZ}
\delta_2^{(F^2)}&=&-2(c_1-\beta_0)-2\beta_0\Big(\f{10}{3}+2\Gamma'(1)+\log{ 4}\Big)
\eea
that inserted in the subleading divergence to order $g^2$ in eq. \eqref{FFFTeps-sep}, by neglecting temporarily the  pure $\log\mu^2$ terms, yields:
\bea\label{divdet}
&&\f{g^2}{2\eps}\bigg(\delta_2^{(F^2)}-2\beta_0\Big(\f{11}{6}+\Gamma'(1)-\log \pi\Big)\bigg)
\nn\\
&&=\f{g^2}{2\eps}\bigg(-2c_1+2\beta_0
-2\beta_0\Big(\f{10}{3}+2\Gamma'(1)+\log{ 4}\Big)-2\beta_0\Big(\f{11}{6}+\Gamma'(1)-\log \pi\Big)\bigg)\nn\\
&&=-\f{g^2c_1}{\eps}+\f{g^2}{2\eps}\bigg(-2\beta_0\Big(\f{25}{6}+3\Gamma'(1)-\log \f{\pi}{4}\Big)\bigg)
\eea
The last line differs from the corresponding divergent term in eq. \eqref{C0S_mr} because of the last contribution proportional to $\beta_0$. 
Hence, the above calculation implies that both a finite and divergent proper contact term $\mu^{-2\eps}\tilde C$ $\Delta^2\delta^{(\td)}(z)$ occur to order $g^2$, with:
 \bea\label{FFplus_primeCt}
 \tilde C= \f{N^2-1}{4\pi^2}\f{g^2}{2\eps}2\beta_0\Big(\f{25}{6}+3\Gamma'(1)-\log \f{\pi}{4}\Big) 
 + \f{N^2-1}{4\pi^2}\bigg(\tilde A_0 +\tilde A_1g^2 +\cdots\bigg)
 \eea
so that the FT of:
\bea\label{FFplus_prime}
 \braket{F^2(z)F^2(0)}' + \mu^{-2\eps}\tilde C\,\Delta^2\delta^{(\td)}(z)
 \eea
reproduces the multiplicatively renormalized OPE coefficient in eq. \eqref{C0S_mr} up to pure $\log\mu^2$ terms,
with $\Delta$ the Laplacian in $\td$ Euclidean space-time dimensions, $\braket{F^2(z)F^2(0)}'$ given by eqs. \eqref{FFpert_eps2} and \eqref{fixCZ2point_norm}, and:
  \bea
  \tilde A_0 &=& 1-\f{31}{6}-3\Gamma'(1)+\log\f{\pi}{4}\nn\\
  \tilde A_1&=&
  \delta_2^{(F^2)}\bigg(\log\f{\sqrt{\pi}}{4}-\f{17}{4}-\f{5}{2}\Gamma'(1)\bigg)
-\beta_0\bigg(\Big(\f{11}{6}+\Gamma'(1)\Big)\log\pi
 +4\Big(\f{5}{3}+\Gamma'(1)\Big)\log{4}\nn\\
 &&\hspace{-0.3truecm}+\f{409}{36}+\f{23}{2}\Gamma'(1)+\f{1}{2}\Gamma^{''}(1)+3\Gamma^{'2}(1)
\bigg)
+\f{1}{16\pi^2}\bigg(-12\zeta_3 +\f{485}{12}-\f{17}{2}\f{N_f}{N}\bigg)
  \eea  
with $\delta_2^{(F^2)}$ in eq. \eqref{d2f2CZ} and $c_1$ in eq. \eqref{c1CZ}. \par
Finally, as a further check, the fully renormalized OPE coefficient $C_0^{(F^2,F^2)}(p)$ in the momentum representation is obtained by adding to the FT of eq. \eqref{FFplus_prime} the counterterm proportional to $Z_{0 \textrm{c.t.}}$ in eq. \eqref{C0S_add}:
\bea\label{FFtot}
C_0^{(F^2,F^2)}(p)=&&\textrm{FT}\big[ \braket{F^2(z)F^2(0)}'\big]+
\textrm{FT}\big[ \mu^{-2\eps}\tilde C\,\Delta^2\delta^{(\td)}(z)\big] \nn\\
&&+\textrm{FT}\big[ \mu^{-2\eps}Z_{0 \textrm{c.t.}}\,\Delta^2\delta^{(\td)}(z)\big]
 \eea
where:
 \bea
 \mu^{-2\eps}Z_{0 \textrm{c.t.}}&=&\f{N^2-1}{4\pi^2}
 \bigg(1-\eps\log\mu^2+\f{1}{2}\eps^2\log^2\mu^2+\cdots\bigg)
 \bigg(-\f{1}{\eps}
 +\f{g^2\beta_0}{\eps^2}+\f{g^2c_1}{\eps}
+\cdots\bigg)\nn\\
&=&\f{N^2-1}{4\pi^2}\bigg(-\f{1}{\eps}
 +\f{g^2\beta_0}{\eps^2}+\f{g^2c_1}{\eps}
+\log\mu^2-\f{g^2\beta_0}{\eps}\log\mu^2\nn\\&&
+\f{g^2\beta_0}{2}\log^2\mu^2
-g^2c_1\log\mu^2 
+\cdots
\bigg)
\eea
Indeed, the identity in eq. \eqref{divdet} provides the complete cancellation of the divergences and pure $\log\mu^2$ terms in eq. \eqref{FFtot}, thus reproducing the fully renormalized $C_{0}^{(F^2,F^2)}(p)$ in eq. \eqref{C0S_tot}.

\section{LET versus perturbative and nonperturbative renormalization} \label{7}

\subsection{Perturbative renormalization} \label{7.1}

 A perturbative computation of the bare LET to order $g^2$ has been worked out for $O=F^2$ with a hard-cutoff regularization \cite{MBL}.
The corresponding bare LET (IC) reads in dimensional regularization (section \ref{5.2}):
\bea
\label{eq:LETCuv_n200}
&& 
\f{\partial}{\partial\log {g}_0} \braket{O(z) O(0)}_0 \bigg|_{\dr}\nonumber \\
&&= \f{1}{2} \int\, \braket{O(z) O(0) F^2(x)}_0 - \braket{O(z)O(0)}_0\braket{F^2(x)}_0\, d^4 x - 2c_{O} \braket{O(z) O(0)}_0\bigg|_{\dr}
\eea
Eq. \eqref{eq:LETCuv_n200} is the most convenient to verify to order $g^2$ in perturbation theory the compatibility of the LET in \cite{MBL} regularized by a hard-cutoff in $d=4$ with our gauge-invariant formulation in dimensional regularization.
Indeed, in \cite{MBL} the occurrence of the finite contact terms in the rhs has not been taken into account because of the hard-cutoff regularization in $d=4$, which is consistent with their automatic subtraction in the rhs of eq. \eqref{eq:LETCuv_n200}.\par
The bare LET expressed in terms of renormalized objects follows by employing eq. \eqref{eq:lhs_gen0}:
\bea \label{LET1C0}
&& Z^{-1}_{F^2}  Z_O^{-2} \f{\partial\braket{O(z)O(0)}}{\partial\log g} -  \f{2\gamma_O(g)}{\eps}  Z_O^{-2} \braket{O(z)O(0)} \bigg|_{\dr} \nn\\
&&=\f{1}{2}  Z_O^{-2} \int \braket{O(z) O(0) F_0^2(x)} - 2 c_O\big(\delta^{(4)}(x-z)+\delta^{(4)}(x)\big)\braket{O(z)O(0)} \nn\\
&&\,\,\,\,\,\,- \braket{O(z)O(0)}\braket{F_0^2(x)} d^4x \bigg|_{\dr}
\eea
Then, the above LET allows us to verify to order $g^2$ in perturbation theory for the special case $O=F^2$  the corresponding version in \cite{MBL},
with the identification $\frac{1}{\epsilon}=\log(\frac{\Lambda^2}{\mu^2})$. Indeed, to this order we can set $Z_{F^2} =Z_O=1$, since both sides of the LET are already of order $g^2$. It follows that all the perturbative arguments in \cite{MBL} -- strictly limited to order $g^2$ -- hold unmodified.

\subsection{Nonperturbative large-$N$ renormalization} \label{7,2}

In the AF nonperturbative case, once the lhs of the LET is expressed in terms of the derivative with respect to $\Lambda_{\scriptscriptstyle{UV}}$, the rhs necessarily contains the insertion
of the RG-invariant object according to eq. \eqref{bareLETLQCD0}:
\bea \label{bareLETLQCD}
&&\braket{O(z)O(0)}\left( 2 \gamma_O(g)  - 2c_O \epsilon- \f{\partial\log \braket{O(z)O(0)}}{\partial\log \Lambda_{\scriptscriptstyle{UV}} }\right) \nn\\
&&=\f{1}{2} \f{\beta(g,\epsilon)}{g} \int \braket{O(z) O(0) F^2(x)} - \braket{O(z)O(0)}\braket{F^2(x)} d^{\td}x 
\eea
It follows that no divergent contact term arises in the limit $\epsilon \rightarrow 0$ both in the lhs and rhs according to eq. \eqref{bareLETLQCDd4}, so that all the arguments involving the  nonperturbative large-$N$ renormalization in $d=4$ dimensions in \cite{MBL} hold unmodified.

\appendix

\section{Anomalous dimension of $F^2$} 
\label{C}

We compute in two different ways the anomalous dimension of $\Tr F^2=\frac{1}{2}F^2$ in massless QCD and $\mathcal{N}=1$ SUSY YM theory: 
First, from the multiplicative renormalization of $\Tr F^2$ in $\overline{MS}$-like schemes. 
Second, from the RG invariance of the trace anomaly that reads in $d=4$ dimensions \cite{Spiridonov}:
\be\label{trace-anom-pg}
T_{\alpha \alpha} = \f{\beta(g)}{g}\Tr F^2
\ee
The two computations are not actually independent. Indeed:
\be
T_{\alpha \alpha} = - \epsilon \Tr F^2_0
\ee
in $\tilde d=4-2\epsilon$ dimensions, with $\Tr F^2_0$ the bare operator. Hence, we get:
\bea
T_{\alpha \alpha} &=& - \epsilon Z^{-1}_{F^2}(g,\epsilon) Z_{F^2}(g,\epsilon) \Tr F^2_0 \nonumber \\
 &=& - \epsilon Z^{-1}_{F^2}(g,\epsilon) \Tr F^2 \nonumber \\
 &=& - \epsilon \left ( 1-\f{\beta(g)}{\epsilon g}\right ) \Tr F^2 \nonumber \\
 &=& \f{\beta(g,\epsilon)}{g} \Tr F^2 
  \eea
where:
\bea
\f{dg}{d\log\mu}=\beta(g,\epsilon)= -\epsilon g+\beta(g)
\eea
is the beta function in $\tilde d=4-2\eps$ dimensions.
It follows that the trace anomaly is a consequence of the multiplicative renormalization of $\Tr F^2$ that also implies the anomalous dimension of $\Tr F^2$ (appendix \ref{C1}).
Vice versa, the anomalous dimension of $\Tr F^2$ follows from the trace anomaly as well (appendix \ref{C2}).

\subsection{Multiplicative renormalization of $F^2$}\label{C1}

In gauge-invariant correlators of massless QCD and $\mathcal{N}=1$ SUSY YM theory $\Tr F^2$ is multiplicatively renormalizable in $\overline{MS}$-like schemes up to operators proportional to the EOM:
\be
\Tr F^2=Z_{F^2}(g,\epsilon) \Tr F^2_0
\ee
where $Z_{F^2}(g,\epsilon)$ \cite{Spiridonov,Z1}:
\be\label{ZF2}
Z_{F^2}(g,\epsilon)=1+g\f{\partial}{\partial g}\log Z_g(g,\epsilon) = \left ( 1-\f{\beta(g)}{\epsilon g}\right )^{-1}
\ee
with:
\bea
g_0=Z_g(g,\epsilon) \mu^\epsilon g(\mu)
\eea
and $\beta(g) = -g\,\f{d\log Z_g}{d\log\mu}$.
Thus, by means of $\f{d}{d\log\mu} = \beta(g,\epsilon )\f{\partial}{\partial g}+\f{\partial}{\partial\log\mu}$ and eq. \eqref{ZF2}, we get the anomalous dimension of $\Tr F^2$:  
\be\label{gF2}
\gamma_{F^2}(g)=-\f{d\log Z_{F^2}}{d\log\mu}=-\beta(g,\epsilon)\f{\partial\log Z_{F^2}}{\partial g}=g\f{\partial}{\partial g}\left (\f{\beta(g,\epsilon)}{g}\right )
= g\f{\partial}{\partial g}\left (\f{\beta(g)}{g}\right )
\ee
Besides, from eq. (\ref{ZF2}) it follows:
\bea\label{Zp}
Z_{F^2}(g,\epsilon)&=& 1-\f{\beta_0g^2}{\epsilon}-
\f{\beta_1g^4}{\epsilon} + \f{\beta_0^2g^4}{\epsilon^2} +O(g^6)
\eea
and:
\bea 
Z_g(g,\epsilon) &=& 1-\f{\beta_0g^2}{2\epsilon}-\f{\beta_1 g^4}{4\epsilon} +\f{3\beta_0^2g^4}{8\epsilon^2} +O(g^6)
\eea 
up to two loops. As a consequence:
\bea \label{Zp1}
Z_g^2(g,\epsilon) &=& 1-2\f{\beta_0g^2}{2\epsilon}-2\f{\beta_1 g^4}{4\epsilon}+2\f{3\beta_0^2g^4}{8\epsilon^2}+\left(\f{\beta_0g^2}{2\epsilon} \right)^2+O(g^6)\nonumber\\
&=&
 1-\f{\beta_0g^2}{\epsilon}-\f{\beta_1 g^4}{2\epsilon} +\f{\beta_0^2g^4}{\epsilon^2} +O(g^6)
\eea
Hence, only to one loop $Z_{F^2}$ coincides with $Z_g^2$ \cite{Z1}. \par

\subsection{$\gamma_{F^2}(g)$ from the trace anomaly} \label{C2}

Eq. (\ref{gF2}) follows from the trace anomaly as well \cite{MBM,scalar}. Because of the RG invariance of the trace anomaly the CS equation reads in $d=4$ dimensions:
\bea
\label{CS-traceanom}
&&\left(x\cdot\f{\partial}{\partial x} + \beta(g)\f{\partial}{\partial g} +4 \right)  \left(\f{\beta(g)}{g}\Tr F^2 (x)\right)  = 0 
\eea
Hence:
\bea
 \left(x\cdot\f{\partial}{\partial x} +
 \beta(g)\f{\partial}{\partial g} +4 + \beta(g)\f{\partial}{\partial g}\log \left( \f{\beta(g)}{g}\right)
\right) \Tr F^2 (x)=0
\eea
that implies:
\bea\label{A10}
 \left(x\cdot\f{\partial}{\partial x} 
+ \beta(g)\f{\partial}{\partial g} +4+
g\f{\partial}{\partial g}\left( \f{\beta(g)}{g}\right) \right) \Tr F^2 (x)  = 0
\eea
Comparing the above equation with the CS equation for $\Tr F^2$:
  \be
\label{CS-F2}
\left(x\cdot\f{\partial}{\partial x} + \beta(g)\f{\partial}{\partial g} + 4 + \gamma_{F^2}(g)\right) \Tr F^2 (x) = 0
\ee
we obtain:
\be\label{gF22}
\gamma_{F^2}(g)=g\f{\partial}{\partial g}\left (\f{\beta(g)}{g}\right )
\ee
that coincides with eq. (\ref{gF2})\footnote{The analogous reasoning in $\tilde d=4-2 \epsilon$ dimensions also leads to eq. (\ref{gF2}).}. The anomalous dimension up to two loops in $\overline{MS}$-like schemes follows:
\be\label{gF2p}
\gamma_{F^2}(g)=-2\beta_0g^2-4\beta_1g^4 + \cdots
\ee
with:
\bea
\beta(g) = -\beta_0g^3-\beta_1g^5+\cdots
\eea
The first coefficient $\gamma^{(F^2)}_0=-2\beta_0$ is universal, i.e. scheme independent, as for the one-loop coefficient of the anomalous dimension of any canonically normalized operator. 
Instead, $\gamma^{(F^2)}_1=-4\beta_1$ is scheme dependent, and it may be changed \cite{MB00} by a finite multiplicative renormalization of $\Tr F^2$ reducing to the identity for $g=0$, contrary to a reparametrization of $g$ that would not affect $\beta_1$ -- the second universal coefficient of the beta function.

\section{Contact terms in the OPE for $F^2$ \cite{Z1,Z3}}
\label{A}

The Euclidean version of the Minkowskian operator $O_1$ \cite{Z1,Z3} reads in our notation:  
\bea\label{eq:O1F2}
O_1&=&-\f{1}{2g_{YM}^2}\Tr{{\cal F}}_{\mu\nu}{{\cal F}}_{\mu\nu}\nn\\
&=&-\f{N}{2g^2}\Tr{{\cal F}}_{\mu\nu}{{\cal F}}_{\mu\nu}\nn\\
&=&-\f{N}{2g^2}\f{g^2}{N}\Tr{F}_{\mu\nu}{F}_{\mu\nu}\nn\\
&=&-\f{1}{2}\Tr{F}_{\mu\nu}{F}_{\mu\nu}\nn\\
&=&-\f{1}{4}F^2
\eea
where $g^2=Ng_{YM}^2$ and we have expressed ${{\cal F}}_{\mu\nu}$ in the Wilsonian normalization in terms of ${F}_{\mu\nu}$ in the canonical one
 (section \ref{3}). 
It follows the fully renormalized OPE coefficient of $O_1$ with itself \cite{Z3} in the momentum representation:
\bea \label{b2}
&&\int O_1(x)O_1(0)  e^{-ip\cdot x} d^{4-2\eps}x \nonumber \\
&&= \int \cdots +\left(C_{1CZ}^{(S)}(x) -\delta^{(4)}(x) \left(\f{Z_{11}^L}{Z_{11}}+ \cdots \right)\right) O_1(0)+ \cdots e^{-ip\cdot x} d^{4-2\eps}x
\eea
with:
 \be
 Z_{11}=\left(1-\f{\beta(\alpha_s)}{\eps} \right)^{-1}
 \ee
 and \cite{Z3}:
\bea\label{eq:Z11}
\f{Z_{11}^L}{Z_{11}}&=& \f{1}{\eps}\left(1-\f{\beta(\alpha_s)}{\eps} \right)^{-1}\alpha_s^2\f{\partial}{\partial\alpha_s}\left(  \f{\beta(\alpha_s)}{\alpha_s}    \right)\nn\\
&=&\f{1}{\eps}\left(1-\f{\beta(\alpha_s)}{\eps} \right)^{-1}\left(\alpha_s\beta'(\alpha_s)-\beta(\alpha_s)    \right)
\eea
where $\beta(\alpha_s)$ reads in our notation: 
\bea
\beta(\alpha_s)&=&\f{d\log\alpha_s}{d\log\mu^2}
=\f{1}{2}\f{1}{\alpha_s}\f{\partial\alpha_s}{\partial g}\f{dg}{d\log\mu}\nn\\
&=&\f{1}{2}\f{4\pi N}{g^2}\f{2g}{4\pi N}\f{dg}{d\log\mu}
=\f{1}{g}\f{dg}{d\log\mu}\nn\\
&=&\f{\beta(g)}{g}
\eea
with $\alpha_s=g_{YM}^2/4\pi = g^2/4\pi N$.
Hence, employing:
\bea
\alpha_s^2\f{\partial}{\partial\alpha_s}\left(  \f{\beta(\alpha_s)}{\alpha_s}    \right)&=&
\f{1}{2}\f{1}{4\pi N}g^3\f{\partial}{\partial g}\left( 4\pi N\f{\beta(g)}{g^3}    \right)
=\f{1}{2}g^3\f{\partial}{\partial g}\left( \f{\beta(g)}{g^3}    \right)\nn\\
&=&\f{1}{2}g^3\f{\partial}{\partial g}\left( \f{1}{g^2}\f{\beta(g)}{g}    \right)
=\f{1}{2}g\f{\partial}{\partial g}\left( \f{\beta(g)}{g}    \right) -\f{\beta(g)}{g}  \nn\\
&=& \f{1}{2} \gamma_{F^2}(g)-\f{\beta(g)}{g}  
\eea
we obtain from eq. \eqref{eq:Z11}: 
\bea\label{eq:Z11usFP}
\f{Z_{11}^L}{Z_{11}}&=& \f{1}{\eps}\left(1-\f{\beta(g)}{\eps g} \right)^{-1}\left(\f{1}{2}g\f{\partial}{\partial g} \left(\f{\beta(g)}{g} \right)  - \f{\beta(g)}{g}\right)
\eea
According to $F^2=-4O_1$ in eq. \eqref{eq:O1F2}, we get the OPE coefficients in both notations:
 \bea
 \int C_1^{(F^2,F^2)}(x) e^{-ip\cdot x} d^{4-2\eps}x =-4 \int C_{1CZ}^{(S)}(x) e^{-ip\cdot x} d^{4-2\eps}x
 \eea
Then, the infinite additive renormalization of $C_1^{(F^2,F^2)}(p)$ is given by:
\bea\label{eq:35us}
4\f{Z_{11}^L}{Z_{11}}
\eea
that coincides with eq. \eqref{CT} for $O=F^2$. 

\section{Integral $I_{\td,\Delta_{F^2},\Delta_{F^2}}$}
\label{AppCC}

Eqs. \eqref{int_mRG_sdr} and \eqref{sec2_int} involve special cases of the integral \cite{Skenderis}:
\bea\label{eq:Igen}
I_{d,\Delta_1,\Delta_2}&=& \int  \f{1}
{ |x|^{\Delta_1}|x-z|^{\Delta_2} }     d^dx \nn\\
&=&(2\pi)^d \,C_{d,\f{\Delta_1}{2},\f{\Delta_2}{2}}\, |z|^{d-\Delta_1-\Delta_2}
\eea
with:
\be\label{eq:Cgen}
C_{d,\f{\Delta_1}{2},\f{\Delta_2}{2}}=\f{ 
\Gamma\big( \f{\Delta_1+\Delta_2-d}{2}\big)  
\Gamma\big( \f{d-\Delta_1}{2}\big)  \Gamma\big( \f{d-\Delta_2}{2}\big) 
}
{(4\pi)^{\f{d}{2}}\Gamma \big(\f{\Delta_1}{2}\big)\Gamma \big(\f{\Delta_2}{2}\big) \Gamma \big(d-\f{\Delta_1}{2}-\f{\Delta_2}{2}\big)}
\ee
For $d\to \td$ and $\Delta_1=\Delta_2\to \Delta_{F^2}$ in eqs. \eqref{eq:Igen} and 
\eqref{eq:Cgen} we get:
\bea\label{eq:Iapp}
I_{\td,\Delta_{F^2},\Delta_{F^2}}
&=& \int  \f{1}
{ |x|^{\Delta_{F^2} }|x-z|^{\Delta_{F^2}} }     d^\td x \nn\\
&=&(2\pi)^\td \,C_{\td,\f{\Delta_{F^2}}{2},\f{\Delta_{F^2}}{2}} 
\, |z|^{\td-2\Delta_{F^2}}
\eea
with:
\be\label{eq:Capp}
 C_{\td,\f{\Delta_{F^2}}{2},\f{\Delta_{F^2}}{2}} = \f{ 
\Gamma\big( \f{2\Delta_{F^2}-\td}{2}\big)  
\Gamma\big( \f{\td-\Delta_{F^2}}{2}\big)^2 
}
{(4\pi)^{\f{\td}{2}}\Gamma \big(\f{\Delta_{F^2}}{2}\big)^2
 \Gamma \big(\td-\Delta_{F^2}\big)}
 \ee
 Eq. \eqref{int_mRG_sdr} is obtained for $\td=4-2\eps$ and $\Delta_{F^2}=4-4\eps$ in eqs. \eqref{eq:Iapp} and \eqref{eq:Capp}:
 \bea\label{Int1_exp}
 I_{4-2\eps,4-4\eps,4-4\eps}&=&(2\pi)^{4-2\eps} \,C_{4-2\eps,2-2\eps,2-2\eps} 
\, |z|^{-4+6\eps}\nn\\
&=&(2\pi)^{4-2\eps}\f{ 
\Gamma( 2-3\eps)  
\Gamma( \eps)^2 
}
{(4\pi)^{2-\eps}\Gamma (2-2\eps)^2
 \Gamma (2\eps)}\, |z|^{-4+6\eps}\nn\\
 &=&\pi^2\pi^{-\eps}\bigg(\f{2}{\eps}+2+2\Gamma'(1)+\cdots\bigg)|z|^{-4+6\eps}\nn\\
 &=&\pi^2\bigg(\f{2}{\eps}+2+2\Gamma'(1)-2\log\pi+\cdots\bigg)|z|^{-4+6\eps}
\eea
where in the $\eps$ expansion of the ratio of $\Gamma$ functions we  have repeatedly employed  $z\Gamma(z)=\Gamma(z+1)$ and the Taylor expansion $\Gamma(1+\eps)=1+\eps\Gamma'(1)+\cdots$.
Eq. \eqref{Int1_exp} multiplied by $\f{\mu^{2\eps}}{|z|^{-4+4\eps}}$ yields as $\eps\to 0$:
 \bea
 \f{\mu^{2\eps}}{|z|^{-4+4\eps}} I_{4-2\eps,4-4\eps,4-4\eps}&=&\pi^2|z\mu|^{2\eps}\bigg(\f{2}{\eps}+2+2\Gamma'(1)-2\log\pi+\cdots\bigg)\nn\\
 &=&\pi^2\bigg(\f{2}{\eps}+4\log|z\mu|+2+2\Gamma'(1)-2\log\pi\bigg)
 \eea
 reported in eq. \eqref{int_mRG_sdr}.
 Eq. \eqref{sec2_int} is obtained for $\td\to d=4$ and $\Delta_{F^2}=4+\gamma_{F^2}$ in eqs. \eqref{eq:Iapp} and \eqref{eq:Capp}:
 \bea\label{Int2_exp}
 I_{4,4+\gamma_{F^2},4+\gamma_{F^2}}&=&(2\pi)^{4} \,C_{4,2+\f{\gamma_{F^2}}{2},2+\f{\gamma_{F^2}}{2}} 
\, |z|^{-4-2\gamma_{F^2}}\nn\\
&=&(2\pi)^{4}\f{ 
\Gamma( 2+\gamma_{F^2})  
\Gamma(\f{-\gamma_{F^2}}{2})^2 
}
{(4\pi)^{2}\Gamma (2+\f{\gamma_{F^2}}{2})^2
 \Gamma (-\gamma_{F^2})}\, |z|^{-4-2\gamma_{F^2}}\nn\\
 &=&\pi^2\bigg(-\f{4}{\gamma_{F^2}}+O\big(\gamma_{F^2}\big)\bigg)|z|^{-4-2\gamma_{F^2}}
\eea
expanded perturbatively in $\gamma_{F^2}$.
Eq. \eqref{Int2_exp} multiplied by $\f{\mu^{-\gamma_{F^2}}}{|z|^{-4-\gamma_{F^2}}}$ yields  perturbatively in $\gamma_{F^2}$:
\bea
\f{\mu^{-\gamma_{F^2}}}{|z|^{-4-\gamma_{F^2}}}I_{4,4+\gamma_{F^2},4+\gamma_{F^2}}&=&\pi^2\bigg(-\f{4}{\gamma_{F^2}}+O\big(\gamma_{F^2}\big)\bigg)|z\mu|^{-\gamma_{F^2}}\nn\\
&=&\pi^2\bigg(-\f{4}{\gamma_{F^2}}+4\log|z\mu|+O\big(\gamma_{F^2}\big)\bigg)
 \eea
 reported in eq. \eqref{sec2_int} for $\gamma_{F^2}=-2\beta_0g^2+\cdots$.

\section{Callan-Symanzik equation in $\td=4-2\eps$ dimensions in the coordinate representation} \label{CS1}

The CS equation \cite{C,S} in a massless QCD-like theory for connected 2-point correlators $G^{(2)}\equiv\braket{O(z) O(0)}'$ at distinct points $z\neq 0$ of a multiplicatively renormalizable gauge-invariant scalar operator $O$ with canonical dimension $\tilde D$ is a consequence of the independence of the bare correlator $G^{(2)}_0\equiv\braket{O(z) O(0)}_0'$:
\be\label{cs0_10}
G^{(2)}_0(z, \eps, g_0)=Z_O^{-2}(\eps,g(\mu))G^{(2)}(z, \mu, g(\mu))
\ee
 from the renormalization scale $\mu$: 
 \be\label{cs0_20}
\mu \f{d} {d\mu}G^{(2)}_0\bigg|_{\eps, g_0}=0
 \ee
with fixed bare parameters $g_0$ and $\eps$ of the dimensionally regularized theory. 
Substituting eq. \eqref{cs0_10} into eq. \eqref{cs0_20} we get the CS equation \cite{C,S,Zub,Pes}:
\be
\label{CS2_eps}
\left(\mu\f{\partial}{\partial \mu} + \beta(g,\eps)\f{\partial}{\partial g} + 2\gamma_{O}(g)\right) G^{(2)}(z, \mu, g(\mu)) = 0
\ee
with:
\be\label{eq:beta}
\beta(g,\eps)=\f{dg}{d\log\mu}\bigg|_{\eps, g_0}=-\eps g+\beta(g)
\ee
and: 
\be
\gamma_{O}(g)=-\f{d\log Z_O}{d\log\mu}\bigg|_{\eps, g_0}
\ee
the anomalous dimension of $O$. As the theory is massless to all orders of perturbation theory we define:
\be
G^{(2)}(z, \mu, g(\mu)) =\f{1}{z^{2\tilde D}} 	\bar G^{(2)}(z\mu, g(\mu)) 
\ee
with the dimensionless correlator $\bar G^{(2)}$ that satisfies eq. \eqref{CS2_eps} as well. Besides:
\be\label{CS2_eps_dl}
\left(z\cdot \f{\partial}{\partial z} + \beta(g,\eps)\f{\partial}{\partial g} + 2\gamma_{O}(g)\right) \bar G^{(2)}(z\mu, g(\mu)) = 0
\ee
as $ \bar G^{(2)} $ depends on $z$ via the product $z\mu$ only. Therefore:
\be\label{CS2_eps_dl0}
\left(z\cdot \f{\partial}{\partial z} + \beta(g,\eps)\f{\partial}{\partial g} + 2 \tilde D+2\gamma_{O}(g)\right) G^{(2)}(z, \mu, g(\mu)) = 0
\ee
The solution of eqs. \eqref{CS2_eps_dl0} and \eqref{CS2_eps} may then be written as:
\bea\label{pert2_eps}
 G^{(2)}(z, \mu, g(\mu)) &=& \frac{1}{z^{2\tilde D}}\,\bar G^{(2)}(z\mu, g(\mu))\nn\\
 &=&\frac{1}{z^{2\tilde D}}\, \mathcal{G}_{2}^{(O)}(\tg(z))\, Z^{(O)2}(\tg(z), g(\mu))
\eea
where $\mathcal{G}_2^{(O)}$ is an RG-invariant function of the running coupling $\tg(z)\equiv \tg(z\mu, g(\mu))$ in $\td$ dimensions that solves:
 \be\label{gz4_eps}
-\f{d \tg(z)}{d \log |z|}=\beta(\tg(z),\eps)
\ee
with the initial condition $\tg(1,g(\mu))=g(\mu)$, and the renormalized multiplicative factor $Z^{(O)}(\tg(z), g(\mu))$:
\begin{equation} \label{def_eps}
Z^{(O)}(\tg(z), g(\mu)) = \exp \int_{g(\mu)} ^{\tg(z)}      \frac{ \gamma_O (g) } {\beta(g,\eps)} dg 
\end{equation}
solves:
\be
\gamma_O (g(\mu))=-\f{d \log Z^{(O)}}{d \log\mu}
\ee
Moreover, it follows from the RG invariance of $\tg(z)$:
\bea\label{betar_eps1}
0&=&\f{d\tg(z)}{d\log\mu}\nn\\
&=&\f{\partial \tg(z)}{\partial \log\mu}+\f{\partial \tg(z)}{\partial g(\mu)}\f{dg(\mu)}{d\log\mu}\nn\\
&=&\f{d \tg(z)}{d \log|z|}+\f{\partial \tg(z)}{\partial g(\mu)}\f{dg(\mu)}{d\log\mu}\nn\\
&=&-\beta(\tg(z),\eps)+\f{\partial \tg(z)}{\partial g(\mu)}\beta(g(\mu),\eps)
\eea
according to eq. \eqref{gz4_eps} and the fact that $\tg(z)$ depends on $z$ via the product $z\mu$ only. As a consequence:
\bea\label{betar_eps}
\f{\partial \tg(z)}{\partial g(\mu)} &=&
\f{\beta(\tg(z),\eps)}{\beta(g(\mu),\eps)}
\eea

\subsection{Asymptotics of the running coupling} \label{D.1} 

 In $\td=4-2\eps$ dimensions the beta function:
\bea
\beta(g,\eps)&=&-\eps g + \beta(g)\nn\\
&=&-\eps g -\beta_0 g^3 -\beta_1 g^5 +\cdots
\eea
has a UV zero at $g=0$.  
The running coupling satisfies: 
\bea\label{int_eps}
&&\int_{g(\mu)}^{\tg(z)}\f{dg}
{\beta(g,\eps)} = -\int_{\mu^{-1}}^{|z|}d\log |z|
\eea 
We evaluate asymptotically eq. \eqref{int_eps} including the leading order of the perturbative expansion of $\beta(g)=-\beta_0g^3+\cdots$, with $\beta_0>0$:
\bea\label{AFeps_running}
&&\int_{g(\mu)}^{\tg(z)}\f{dg}
{-\eps g - \beta_0g^3+\cdots} =  -\int_{\mu^{-1}}^{|z|}d\log |z|
\eea
where both length scales, $|z|=\sqrt{z^2}$ and $\mu^{-1}$, are assumed to be close to zero in order for $\tg(z)$ and $g(\mu)$ to stay in a neighborhood of $g=0$.  
 By employing:
\be
\int_{g(\mu)}^{\tg(z)}\f{dg}
{g(\eps  + \beta_0g^2)} = \f{1}{2\eps}\log\f{g^2}{\eps+\beta_0g^2}\bigg|_{g(\mu)}^{\tg(z)}
\ee
eq. \eqref{AFeps_running} yields:
\bea\label{grun_eps_log}
&&\log\left( \f{\tg^2(z)}{g^2(\mu)}\,\f{\eps  + \beta_0g^2(\mu)+\cdots} {\eps  + \beta_0\tg^2(z)+\cdots}\right)=2\eps\log|z\mu|
\eea
 After exponentiating both sides:
\bea\label{ggeps}
&& \f{\tg^2(z)}{g^2(\mu)} = |z\mu|^{2\eps}\,\f{\eps  + \beta_0\tg^2(z)+\cdots}{\eps  + \beta_0g^2(\mu)+\cdots}
\eea
we obtain:
\be\label{ggeps0}
\f{\eps}{\tg^2(z)}+\beta_0+\cdots = |z\mu|^{-2\eps}\left( \f{\eps}{g^2(\mu)}+\beta_0+\cdots\right)
\ee
that implies: 
\be
\f{g^2(\mu)}{\tg^2(z)}=|z\mu|^{-2\eps}\left(1-\beta_0g^2(\mu)\f{|z\mu|^{2\eps}-1}{\eps} +\cdots\right)
\ee
whose inverse is:
\be\label{sol_eps_AF}
\f{\tg^2(z)}{g^2(\mu)}=\f{|z\mu|^{2\eps}}{
1-\beta_0g^2(\mu)\f{|z\mu|^{2\eps}-1}{\eps} +\cdots}
\ee
For $\eps\to 0$ eq. \eqref{sol_eps_AF} provides the solution in $d=4$ dimensions (appendix \ref{C.1}): 
\bea
\lim_{\eps\to 0}\f{\tg^2(z)}{g^2(\mu)}&=&\lim_{\eps\to 0}
\f{1+2\eps\log|z\mu|+\cdots}{1-2\beta_0g^2(\mu)\log|z\mu|+\cdots}\nn\\
&=& \f{1}{1-2\beta_0 g^2(\mu)\log|z\mu|+\cdots}
\eea 
 Instead, for fixed $\eps>0$ and around the $\f{1}{\tg(z)}$ singularity as $\tg(z)\to 0$, i.e. for large $\log|z\mu|$, the asymptotics of the solution in eq. \eqref{sol_eps_AF} reads:
\be
\tg^2(z)\sim \f{\eps}{\beta_0}|z\mu|^{2\eps}
\ee
 Moreover, eq. \eqref{grun_eps_log}:
\bea\label{grun_eps_log0}
&&-\f{1}{2\eps}\log \beta_0\bigg(1+\f{\eps}{\beta_0\tg^2(z)}+\cdots\bigg)-\log|z|=\nn\\
&&~~~~~~~~~~-\f{1}{2\eps}\log \beta_0\bigg(1+\f{\eps}{\beta_0g^2(\mu)}+\cdots\bigg)+\log\mu
\eea
implies that both sides are independent of $|z|$ and $\mu^{-1}$, i.e. RG-invariant, though scheme dependent. After subtracting the infinite constant $-\f{1}{2\eps}\log \beta_0$  from both sides and introducing a scheme-dependent integration constant $C$, eq. \eqref{grun_eps_log0} thus defines the RG-invariant mass scale $\Lrgi$ in $\td=4-2\eps$ dimensions:
 \bea
 \log\Lrgi&=&-\f{1}{2\eps}\log \bigg(1+\f{\eps}{\beta_0\tg^2(z)}+\cdots\bigg)-\log|z|-\f{C}{2\beta_0}
 \eea
 whose exponential reads:
 \bea\label{LUV_eps}
 \Lrgi&=&|z|^{-1}\bigg(1+\f{\eps}{\beta_0\tg^2(z)}+\cdots\bigg)^{-\f{1}{2\eps}} \exp\bigg(-\f{C}{2\beta_0}\bigg)
 \eea
 Solving eq. \eqref{LUV_eps} in terms of $\tg(z)$ as $\eps\to 0$ and asymptotically as $|z|\to 0^+$ we get:
 \bea\label{grunLUV_eps}
 \tg^2(z)&\sim&\f{1}{\beta_0} \f{1}{\f{\big(-1+ |z\Lrgi|^{-2\eps}+\cdots\big)e^{-\f{\eps C}{\beta_0}}}{\eps}}\nn\\
 &\sim&\f{1}{\beta_0}\f{1}{-2\log|z|\Lrgi-\f{C}{\beta_0}+\cdots}\nn\\
  &\sim&\f{1}{-2\beta_0\log|z|\Lrgi}\bigg(1-\f{C}{2\beta_0\log|z|\Lrgi}+\cdots\bigg)
 \eea
 Eqs. \eqref{LUV_eps} and \eqref{grunLUV_eps} yield as $\eps\to 0$ the corresponding eqs. \eqref{C0} and \eqref{alfa2} in $d=4$ dimensions.

\subsection{Asymptotics of $\braket{O(z)O(0)}'$}

 In perturbation theory\footnote{In the present paper the convention about the sign of the coefficients $\gamma^{(O)}_{i}$ agrees with \cite{MBM,MBN,BB}, but it is opposite to the standard one \cite{MB00}.}:
\bea \label{a}
\gamma_{O}(g)= -\gamma^{(O)}_{0} g^2 - \gamma^{(O)}_{1} g^4+ \cdots
\eea 
 with the first coefficient $\gamma^{(O)}_{0} $ universal, i.e. scheme independent. 
We employ $\gamma_{O}(g)$ to leading order 
to evaluate asymptotically the renormalized multiplicative factor $Z^{(O)}(\tg(z), g(\mu))$ for small coupling, i.e. $\tg(z), g(\mu)\to 0$:
\bea\label{ZO_IR_AF_eps}
Z^{(O)}(\tg(z), g(\mu))&\sim& \exp \int_{g(\mu)}^{\tg(z)}
\f{-\gamma_0^{(O)} g^2+\cdots }{-\eps g -\beta_0 g^3 +\cdots  }\,dg\nn\\
&\sim& \exp \bigg( \f{\gamma_0^{(O)}}{2}\int_{g^2(\mu)}^{\tg^2(z)}
\f{1+\cdots }{\eps  +\beta_0 g^2 +\cdots  }\,dg^2\bigg)\nn\\
&\sim& \exp \bigg( \f{\gamma_0^{(O)}}{2\beta_0}\log\left(\f{\eps  + \beta_0\tg^2(z)+\cdots}{\eps  + \beta_0g^2(\mu)+\cdots}\right)\bigg)\nn\\
&\sim& \exp \bigg( \f{\gamma_0^{(O)}}{2\beta_0}\log\left(\f{\tg^2(z)}{g^2(\mu)}|z\mu|^{-2\eps}\right)\bigg)\nn\\
&\sim&\bigg(\f{\tg^2(z)}{g^2(\mu)}|z\mu|^{-2\eps}\bigg)^{\f{\gamma_0^{(O)}}{2\beta_0}}
\eea
where we have employed eq. \eqref{ggeps}.
It follows the short-distance asymptotics of the 2-point correlator in eq. \eqref{pert2_eps} as $\tg(z), g(\mu)\to 0$:
\bea
\langle O(z) O(0) \rangle' &\sim& \f{\mathcal{G}^{(O)}_2(0)}{z^{2\tilde D}} \bigg(
\f{\tg^2(z)}{g^2(\mu)}|z\mu|^{-2\eps}\bigg)^{\f{\gamma_0^{(O)}}{\beta_0}}
\eea

\subsection{Asymptotics of $\braket{F^2(z)F^2(0)}'$}

 $Z^{(F^2)}$ admits a closed form in terms of the beta function according to eqs. \eqref{gF22} and \eqref{def_eps}:
\bea \label{zeps}
Z^{(F^2)}(\tilde g(z), g(\mu)) &=& \exp \int_{g(\mu)} ^{\tg(z)}      \frac{ \gamma_{F^2} (g) } {\beta(g,\eps)} dg \nonumber \\
                                                  &=& \exp \int_{g(\mu)} ^{\tg(z)}      \frac{\f{\partial}{\partial g}\left(\f{\beta(g,\eps)}{g}\right) } {\frac{\beta(g,\eps)}{g}} dg  \nonumber \\
                                                  &=& \f{ \beta(\tg(z),\eps) }{\tg(z) }  \f{g(\mu) } { \beta(g(\mu),\eps)}
\eea
 Correspondingly, we get for the solution of the CS equation:
\be
\label{CS_2ptF20}
 \braket{F^2(z) F^2(0)}' = \f{\mathcal{G}^{(F^2)}_2(\tg(z))}{z^{8-4\eps}}
\left (\f{ \beta(\tg(z),\eps) }{\tg(z) }\right )^2 \left ( \f{g(\mu) } { \beta(g(\mu),\eps)} \right)^2
 \ee
with $\tilde D=\td=4-2\eps$ the canonical dimension of $F^2$. 
For small coupling we evaluate asymptotically eq. \eqref{zeps}:
\bea\label{ZF_IR_AF_eps}
Z^{(F^2)}(\tg(z), g(\mu))&=& \f{\beta(\tg(z),\eps)}{\tg(z)}\f{g(\mu)}{\beta(g(\mu),\eps)}\nn\\
&\sim&\f{\eps  + \beta_0\tg^2(z)+\cdots}{\eps  + \beta_0g^2(\mu)+\cdots}\nn\\
&\sim&\f{\tg^2(z)}{g^2(\mu)}|z\mu|^{-2\eps}
\eea
that, indeed, for $\gamma_0^{(F^2)}=2\beta_0$ coincides with eq. \eqref{ZO_IR_AF_eps}. 
The  short-distance asymptotics of eq. \eqref{CS_2ptF20} as $\tg(z), g(\mu)\to 0$ follows:
\bea
\langle F^2(z) F^2(0) \rangle' &\sim& \f{\mathcal{G}^{(F^2)}_2(0)}{z^{8-4\eps}} 
\f{\tg^4(z)}{g^4(\mu)}|z\mu|^{-4\eps}
\eea

\subsection{Change of renormalization scheme}

Replacing $z\to \tilde\mu^{-1}$ in eq. \eqref{sol_eps_AF} and expanding for small $g(\mu)$, we obtain: 
\be
g^2(\tilde\mu)=\bigg(\f{\mu}{\tilde\mu}\bigg)^{2\eps}g^2(\mu)\bigg(1+2\beta_0g^2(\mu)\log
\f{\mu}{\tilde\mu}+\cdots\bigg)
\ee
Hence, the change of scheme $\tilde\mu=\alpha\mu$ yields for the coupling:
\bea\label{grun-RGeps}
g^2(\tilde\mu=\alpha\mu)&=&\alpha^{-2\eps}g^2(\mu)\big(1-2\beta_0g^2(\mu)\log\alpha +\cdots\big)\nn\\
&=&\alpha^{-2\eps}g^2(\mu)\big(1 +\cdots\big)
\eea
where the dots contain $O(g^2)$ contributions. 
We determine the scheme dependence of the complete 2-point correlator entering the FT in eq. \eqref{FFtot}  and  the 2-point correlator at distinct points in eq. \eqref{FFpert_eps2} by replacing $\mu\to \tilde\mu=\alpha\mu$
and rewriting $g^2\to g^2(\tilde\mu=\alpha\mu)$ in terms of $g(\mu)=g$ by means of eq. \eqref{grun-RGeps}. 
For the correlator at distinct points in eq. \eqref{FFpert_eps2}, we obtain to order $g^2$:
\bea\label{FTshiftcoordsep}
&&\braket{F^2(z)F^2(0)}'\big|_{\tilde\mu}\nn\\
&&=
\mathcal{G}^{(F^2)}_2(0)\bigg\{\f{1}{z^{8-4\eps}}\left(1-\f{2g^2(\tilde\mu)\beta_0}{\eps}\right)
 +\f{\tilde\mu^{2\eps}}{z^{8-6\eps}}g^2(\tilde\mu)\left(\delta_2^{(F^2)}
+\f{2\beta_0}{\eps}\right)+\cdots
\bigg\}\nn\\
&&= \mathcal{G}^{(F^2)}_2(0)\bigg\{\f{1}{z^{8-4\eps}}\left(1-\f{2g^2\beta_0\alpha^{-2\eps}}{\eps}\right)
+\f{\mu^{2\eps}\alpha^{2\eps}}{z^{8-6\eps}}g^2\alpha^{-2\eps}\left(\delta_2^{(F^2)}
+\f{2\beta_0}{\eps}\right)+\cdots
\bigg\}\nn\\
&&= \mathcal{G}^{(F^2)}_2(0)\bigg\{\f{1}{z^{8-4\eps}}\left(1-\f{2g^2\beta_0}{\eps}\right)
+\f{\mu^{2\eps}}{z^{8-6\eps}}g^2\left(\delta_2^{(F^2)}
+\f{2\beta_0}{\eps}\right)
\bigg\} \nn\\
&&~~+\f{\mathcal{G}^{(F^2)}_2(0) } {z^{8-4\eps}}\f{2g^2\beta_0}{\eps}(1-\alpha^{-2\eps})+\cdots
\nn\\
&&=
\braket{F^2(z)F^2(0)}'\big|_{\mu}+\f{\mathcal{G}^{(F^2)}_2(0) } {z^{8-4\eps}}\f{2g^2\beta_0}{\eps}(1-\alpha^{-2\eps})+\cdots
\nn\\
&&=
\braket{F^2(z)F^2(0)}'\big|_{\mu}+
\f{48(N^2-1) } {\pi^4 z^{8-4\eps}}\f{2g^2\beta_0}{\eps}(1-\alpha^{-2\eps})
+\cdots
\eea
where we have employed eq. \eqref{fixCZ2point_norm} in the last equality. 
For the complete correlator entering the FT in eq. \eqref{FFtot} we obtain to order $g^2$:
\bea
 &&\braket{F^2(z)F^2(0)}\big|_{\tilde\mu}\nn\\
 &&=\braket{F^2(z)F^2(0)}'\big|_{\tilde\mu} + 
 \f{N^2-1}{4\pi^2}\Delta^2\delta^{(\td)}(z)\tilde\mu^{-2\eps}\bigg\{
  -\f{1}{\eps}
 +\f{g^2(\tilde\mu)\beta_0}{\eps^2}\nn\\
&&~-\f{g^2(\tilde\mu)}{\eps}\bigg(
\f{1}{16\pi^2}\Big(\f{17}{2}-\f{5}{3}\f{N_f}{N}\Big)
-\beta_0\Big(\f{25}{6}+3\Gamma'(1)-\log \f{\pi}{4}\Big) \bigg)
+ \tilde A_0 +\tilde A_1g^2(\tilde\mu) \bigg\}+\cdots\nn\\
&&=\braket{F^2(z)F^2(0)}'\big|_{\mu}+
\f{48(N^2-1) } {\pi^4 z^{8-4\eps}}\f{2g^2\beta_0}{\eps}(1-\alpha^{-2\eps})
\nn\\&&~
+ 
 \f{N^2-1}{4\pi^2}\Delta^2\delta^{(\td)}(z)\mu^{-2\eps}\alpha^{-2\eps}
 \bigg\{
  -\f{1}{\eps}
 +\f{g^2\alpha^{-2\eps}
 \beta_0}{\eps^2}\nn\\
 &&~
-\f{g^2\alpha^{-2\eps}}{\eps}\bigg(
\f{1}{16\pi^2}\Big(\f{17}{2}-\f{5}{3}\f{N_f}{N}\Big)
-\beta_0\Big(\f{25}{6}+3\Gamma'(1)-\log \f{\pi}{4}\Big) \bigg)\nn\\
 &&~
+ \tilde A_0 +\tilde A_1g^2\alpha^{-2\eps} \bigg\}+\cdots\nn\\
&&
=\braket{F^2(z)F^2(0)}\big|_{\mu}+
\f{48(N^2-1) } {\pi^4 z^{8-4\eps}}\f{2g^2\beta_0}{\eps}(1-\alpha^{-2\eps})
\nn\\
&&~+ 
\f{N^2-1}{4\pi^2}\Delta^2\delta^{(\td)}(z)\mu^{-2\eps}
\bigg\{
\f{1-\alpha^{-2\eps}}{\eps}
+(1-\alpha^{-4\eps})
\bigg[
-\f{g^2\beta_0}{\eps^2}\nn\\
&&~+\f{g^2}{\eps}\bigg(
\f{1}{16\pi^2}\Big(\f{17}{2}-\f{5}{3}\f{N_f}{N}\Big)
-\beta_0\Big(\f{25}{6}+3\Gamma'(1)-\log \f{\pi}{4}\Big) \bigg)
-\tilde A_1g^2
\bigg]
\bigg\}+\cdots
 \eea
Hence, the scheme dependence of its FT in eq. \eqref{FFtot} to order $g^2$ reads as $\eps\to 0$:
\bea\label{FTshift}
&&\textrm{FT}\big[\braket{F^2(z)F^2(0)}\big|_{\tilde\mu}\big]
=\textrm{FT}\big[\braket{F^2(z)F^2(0)}\big|_{\mu}\big]\nn\\
&&~~+\f{N^2-1}{4\pi^2}p^4\bigg\{
\f{2g^2\beta_0}{\eps^2}\bigg(\f{\pi p^2}{4}\bigg)^{-\eps}
\Big(1+\eps\Big(\f{31}{6}+3\Gamma'(1)\Big)+\cdots\Big)(1-\alpha^{-2\eps})\bigg\}\nn\\
&&~~+
\f{N^2-1}{4\pi^2}p^4\mu^{-2\eps}\bigg\{
\f{1-\alpha^{-2\eps}}{\eps}
+(1-\alpha^{-4\eps})
\bigg[
-\f{g^2\beta_0}{\eps^2}\nn\\
&&~~+\f{g^2}{\eps}\bigg(
\f{1}{16\pi^2}\Big(\f{17}{2}-\f{5}{3}\f{N_f}{N}\Big)
-\beta_0\Big(\f{25}{6}+3\Gamma'(1)-\log \f{\pi}{4}\Big) \bigg)
-\tilde A_1g^2
\bigg]
\bigg\}+\cdots\nn\\
&&=\textrm{FT}\big[\braket{F^2(z)F^2(0)}\big|_{\mu}\big]+
\f{N^2-1}{4\pi^2}p^4\bigg\{
\nn\\
&&~~
\f{2g^2\beta_0}{\eps^2}\bigg(2\eps \log\alpha-2\eps^2\log^2\alpha-2\eps^2\log\alpha\Big(\log p^2+\log \f{\pi}{4}-\f{31}{6}-3\Gamma'(1)\Big)\bigg)
\nn\\
&&~~
+2\log\alpha
-\f{g^2\beta_0}{\eps^2}\Big(4\eps\log\alpha-8\eps^2\log^2\alpha-4\eps^2\log\alpha\log\mu^2\Big)
\nn\\
&&~~
+4 g^2\log\alpha\bigg(
\f{1}{16\pi^2}\Big(\f{17}{2}-\f{5}{3}\f{N_f}{N}\Big)
-\beta_0\Big(\f{25}{6}+3\Gamma'(1)-\log \f{\pi}{4}\Big) \bigg)\bigg\}+\cdots
\nn\\
&&=\textrm{FT}\big[\braket{F^2(z)F^2(0)}\big|_{\mu}\big]+
\f{N^2-1}{4\pi^2}p^4\bigg\{
2\log\alpha
-4g^2\beta_0\log\alpha \log\f{p^2}{\mu^2}\nn\\
&&~~+4g^2\beta_0\log^2\alpha
+2g^2\log\alpha\f{1}{16\pi^2}\Big(\f{73}{3}-\f{14}{3}\f{N_f}{N}\Big)
\bigg\}+\cdots
\eea
where we have employed eqs. \eqref{FT1eps} and \eqref{divdet}. Eq. \eqref{FTshift} manifestly agrees with the scheme dependence obtained by replacing $\mu\to\tilde\mu=\alpha\mu$ in $C_0^{(F^2,F^2)}(p)$ in eq. \eqref{C0S_tot}.

\subsection{Derivative of the 2-point correlator $\f{\partial\log\braket{O(z)O(0)}'}{\partial\log g}$ in the lhs of the LET} \label{D.6}

The lhs of the LET (IC) in $\td=4-2\eps$ dimensions for $z\neq 0$ reads:
\be\label{lhsRGdep}
\textrm{lhs}=\braket{O(z)O(0)}^\prime\f{\partial\log\braket{O(z)O(0)}^\prime}{\partial\log g}
 \ee
 where $g=g(\mu)$ and the 2-point correlator is the solution in eq. \eqref{pert2_eps} of the
  CS equation in $4-2\eps$ dimensions. By explicit computation we obtain: 
 \bea\label{OOdlog}
 \f{\partial\log\braket{O(z)O(0)}'}{\partial\log g}&=&\f{\partial\log  \mathcal{G}_{2}^{(O)}(\tg(z))}{\partial\log g}+2\f{\partial\log Z^{(O)}(\tg(z), g(\mu)) }{\partial\log g}\nn\\
 &=&g\f{\partial\log  \mathcal{G}_{2}^{(O)}(\tg(z))}{\partial g}+2g\f{\partial}{\partial g}
 \int_{g}^{\tg(z)}\f{\gamma_O(g')}{\beta(g',\eps)}\,dg'\nn\\
 &=&g\f{\partial\log  \mathcal{G}_{2}^{(O)}(\tg(z))}{\partial g}+2g
 \bigg(\f{\partial\tg(z)}{\partial g}\f{\gamma_O(\tg(z))}{\beta(\tg(z),\eps)}-\f{\gamma_O(g)}{\beta(g,\eps)}\bigg)\nn\\
 &=&g\f{\partial\log  \mathcal{G}_{2}^{(O)}(\tg(z))}{\partial g}+2g
 \bigg(\f{\beta(\tg(z),\eps)}{\beta(g,\eps)}\f{\gamma_O(\tg(z))}{\beta(\tg(z),\eps)}-\f{\gamma_O(g)}{\beta(g,\eps)}\bigg)\nn\\
&=&g\f{\partial\log  \mathcal{G}_{2}^{(O)}(\tg(z))}{\partial g}+\f{2g}{\beta(g,\eps)}\big(\gamma_O(\tg(z))-\gamma_O(g)\big)\nn\\
&=&g\f{\partial\log  \mathcal{G}_{2}^{(O)}(\tg(z))}{\partial g}-\f{2Z_{F^2}}{\eps}\big(\gamma_O(\tg(z))-\gamma_O(g)\big)
 \eea
where -- in the order -- we have employed eq. \eqref{def_eps} for $Z^{(O)}$, eq. \eqref{betar_eps} for the ratio of the running couplings, eq. \eqref{ZF2} and $\beta(g,\eps)=-\eps g+\beta(g)$ that imply $Z_{F^2}^{-1}=-\f{\beta(g,\eps)}{\eps g}$.
To order $g^2$ in perturbation theory:
\bea
\mathcal{G}_{2}^{(O)}(\tg(z))&=&\mathcal{G}_{2}^{(O)}(0)(1+\delta_2^{(O)}\tg^2(z)+\cdots)\nn\\&=&\mathcal{G}_{2}^{(O)}(0)(1+\delta_2^{(O)}|z\mu|^{2\eps}g^2+\cdots)
\eea
and:
\bea
\gamma_O(\tg(z))-\gamma_O(g)&=&-\gamma_0^{(O)}(\tg^2(z)-g^2)+\cdots\nn\\
&=&-\gamma_0^{(O)}g^2(|z\mu|^{2\eps}-1)+\cdots
\eea
so that eq. \eqref{OOdlog}  reads to order $g^2$ as $\eps\to 0$:
\bea\label{OOdlogg2}
\f{\partial\log\braket{O(z)O(0)}'}{\partial\log g}\bigg|_{\textrm{to order }g^2}&=&
2 \delta_2^{(O)}|z\mu|^{2\eps}g^2+\f{2}{\eps}\gamma_0^{(O)}g^2(|z\mu|^{2\eps}-1)\nn\\
&=&2 \delta_2^{(O)}g^2+4\gamma_0^{(O)}g^2 \log|z\mu|+\cdots
\eea
where the dots stand for $O(\eps)$ contributions.
For $O=F^2$, eq. \eqref{OOdlogg2} yields:
\bea\label{FFdlogg2}
\f{\partial\log\braket{F^2(z)F^2(0)}'}{\partial\log g}\bigg|_{\textrm{to order }g^2}&=&
2 \delta_2^{(F^2)}g^2+8\beta_0g^2 \log|z\mu|+\cdots
\eea
where we have employed $\gamma_0^{(F^2)}=2\beta_0$.

\subsection{Scheme dependence of the lhs of the LET for $O=F^2$ to order $g^2$}

The scheme dependence of the derivative in eq. \eqref{FFdlogg2} 
is implied by the one of the correlator at distinct points in eq. \eqref{FTshiftcoordsep}. For $\tilde\mu=\alpha\mu$ and setting $g(\mu)=g$, the derivative of eq. \eqref{FTshiftcoordsep} yields:
  \bea\label{FFsepcoordder}
&&\f{\partial\log\braket{F^2(z)F^2(0)}'\big|_{\tilde\mu}}{\partial\log g(\tilde\mu)}\nn\\
&&~~~~=\f{\partial}{\partial\log g}\log\bigg( 
\braket{F^2(z)F^2(0)}'\big|_{\mu}
+\f{\mathcal{G}^{(F^2)}_2(0) } {z^{8-4\eps}}\f{2\beta_0g^2}{\eps}(1-\alpha^{-2\eps})
+\cdots\bigg)\nn\\
&&~~~~=
\f{\partial}{\partial\log g}\bigg\{\log\braket{F^2(z)F^2(0)}'\big|_{\mu}\nn\\
&&~~~~~~+\log\bigg( 1+
\f{1}{\braket{F^2(z)F^2(0)}'\big|_{\mu}}
\f{\mathcal{G}^{(F^2)}_2(0) } {z^{8-4\eps}}\f{2\beta_0g^2}{\eps}(1-\alpha^{-2\eps})
+\cdots\bigg)\bigg\}\nn\\
&&~~~~=\f{\partial\log\braket{F^2(z)F^2(0)}'\big|_{\mu}}{\partial\log g}\nn\\
&&~~~~~~+\f{\partial}{\partial\log g}
\log\bigg( 1+
\f{1}{\f{\mathcal{G}^{(F^2)}_2(0)} {z^{8-4\eps}}(1+O(g^2))}
\f{\mathcal{G}^{(F^2)}_2(0) } {z^{8-4\eps}}\f{2\beta_0g^2}{\eps}(1-\alpha^{-2\eps})
+\cdots\bigg)\nn\\
&&~~~~=\f{\partial\log\braket{F^2(z)F^2(0)}'\big|_{\mu}}{\partial\log g}
+\f{\partial}{\partial\log g}
\log\bigg( 1+ 
\f{2\beta_0g^2}{\eps}(1-\alpha^{-2\eps})
+\cdots\bigg)\nn\\
&&~~~~=\f{\partial\log\braket{F^2(z)F^2(0)}'\big|_{\mu}}{\partial\log g}
+ \f{4\beta_0g^2}{\eps}(1-\alpha^{-2\eps})
+\cdots\nn\\
&&~~~~=\f{\partial\log\braket{F^2(z)F^2(0)}'\big|_{\mu}}{\partial\log g}
+ 8\beta_0g^2\log\alpha
+\cdots
\eea
 where we have employed to the relevant order as $\eps\to 0$:
 \be
 \f{\partial}{\partial\log g(\tilde\mu)}= \f{\partial}{\partial\log g}\f{\partial\log g}{\partial\log g(\tilde\mu)}=\f{\partial}{\partial\log g}\big(1+\cdots\big)
 \ee
The result in eq. \eqref{FFsepcoordder} can also be straightforwardly obtained by replacing $\mu\to\tilde\mu=\alpha\mu$ in eq. \eqref{FFdlogg2}.
Hence, after factoring out the $2$-point correlator $\braket{F^2(z)F^2(0)}'$ in eq. \eqref{lhsRGdep} for $O=F^2$, the remaining scheme dependence of the lhs of the LET to order $g^2$ amounts to:
\bea\label{RGdep2}
\f{\partial\log\braket{F^2(z)F^2(0)}'\big|_{\tilde\mu}}{\partial\log g(\tilde\mu)}\bigg|_{\textrm{to order }g^2}&=&
\f{\partial\log\braket{F^2(z)F^2(0)}'\big|_{\mu}}{\partial\log g}\bigg|_{\textrm{to order }g^2}
+ 8\beta_0g^2\log\alpha\nn\\
\eea

\subsection{Scheme dependence of the rhs of the LET for $O=F^2$ to order$\,g^2$}

For $O=F^2$, the rhs of the LET (IC) to order $g^2$ in $\td=4-2\eps$ dimensions reads for $z\neq 0$:
\bea\label{rhsFRG}
\textrm{rhs}&=&\f{1}{2}Z_{F^2} \int \braket{F^2(z) F^2(0) F_0^2(x)}d^{\td} x\nn\\
&&\,\,\,\,\,\, - \f{1}{2}Z_{F^2}\int 4\big(\delta^{(\td)}(x-z)+\delta^{(\td)}(x)\big)\braket{F^2(z)F^2(0)}' d^{\td} x\nn\\
&&\,\,\,\,\,\, +Z_{F^2} \f{2\gamma_{F^2}(g)}{\eps} \braket{F^2(z)F^2(0)}'
\,\bigg|_{\textrm{up to order } g^2}\nn\\
&=&\f{1}{2}\int  \braket{F^2(z)F^2(0)F^2(x)}' d^{\td}x\nn\\
&&+\f{1}{2}
\int 
\tilde B_0^{(F^2)}g^2
\Big(\delta^{(\td)}(x)+\delta^{(\td)}(x-z)\Big)\braket{F^2(z)F^2(0)}' d^{\td}x
\nn\\
&&\,\,\,\,\,\, + \f{2\gamma_{F^2}(g)}{\eps} \braket{F^2(z)F^2(0)}'
\,\bigg|_{\textrm{up to order } g^2}\nn\\
&=&2\beta_0g^2
\bigg(\f{2}{\eps}+4\log|z\mu|-2\log\pi+2+2\Gamma'(1)\bigg)\braket{F^2(z)F^2(0)}'\nn\\
&&+\tilde B_0^{(F^2)}g^2\braket{F^2(z)F^2(0)}' - \f{4\beta_0g^2}{\eps} \braket{F^2(z)F^2(0)}'\,\bigg|_{\textrm{to order } g^2}
\eea
where we have employed eq. \eqref{ICbarecorr_sdr} for $O=F^2$:
\bea
&&\braket{F^2(z) F^2(0) F_0^2(x)}\bigg|_{\textrm{up to order }  g^2} \nn\\
&&\hspace{2.5cm}=(4 +\tilde B_0^{(F^2)}g^2)(\delta^{(\td)}(x-z)+\delta^{(\td)}(x)) \braket{F^2(z)F^2(0)}'\nn\\
&&\hspace{2.5cm}~~+Z^{-1}_{F^2} \braket{F^2(z) F^2(0) F^2(x)}'  \bigg|_{\textrm{up to order }  g^2}
\eea
and eqs. \eqref{int_mRG_sdr} and \eqref{eq:eps_sdr}, with $\gamma_{F^2}(g)=-2\beta_0g^2+\cdots$. 
After factoring out $\braket{F^2(z)F^2(0)}'$ in eq. \eqref{rhsFRG}, the scheme dependence of the rhs of the LET to order $g^2$ amounts to:
\bea
&&2\beta_0g^2
\bigg(\f{2}{\eps}+4\log|z\mu|-2\log\pi+2+2\Gamma'(1)\bigg)+\tilde B_0^{(F^2)}g^2- \f{4\beta_0g^2}{\eps}\nn\\
&&=2\beta_0g^2
\bigg(4\log|z\mu|-2\log\pi+2+2\Gamma'(1)\bigg)+\tilde B_0^{(F^2)}g^2
\eea
where the divergent contact term has cancels the divergence of the integrated contribution at distinct points. 
Correspondingly, the result in the $\tilde\mu$-scheme is related to the one in the $\mu$-scheme, with $\tilde\mu=\alpha\mu$ and $g(\mu)=g$, as $\eps\to 0$ as follows:
\bea\label{RGschemedepRHS}
&&2\beta_0g^2(\tilde\mu)
\bigg(4\log|z\tilde\mu|-2\log\pi+2+2\Gamma'(1)\bigg)+\tilde B_0^{(F^2)}g^2(\tilde\mu)\nn\\
&&= 2\beta_0g^2
\bigg(4\log|z\mu|+4\log\alpha-2\log\pi+2+2\Gamma'(1)\bigg)+\tilde B_0^{(F^2)}g^2\nn\\
&&= 2\beta_0g^2
\bigg(4\log|z\mu|-2\log\pi+2+2\Gamma'(1)\bigg)+\tilde B_0^{(F^2)}g^2
+8\beta_0g^2\log\alpha
\eea
where we have employed eq. \eqref{grun-RGeps} for $\eps\to 0$.

\subsection{Matching the finite terms in the LET  for $O=F^2$ to order $g^2$} \label{D.9}

Importantly, we verify that the matching of the lhs and rhs of the LET is scheme independent. Indeed, after factoring out $\braket{F^2(z)F^2(0)}'$
on both sides of the LET, the scheme dependence of the lhs is given by eq. \eqref{RGdep2}, i.e. it consists of the shift by the term $8\beta_0 g^2\log\alpha$ that equates the one of the rhs of the LET in eq. \eqref{RGschemedepRHS}. \par
Hence, for $O=F^2$ the LET implies the following scheme-independent matching to order $g^2$ as $\eps\to 0$:
\bea
&&\f{\partial\log\braket{F^2(z)F^2(0)}'}{\partial\log g}\bigg|_{\textrm{to order }  g^2}\nn\\
&&\hspace{1.8cm}=
2\beta_0g^2
\bigg(4\log|z\mu|-2\log\pi+2+2\Gamma'(1)\bigg)+\tilde B_0^{(F^2)}g^2
\eea
that, by means of eq. \eqref{FFdlogg2}, reads:
\bea
&&2 \delta_2^{(F^2)}g^2+8\beta_0g^2 \log|z\mu|\nn\\
&&\hspace{1.8cm}=
2\beta_0g^2
\bigg(4\log|z\mu|-2\log\pi+2+2\Gamma'(1)\bigg)+\tilde B_0^{(F^2)}g^2
\eea
The terms proportional to $\log|z\mu|$ indeed match. The matching of the remaining finite terms implies the relation: 
\bea\label{LETcons}
\delta_2^{(F^2)}=\f{1}{2}\tilde B_0^{(F^2)}+\beta_0
\bigg(2+2\Gamma'(1)-2\log\pi\bigg)
\eea
between the  coefficients $\delta_2^{(F^2)}$ and $\tilde B_0^{(F^2)}$ computed in the same RG scheme. \par
The LET thus relates in a scheme-independent way two scheme-dependent quantities that enter different OPE coefficients computed in perturbation theory.  Specifically, $\delta_2^{(F^2)}$ -- hence the coefficient $2(c_1-\beta_0)$ of the subleading $\log\f{p^2}{\mu^2}$ to order $g^2$ -- that enters  $C_0^{(F^2,F^2)}(p)$ computed in \cite{Z1,Kataev,Z3}
and $\tilde B_0^{(F^2)}$ -- hence $B_{1,1}$ -- that enters $C_1^{(F^2,F^2)}(p)$ computed in \cite{Z1,Z2}. \par
The value of $\tilde B_0^{(F^2)}$  determined in \cite{Z1,Z2} in the 
 $\overline{MS}$ scheme in our notation reads: 
\bea\label{fixB11}
\tilde B_{0\,CZ}^{(F^2)}&=&-4\tilde B_{1,1}\nn\\
&=&-4 B_{1,1}-4\beta_0\bigg(2+3\Gamma'(1)-\log\f{\pi}{4}\bigg)\nn\\
&=&-4\f{1}{(4\pi)^2}\bigg(-\f{49}{9}+\f{10}{9}\f{N_f}{N}\bigg)-4\beta_0\bigg(2+3\Gamma'(1)-\log\f{\pi}{4}\bigg)
\eea
where in the last line we have employed the value of $B_{1,1}$ reported in eq. \eqref{relB}.
The value of $\delta_2^{(F^2)}$ determined in \cite{Z1,Z3} in the $\overline{MS}$ scheme in our notation reads: 
\bea\label{con2}
\delta_{2\,CZ}^{(F^2)}&=& -2(c_1-\beta_0)-2\beta_0\bigg(\f{10}{3}+2\Gamma'(1)+\log{4}\bigg)\nn\\
&=&\f{1}{(4\pi)^2}\bigg(\f{73}{3}-\f{14}{3}\f{N_f}{N}\bigg)
-2\beta_0\bigg(\f{10}{3}+2\Gamma'(1)+\log{4}\bigg)\nn\\
&=&-2\f{1}{(4\pi)^2}\bigg(-\f{49}{9}+\f{10}{9}\f{N_f}{N}\bigg)
-2\beta_0\bigg(\f{3}{2}+2\Gamma'(1)+\log{4}\bigg)\nn\\
&=&-2B_{1,1}-2\beta_0\bigg(\f{3}{2}+2\Gamma'(1)+\log{4}\bigg)\nn\\
&=&\f{1}{2}\tilde B_{0\,CZ}^{(F^2)}+2\beta_0\bigg(2+3\Gamma'(1)-\log\f{\pi}{4}\bigg)-2\beta_0\bigg(\f{3}{2}+2\Gamma'(1)+\log{4}\bigg)\nn\\
&=&\f{1}{2}\tilde B_{0\,CZ}^{(F^2)}+\beta_0\bigg( 1+2\Gamma'(1)-2\log\pi\bigg)
\eea
where we have employed eqs. \eqref{d2f2CZ}, \eqref{c1CZb}, \eqref{relB} and \eqref{fixB11}.
Eq. \eqref{con2} differs from the LET constraint in eq. \eqref{LETcons} for a term $\beta_0$:
 \be
 \bigg(\delta_{2\,CZ}^{(F^2)}-\f{1}{2}\tilde B_{0\,CZ}^{(F^2)}\bigg)=
\bigg(\delta_2^{(F^2)}-\f{1}{2}\tilde B_0^{(F^2)}\bigg)-\beta_0
\ee
whereas all terms with $\Gamma'(1)$, $\log\pi$ and $\log 4$ do agree.
Tentatively, we suggest two alternative explanations. \par
Either, eq. \eqref{con2} for the coefficient in $C_0^{(F^2,F^2)}(p)$ in \cite{Z1,Z3} or eq. \eqref{fixB11} for the coefficient in $C_1^{(F^2,F^2)}(p)$ in \cite{Z1,Z2} need a correction. 
Or, our ansatz conformal in the limit $\epsilon \rightarrow 0$ for the solution of the LET to order $g^2$ -- despite reproducing exactly the OPE to the relevant order -- introduces an error of order $\epsilon$ that somehow generates a finite discrepancy once multiplied by some divergence in the rhs of the LET.

\section{Callan-Symanzik equation in $d=4$ dimensions in the coordinate representation} \label{CS0}

 The CS equation in $d=4$ dimensions 
for $G^{(2)}\equiv\braket{O(z) O(0)}'$ at distinct points $z\neq 0$ and a multiplicatively renormalizable gauge-invariant scalar operator $O$ with canonical dimension $D$ 
follows from eq. \eqref{CS2_eps}
as $\eps\to 0$:
\be
\label{CS2_eps0}
\left(\mu\f{\partial}{\partial \mu} + \beta(g)\f{\partial}{\partial g} + 2\gamma_{O}(g)\right) G^{(2)}(z, \mu, g(\mu)) = 0
\ee
whose solution reads:
\bea\label{pert2_eps0}
 G^{(2)}(z, \mu, g(\mu)) &=& \frac{1}{z^{2D}}\,\bar G^{(2)}(z\mu, g(\mu))\nn\\
 &=&\frac{1}{z^{2D}}\, \mathcal{G}_{2}^{(O)}(g(z))\, Z^{(O)2}(g(z), g(\mu))
\eea
  in terms of the RG-invariant function $\mathcal{G}_2^{(O)}$ of the running coupling $g(z)\equiv g(z\mu, g(\mu))$ that solves: 
 \be\label{gz4}
-\f{d g(z)}{d \log|z|}=\beta(g(z))
\ee
with the initial condition $g(1,g(\mu))=g(\mu)$ and the renormalized multiplicative factor $Z^{(O)}(g(z), g(\mu))$:
\begin{equation} \label{def_eps0}
Z^{(O)}(g(z), g(\mu)) = \exp \int_{g(\mu)} ^{g(z)}      \frac{ \gamma_O (g) } {\beta(g)} dg 
\end{equation}
that solves:
\be
\gamma_O (g(\mu))=-\f{d \log Z^{(O)}}{d \log\mu}
\ee
Moreover, the RG invariance of $g(z)$ implies:
\bea\label{betar_eps01}
0&=&\f{dg(z)}{d\log\mu}\nn\\
&=&\f{\partial g(z)}{\partial \log\mu}+\f{\partial g(z)}{\partial g(\mu)}\f{dg(\mu)}{d\log\mu}\nn\\
&=&\f{d g(z)}{d \log|z|}+\f{\partial g(z)}{\partial g(\mu)}\f{dg(\mu)}{d\log\mu}\nn\\
&=&-\beta(g(z))+\f{\partial g(z)}{\partial g(\mu)}\beta(g(\mu))
\eea
according to eq. \eqref{gz4} and the fact that $g(z)$ depends on $z$ via the product $z\mu$ only. As a consequence:
\bea\label{betar_eps0}
\f{\partial g(z)}{\partial g(\mu)} &=&
\f{\beta(g(z))}{\beta(g(\mu))}
\eea

\subsection{Asymptotics of the running coupling} \label{C.1}

The beta function has a zero at $g=0$:
\bea \label{beta00}
\beta(g)=-\beta_0 g^3 - \beta_1 g^5 -\beta_2 g^7 +  \cdots
\eea
and for $\beta_0 > 0$ the theory is AF at short distances.
 From the integral of the inverse of the beta function: 
\bea\label{intUV}
\int_{g(\mu)}^{g(z)}\f{dg}
{-\beta_{0} g^3 - \beta_1 g^5    +\cdots } = -\int_{\mu^{-1}}^{|z|}d\log |z|
\eea
where the length scales, $|z|=\sqrt{z^2}$ and $\mu^{-1}$, are assumed to be close to zero in order for $g(z)$ and $g(\mu)$ to stay in a neighborhood of $g=0$, we obtain:
\bea\label{exp-grun}
\frac{1}{g^2(z)}-\f{1}{g^2(\mu)}=-2\beta_0\log|z\mu|-2\f{\beta_1}{\beta_0}\log\left (\f{g(z)}{g(\mu)}\right)+\cdots
\eea 
or, equivalently:
\bea \label{rgi_exp_uv}
&&\f{1}{g^2(z)}+2\beta_0\log |z|+2\f{\beta_1}{\beta_0}\log {g(z)}+C+\cdots \nonumber \\
&&\hspace{0.5truecm}= \f{1}{g^2(\mu)}-2\beta_0\log \mu +2\f{\beta_1}{\beta_0}\log {g(\mu)}+C+\cdots
\eea
where the scheme-dependent integration constant $C$ arises from the indefinite version of the integral in eq. \eqref{intUV}. 
The above equation implies the existence of an RG-invariant mass scale $\Lrgi$:
\bea\label{C0}
\Lrgi &=& \mu \exp\left (-\f{1}{2\beta_0g^2(\mu)} \right) g^{-\f{\beta_1}{\beta_0^2}}(\mu) \exp\left( -\f{C}{2\beta_0} +\cdots \right) \nonumber \\
&=& |z|^{-1} \exp\left (-\f{1}{2\beta_0g^2(z)} \right) g^{-\f{\beta_1}{\beta_0^2}}(z) \exp\left( -\f{C}{2\beta_0} +\cdots \right)
\eea
with the dots a scheme-dependent series in even powers of $g(\mu)$ and $g(z)$ respectively. 
Solving eq. \eqref{C0} asymptotically as $|z|\to 0^+$ we get:
\bea \label{alfa2}
g^2(z) &\sim& \dfrac{1}{-2\beta_0 \log({|z|\Lrgi })} 
\bigg(1+\frac{\beta_1}{2\beta_0^2}\frac{\log(-2\beta_0\log({|z|\Lrgi }))}{\log({|z|\Lrgi })}\nn\\
&& -\f{C}{2\beta_0\log  ({|z|\Lrgi }  )  } 
+ \cdots\bigg)
\eea
Incidentally, the $\overline{MS}$ scheme is defined by choosing $C=(\beta_1/\beta_0)\log\beta_0$ that cancels the last term in eq. \eqref{alfa2} against the term $\frac{\beta_1}{2\beta_0^2}\f{\log2\beta_0}{\log({|z|\Lrgi } ) }$ in the first line.
The universal asymptotics of the running coupling as $ |z|\to 0^+$ follows:
\be \label{alfa}
g^2(z) \sim \dfrac{1}{-2\beta_0 \log({|z|\Lrgi })} 
\left(
1+\frac{\beta_1}{2\beta_0^2}\frac{\log(-2\beta_0\log({|z|\Lrgi } ))}{\log({|z|\Lrgi })} 
\right)
\ee 
Perturbatively to order $g^4(\mu)$ we obtain from eq. \eqref{exp-grun}:
\bea
\label{g2loop_large}
\f{g^2(\mu)}{g^2(z)}& =&  1 - g^2(\mu)2\beta_0\log|z\mu| +g^2(\mu)\f{\beta_1}{\beta_0}
\log\bigg(\f{g^2(\mu)}{g^2(z)} \bigg)+\cdots\nn\\
&=& 1 - g^2(\mu)2\beta_0\log|z\mu|-g^4(\mu)2\beta_1\log|z\mu|+\cdots
\eea
for $\log|z\mu|$ of order one,
where we have employed in the second equality:
\be
\log\bigg(\f{g^2(\mu)}{g^2(z)} \bigg)=\log(1- g^2(\mu)2\beta_0\log|z\mu|+\cdots)=-g^2(\mu)2\beta_0\log|z\mu|+\cdots
\ee
Eq. \eqref{g2loop_large} yields:
\be
\label{g2loop}
g^2(z) =  g^2(\mu)\left(1 + g^2(\mu)2\beta_0\log|z\mu| + g^4(\mu)(2\beta_1\log|z\mu| 
 + 4\beta_0^2\log^2|z\mu|)+\cdots\right)
\ee

\subsection{Asymptotics of $\braket{O(z)O(0)}'$}\label{D.2AF}

 From eq. \eqref{def_eps0} it follows asymptotically at short distances by means of eq. (\ref{a}):
\bea \label{zeta00}
&&Z^{(O)}(g(z), g(\mu)) \sim \left(\frac{g(z)}{g(\mu)}\right)^{\frac{\gamma_0^{(O)}}{\beta_0}} \exp \bigg( \frac{\gamma_1^{(O)}  \beta_0 - \gamma_{0}^{(O)} \beta_1}{2 \beta_0^2} (g^2(z)-g^2(\mu))+\cdots \bigg)\nn\\
&&\sim \left(\frac{1}{-g^2(\mu)2\beta_0 \log({|z|\Lrgi })}
\left(
1+\frac{\beta_1}{2\beta_0^2}\frac{\log(-\beta_0\log({|z|\Lrgi } ))}{\log({|z|\Lrgi })} 
\right)\right)^{\frac{\gamma_0^{(O)}}{2\beta_0}} Z^{(O)'}(g(\mu))\nn\\
\eea
where:
\be\label{ZOF}
 Z^{(O)'}(g(\mu))=\exp \bigg(- \frac{\gamma_1^{(O)}  \beta_0 - \gamma_{0}^{(O)} \beta_1}{2 \beta_0^2} g^2(\mu)+\cdots \bigg)
 \ee
and we have employed eq. \eqref{alfa} for $g(z)$.
It follows from eq. \eqref{zeta00} the asymptotics  at short distances of the 2-point correlator\cite{MBM,MBN,BB}: 
\bea
\label{CSS0}
 \langle O(z) O(0) \rangle' &\sim& \f{\mathcal{G}^{(O)}_2(0)}{z^{2D}} \left( \f{g(z)}{g(\mu)} \right)^{\f{2 \gamma^{(O)}_{0}}{\beta_0}} 
\exp \bigg( \frac{\gamma_1^{(O)}  \beta_0 - \gamma_{0}^{(O)} \beta_1}{ \beta_0^2} (g^2(z)-g^2(\mu))+\cdots \bigg)
\nn\\
 & \sim& \f{\mathcal{G}^{(O)}_2(0)}{z^{2D}}
 \bigg(\frac{1}{-g^2(\mu)2\beta_0 \log({|z|\Lrgi })}
 \nn\\
&&\bigg(
1+\frac{\beta_1}{2\beta_0^2}\frac{\log(-\beta_0\log({|z|\Lrgi } ))}{\log({|z|\Lrgi })} 
\bigg)\bigg)^{\f{ \gamma^{(O)}_{0}}{\beta_0}} Z^{(O)'2}(g(\mu))
 \eea
Perturbatively, from eqs. \eqref{zeta00} and \eqref{g2loop} we get:
\bea \label{zeta00_small}
Z^{(O)}(g(z), g(\mu)) &=& 1+g^2(\mu) \gamma_0^{(O)}\log|z\mu|+g^4(\mu)\bigg(\gamma_1^{(O)}\log|z\mu| \nn\\
&&+\big(\gamma_0^{(O)}\beta_0+\f{\gamma_0^{(O)2}}{2} \big)\log^2|z\mu|
\bigg)+
\cdots
\eea

\subsection{Asymptotics of $\braket{F^2(z)F^2(0)}'$}\label{D.3}

 $Z^{(F^2)}$ admits the closed form:
\bea \label{z}
Z^{(F^2)}(g(z), g(\mu)) &=& \exp \int_{g(\mu)} ^{g(z)}      \frac{ \gamma_{F^2} (g) } {\beta(g)} dg \nonumber \\
                                                  &=& \exp \int_{g(\mu)} ^{g(z)}      \frac{\f{\partial}{\partial g}\left(\f{\beta(g)}{g}\right) } {\frac{\beta(g)}{g}} dg  \nonumber \\
                                                  &=& \f{ \beta(g(z)) }{g(z) }  \f{g(\mu) } { \beta(g(\mu))}
\eea
 that follows from eq. \eqref{zeps} as $\eps\to 0$.
Correspondingly:
\be
\label{CS_2ptF2}
 \braket{F^2(z) F^2(0)}' = \f{\mathcal{G}^{(F^2)}_2(g(z))}{z^{8}}
\left (\f{ \beta(g(z)) }{g(z) }\right )^2 \left ( \f{g(\mu) } { \beta(g(\mu))} \right)^2
 \ee
whose asymptotics at short distances reads: 
\bea
&&\braket{ F^2(z) F^2(0)}' \sim \f{\mathcal{G}^{( F^2)}_2(0)}{z^{8}}g^4(z)\beta_0^2
\left( \f{g(\mu) } { \beta(g(\mu))} \right)^2\nn\\
&&\hspace{1.5truecm}\sim\f{\mathcal{G}^{( F^2)}_2(0)}{z^{8}}
\left(\f{1}{-2\beta_0\log(|z|\Lrgi)}
\left(
1+\frac{\beta_1}{2\beta_0^2}\frac{\log(-\beta_0\log({|z|\Lrgi } ))}{\log({|z|\Lrgi })} 
\right)
\right)^2\nn\\
&&\hspace{1.8truecm}\left ( \f{\beta_0g(\mu) } { \beta(g(\mu))} \right)^2
\eea
by means of eq. \eqref{alfa} and: 
\bea
\f{ \beta(g(z)) }{g(z) } &=& -\beta_0g^2(z)\bigg(1+\f{\beta_1}{\beta_0}g^2(z)+\cdots\bigg)
\eea
It agrees with eq. \eqref{CSS0} for $O=F^2$, $D=4$, $\gamma_0^{(F^2)}=2\beta_0$ due to:
\bea
\f{g(\mu) } { \beta(g(\mu)) }&=& \f{1}{-\beta_0g^2(\mu)}\bigg(1-\f{\beta_1}{\beta_0}g^2(\mu)+\cdots\bigg)
\eea
that implies:
\bea
\left(\f{\beta_0g(\mu) } { \beta(g(\mu)) }\right)^2&=& \f{1}{g^4(\mu)}\bigg(1-2\f{\beta_1}{\beta_0}g^2(\mu)+\cdots\bigg)\nn\\
&=& \f{1}{g^4(\mu)}Z^{(F^2)'2}(g(\mu))
\eea
with $Z^{(F^2)'}(g(\mu))$ in eq. \eqref{ZOF} for $O=F^2$ and $\gamma_1^{(F^2)}=4\beta_1$.\par
Perturbatively,  by means of  eq. \eqref{g2loop}
we obtain:
\bea
Z^{(F^2)}(g(z), g(\mu)) &=&\left(\f{ \beta(g(z)) }{g(z) }\right)\left(\f{g(\mu) } { \beta(g(\mu)) }\right) \nn\\
&&\hspace{-2.5truecm}=\f{g^2(z)}{g^2(\mu)}
\bigg(1+\f{\beta_1}{\beta_0}(g^2(z)-g^2(\mu))+\cdots\bigg)\nn\\
&&\hspace{-2.5truecm}=1+g^2(\mu)2\beta_0\log|z\mu|+g^4(\mu)(4\beta_1\log|z\mu|+4\beta_0^2\log^2|z\mu|)+\cdots
\eea
that agrees with eq. \eqref{zeta00_small} for $O=F^2$, $\gamma_0^{(F^2)}=2\beta_0$ and $\gamma_1^{(F^2)}=4\beta_1$.

\section{Callan-Symanzik equation in $\td=4-2\eps$ dimensions in the momentum representation} \label{F}

\subsection{Callan-Symanzik equation for $C_{1}^{(F^2,O)}(p)$}

The CS equation in dimensional regularization for the fully renormalized OPE coefficient in the momentum representation in eq. \eqref{29}:
\bea\label{29_app}
C_{1}^{(F^2,O)}(p)=Z_{F^2} C_1^{(F_0^2,O)}(p)+Z_{1 \textrm{c.t.}} 
\eea
 follows from the renormalization-scale independence of the bare coefficient: 
\be
 \mu\f{dC_1^{(F_0^2,O)}(p)}{d\mu}\big|_{\eps,g_0}=0
 \ee
  for fixed bare parameters $\eps$ and $g_0$. In the massless case, employing  
 $\mu \f{ d}{d\mu}=\mu\f{\partial}{\partial \mu} +\beta(g,\epsilon) \f{\partial}{\partial g}$ with $\beta(g,\eps)=-\eps g +\beta(g)$, we obtain: 
\bea \label{CSmomreg_mu}
\bigg(\mu \frac{\partial}{\partial \mu} +\beta(g,\epsilon) \f{\partial}{\partial g}+ \gamma_{F^2}(g)\bigg) C_1^{(F^2,O)}(p,\mu,g(\mu)) = \gamma_{1 \textrm{c.t.}}(g) 
\eea
where:
\bea\label{gamma1ct}
\gamma_{1 \textrm{c.t.}}(g) &=&\mu\f{dZ_{1 \textrm{c.t.}}}{d\mu} +\gamma_{F^2}Z_{1 \textrm{c.t.}}\nn\\
&=&\mu\f{dg}{d\mu}\f{\partial Z_{1 \textrm{c.t.}}}{\partial g}
 +\gamma_{F^2}Z_{1 \textrm{c.t.}}\nn\\
&=&\beta(g,\epsilon)\f{\partial Z_{1 \textrm{c.t.}}}{\partial g}+\gamma_{F^2}Z_{1 \textrm{c.t.}}
\eea
 Given that $Z_{1 \textrm{c.t.}}$ is a series of pure poles in $\eps$ in $\overline{MS}$-like schemes, $\gamma_{1 \textrm{c.t.}}$ in eq. \eqref{gamma1ct} may contain poles in $\eps$ in addition to $O(\eps^0)$ terms, but no 
$O(\eps)$ terms, since in the rhs of eq. \eqref{gamma1ct} all the terms that contain factors of $\eps$
are multiplied by poles.
Hence, $\gamma_{1 \textrm{c.t.}}$ is $\eps$ independent if and only if it is not divergent as $\eps \rightarrow 0$. 
Yet, the rhs of eq. \eqref{CSmomreg_mu} must be finite as $\eps\to 0$
consistently with the finiteness as $\eps\to 0$ of the fully renormalized $C_1^{(F^2,O)}$ in the lhs. It follows that $\gamma_{1 \textrm{c.t.}}$ is finite as $\eps \rightarrow 0$ and, therefore, $\eps$ independent. Since the solution of eq. \eqref{CSmomreg_mu}
solely depends on the ratio $\f{p}{\mu}$, we write:
\bea \label{CSmomreg_p}
\bigg(p \cdot \frac{\partial}{\partial p} -\beta(g,\epsilon) \f{\partial}{\partial g}- \gamma_{F^2}(g)\bigg) C_1^{(F^2,O)}(p,\mu,g(\mu)) = -\gamma_{1 \textrm{c.t.}}(g) 
\eea
The CS equation in $d=4$ dimensions is obtained from eqs. \eqref{CSmomreg_mu} and \eqref{CSmomreg_p} by replacing $\beta(g,\eps)$ with $\beta(g)$. Hence, for $d=4$:
\bea \label{CSmomreg_p_4}
\bigg(p \cdot \frac{\partial}{\partial p} -\beta(g) \f{\partial}{\partial g}- \gamma_{F^2}(g)\bigg) C_1^{(F^2,O)}(p,\mu,g(\mu)) = -\gamma_{1 \textrm{c.t.}}(g) 
\eea

\subsection{General solution for $C_1^{(F^2,O)}(p)$ } \label{F1}

The general solution of eq. \eqref{CSmomreg_mu} or \eqref{CSmomreg_p} is:
\bea\label{CSmomregsol}
C_1^{(F^2,O)}(p,\mu,g(\mu))&=&Z^{(F^2)}(\tilde g(p),g(\mu))\Big({\cal G}^{(O)}(\tilde g(p))+\Delta^{(O)}(\tilde g(p),g(\mu))\Big)
\eea
where:
\bea\label{ZFpmom}
Z^{(F^2)}(\tilde g(p),g(\mu))&=&\exp{\int_{g(\mu)}^{\tilde g(p)} \f{\gamma_{F^2}(g)}{\beta(g,\epsilon) }dg}\nn\\
&=&\f{\beta(\tilde g(p),\eps)}{\tilde g(p)} \f{g(\mu)}{\beta(g(\mu),\eps)}
\eea
with the second equality implied by eq. \eqref{gF2}, and:
\bea\label{DeltaOp}
\Delta^{(O)}(\tilde g(p),g(\mu))&=& - \int_{g(\mu)}^{\tilde g(p)} \f{\gamma_{1 \textrm{c.t.}}(g) }{\beta(g,\epsilon) } \exp{\bigg(-\int_{g}^{\tilde g(p)} \f{\gamma_{F^2}(g')}{\beta(g',\epsilon) }dg'\bigg)}\, dg
\nn\\
&=& - \int_{g(\mu)}^{\tilde g(p)} \f{\gamma_{1 \textrm{c.t.}}(g) }{\beta(g,\epsilon) }
{Z^{(F^2)}}^{-1}(\tilde g(p),g)\,dg
\eea
${\cal G}^{(O)}(\tilde g(p))$ in eq. \eqref{CSmomregsol} is a function of the running coupling
$\tilde g(p)$ in the momentum representation in $\td=4-2\eps$ dimensions. The latter can be obtained from $\tilde g(z)$ in appendix \ref{CS1} by the substitution $z\to 1/p$.
By means of eqs. \eqref{ZFpmom} and \eqref{DeltaOp}, eq. \eqref{CSmomregsol} reads:
\bea\label{CSmomregsol2}
C_1^{(F^2,O)}(p,\mu,g(\mu))=\f{g(\mu)}{\beta(g(\mu),\eps)}\bigg(
{\cal G}^{(O)}(\tilde g(p))\f{\beta(\tilde g(p),\eps)}{\tilde g(p)} 
- \int_{g(\mu)}^{\tilde g(p)} \f{\gamma_{1 \textrm{c.t.}}(g) }{g }\, dg\bigg)
\eea
The solution of the CS eq. \eqref{CSmomreg_p_4} in $d=4$ dimensions is obtained by replacing $\beta(g,\eps)$ with $\beta(g)$ and $\tilde g(p)$ with $g(p)$ in eq. \eqref{CSmomregsol2}, where 
$g(p)$ can be  obtained from $g(z)$ in appendix \ref{CS0} by the substitution $z\to 1/p$.
Hence, for $d=4$:
\bea\label{CSmomregsol4D}
C_1^{(F^2,O)}(p,\mu,g(\mu))=\f{g(\mu)}{\beta(g(\mu))}\bigg(
{\cal G}^{(O)}(g(p))\f{\beta(g(p))}{g(p)} 
- \int_{g(\mu)}^{ g(p)} \f{\gamma_{1 \textrm{c.t.}}(g) }{g }\, dg\bigg)
\eea
Employing eq. \eqref{CT}:
\begin{equation}
C_{1}^{(F^2,O)}(p)=Z_{F^2} C_{1}^{(F_0^2,O)}(p)+Z_{F^2} \f{2\gamma_{O}(g)-2c_O \frac{\beta(g)}{g}}{\eps}
\end{equation}
we obtain:
\be
Z_{1 \textrm{c.t.}}=Z_{F^2} \f{2\gamma_{O}(g)-2c_O \frac{\beta(g)}{g}}{\eps}
\ee
and from eq. \eqref{gamma1ct}:
\bea
&&\gamma_{1 \textrm{c.t.}}(g)= \beta(g,\epsilon)\f{\partial Z_{1 \textrm{c.t.}}}{\partial g}+\gamma_{F^2}Z_{1 \textrm{c.t.}}\nn\\
&&~~=\beta(g,\epsilon)\bigg\{ \f{\partial Z_{F^2}}{\partial g}\f{2\gamma_{O}(g)-2c_O \frac{\beta(g)}{g}}{\eps} +\f{2Z_{F^2}}{\eps}\bigg(\f{\partial \gamma_O}{\partial g}-c_O\f{\partial}{\partial g}\bigg (\f{\beta(g)}{g}\bigg)\bigg)\bigg\}\nn\\
&&~~~+Z_{F^2}g\f{\partial}{\partial g}\bigg (\f{\beta(g)}{g}\bigg)\f{2\gamma_{O}(g)-2c_O \frac{\beta(g)}{g}}{\eps}\nn\\
&&~~=-Z_{F^2} g\f{\partial}{\partial g}\bigg (\f{\beta(g)}{g}\bigg)  \f{2\gamma_{O}(g)-2c_O \frac{\beta(g)}{g}}{\eps}
+\f{2\beta(g,\epsilon)}{\eps}Z_{F^2}\bigg(\f{\partial \gamma_O}{\partial g}-c_O\f{\partial}{\partial g}\bigg (\f{\beta(g)}{g}\bigg)\bigg)\nn\\
&&~~~+Z_{F^2}g\f{\partial}{\partial g}\bigg (\f{\beta(g)}{g}\bigg)\f{2\gamma_{O}(g)-2c_O \frac{\beta(g)}{g}}{\eps}\nn\\
&&~~=\f{2\beta(g,\epsilon)}{\eps}Z_{F^2}\bigg(\f{\partial \gamma_O}{\partial g}-c_O\f{\partial}{\partial g}\bigg (\f{\beta(g)}{g}\bigg)\bigg)\nn\\
&&~~=-2g\bigg(\f{\partial\gamma_O}{\partial g}-c_O\f{\partial}{\partial g}\bigg (\f{\beta(g)}{g}\bigg)\bigg)
\eea
 that is, indeed, manifestly $\eps$ independent by employing $\beta(g,\eps)=-\eps g+\beta(g)$, eqs. \eqref{gF2}, \eqref{ZF2} and:
\be
\beta(g,\epsilon) \f{\partial Z_{F^2}}{\partial g}=-Z_{F^2}g\f{\partial}{\partial g}\bigg (\f{\beta(g,\eps)}{g}\bigg)=-Z_{F^2}g\f{\partial}{\partial g}\bigg (\f{\beta(g)}{g}\bigg)
\ee
Then, the solution in eq. \eqref{CSmomregsol2} reads: 
\bea
C_1^{(F^2,O)}(p,\mu,g(\mu))&=&2c_O +\big({\cal G}^{(O)}(\tilde g(p))-2c_O\big)\f{\beta(\tilde g(p),\eps)}{\tilde g(p)} \f{g(\mu)}{\beta(g(\mu),\eps)}\nn\\
&&+2g(\mu)\f{\gamma_O(g(\mu))}{\beta(g(\mu),\eps)}\bigg(\f{\gamma_O(\tilde g(p))}{\gamma_O(g(\mu))}-1 \bigg)
\eea
where the identity:
\be
\f{\beta(\tilde g(p),\eps)}{\tilde g(p)} -\f{\beta( g(\mu),\eps)}{g(\mu)}= \f{\beta(\tilde g(p))}{\tilde g(p)} -\f{\beta( g(\mu))}{g(\mu)}
\ee
has been conveniently employed, 
and the solution in $d=4$ dimensions in eq. \eqref{CSmomregsol4D} reads:
\bea
C_1^{(F^2,O)}(p,\mu,g(\mu))&=&2c_O +\big({\cal G}^{(O)}(g(p))-2c_O\big)\f{\beta(g(p))}{g(p)} \f{g(\mu)}{\beta(g(\mu))}\nn\\
&&+2g(\mu)\f{\gamma_O(g(\mu))}{\beta(g(\mu))}\bigg(\f{\gamma_O(g(p))}{\gamma_O(g(\mu))}-1 \bigg)
\eea

\subsection{Specializing to $C_1^{(F^2,F^2)}(p)$  } \label{F2}

For $O=F^2$ the fully renormalized OPE coefficient in eq. \eqref{29_app} has been explicitly computed in perturbation theory up to order $g^4$ \cite{Z1,Z2}. Moreover, in this case the additive renormalization, i.e. $Z_{1 \textrm{c.t.}}$ in eq. \eqref{29_app}, has also been determined to all orders in perturbation theory \cite{Z3}:
\begin{equation}\label{Cf_app}
C_{1}^{(F^2,F^2)}(p)=Z_{F^2} C_1^{(F^2_0,F^2)}(p)+Z_{F^2} \f{2g\f{\partial}{\partial g}\left (\f{\beta(g)}{g}\right )-4 \frac{\beta(g)}{g}}{\eps} 
\end{equation}
where \cite{Z1,Z2}:
\bea\label{Cf_app_Z}
Z_{F^2} C_1^{(F_0^2,F^2)}(p)&=&4-4 B_{1,1}g^2-4 B_{1,2}g^4+\f{4\beta_1g^4}{\eps} \nn\\
&&-4\beta_0g^2\log\f{p^2}{\mu^2}
+4\beta_0^2g^4\log^2\f{p^2}{\mu^2}+4B_{1,3}g^4\log\f{p^2}{\mu^2} +\cdots
\eea
and \cite{Z3}:
\bea\label{ZFall}
Z_{F^2} \f{2g\f{\partial}{\partial g}\left (\f{\beta(g)}{g}\right )-4 \frac{\beta(g)}{g}}{\eps}= -\f{4\beta_1g^4}{\eps}+ \cdots
\eea
so that to order $g^4$:
\bea\label{C1totapp}
C_{1}^{(F^2,F^2)}(p)&=&4-4 B_{1,1}g^2-4 B_{1,2}g^4 \nn\\
&&-4\beta_0g^2\log\f{p^2}{\mu^2}
+4\beta_0^2g^4\log^2\f{p^2}{\mu^2}+4B_{1,3}g^4\log\f{p^2}{\mu^2} +\cdots
\eea
The fully renormalized coefficient $C_1^{(F^2,F^2)}(p)$ in eq. \eqref{C1totapp}
must be a solution of eq. \eqref{CSmomreg_p_4} and, equivalently, eq. \eqref{CSmomreg_mu} for $\eps\to 0$.
Indeed, we straightforwardly verify it inserting eq. \eqref{C1totapp} into eq. \eqref{CSmomreg_p_4} and employing for $O=F^2$: 
\bea\label{gamma1Fp}
\gamma_{1 \textrm{c.t.}}(g)&=&
(-\eps g+\beta(g))\f{\partial Z_{1 \textrm{c.t.}}}{\partial g}+\gamma_{F^2}Z_{1 \textrm{c.t.}}\nn\\
&=&-\eps g \f{\partial}{\partial g}\bigg(-\f{4\beta_1g^4}{\eps}\bigg)+\cdots\nn\\
&=&16\beta_1g^4+\cdots
\eea
that follows from eq. \eqref{gamma1ct} with $Z_{1 \textrm{c.t.}}$ given by eq. \eqref{ZFall}. \par
We also verify that $C_{1}^{(F^2,F^2)}(p)$ in eq. \eqref{C1totapp} is of the form in eq. \eqref{CSmomregsol4D} with the RG-invariant coefficient for $O=F^2$:
\bea\label{GFp}
{\cal G}^{(F^2)}(g(p))&=& 4 -4B_{1,1}g^2(p) -4B_{1,2}g^4(p)+\cdots
\eea
Indeed, for $O=F^2$ the perturbative expansion of eq. \eqref{CSmomregsol4D} reads:
\bea\label{FFCSsolpert}
&&C_1^{(F^2,F^2)}(p,\mu,g(\mu))=\f{g(\mu)}{\beta(g(\mu))}\bigg\{
\f{\beta(g(p))}{g(p)} {\cal G}^{(F^2)}(g(p))
- \int_{g(\mu)}^{ g(p)} \f{\gamma_{1 \textrm{c.t.}}(g) }{g }\, dg\bigg\}\nn\\
&&=\f{-1}{\beta_0g^2}\Big(1-\f{\beta_1}{\beta_0}g^2\Big)\bigg\{
-\beta_0g^2(p)\Big(1+\f{\beta_1}{\beta_0}g^2(p)\Big)
\Big(4-4B_{1,1}g^2(p) -4B_{1,2}g^4(p)\Big)\nn\\
&&~~~~-4\beta_1 g^4\Big(\f{g^4(p)}{g^4} -1\Big)
\bigg\}+\cdots\nn\\
&&=\f{g^2(p)}{g^2}\bigg(1+\f{\beta_1}{\beta_0}g^2\Big(\f{g^2(p)}{g^2} -1\Big)
 \bigg)\Big(4-4B_{1,1}g^2(p) -4B_{1,2}g^4(p)\Big)\nn\\
 &&~~~~+4\f{\beta_1}{\beta_0}g^2\Big(\f{g^4(p)}{g^4} -1\Big)+\cdots\nn\\
 &&=\Big(1-g^2 \beta_0 \log\f{p^2}{\mu^2}+g^4\big( -2\beta_1\log\f{p^2}{\mu^2}+\beta_0^2\log^2\f{p^2}{\mu^2}\big)\Big)
 \nn\\&&~~~~ 
 \Big(4-4B_{1,1}g^2\big( 1-g^2 \beta_0 \log\f{p^2}{\mu^2} \big) -4B_{1,2}g^4\Big)\nn\\
&&~~~~ -8\beta_1g^4\log\f{p^2}{\mu^2}+\cdots\nn\\
&&=4-4B_{1,1}g^2 -4B_{1,2}g^4\nn\\
&&~~~~-4\beta_0g^2\log\f{p^2}{\mu^2}
+4\beta_0^2g^4\log^2\f{p^2}{\mu^2} +(-8\beta_1+8\beta_0B_{1,1})g^4\log\f{p^2}{\mu^2}\nn\\
&&~~~~ -8\beta_1g^4\log\f{p^2}{\mu^2}+\cdots\nn\\
&&=4-4B_{1,1}g^2 -4B_{1,2}g^4\nn\\
&&~~~~-4\beta_0g^2\log\f{p^2}{\mu^2}
+4\beta_0^2g^4\log^2\f{p^2}{\mu^2} +4B_{1,3}g^4\log\f{p^2}{\mu^2}+\cdots
\eea
where we have employed in the second equality $\beta(g)=-\beta_0g^3-\beta_1g^4+\cdots$, eqs. \eqref{gamma1Fp} and \eqref{GFp}, the perturbative expansions for the running coupling $g(p)$ in terms of $g$: 
\bea\label{g24mom}
g^2(p) &=&g^2\Big(1-g^2 \beta_0 \log\f{p^2}{\mu^2}+g^4\big( -\beta_1\log\f{p^2}{\mu^2}+\beta_0^2\log^2\f{p^2}{\mu^2}\big)+\cdots \Big)\nn\\
g^4(p) &=&g^4\Big(1-g^2 2\beta_0 \log\f{p^2}{\mu^2}+g^4\big( -2\beta_1\log\f{p^2}{\mu^2}+3\beta_0^2\log^2\f{p^2}{\mu^2}\big)+\cdots \Big)
\eea
in the fourth equality
and in the last equality $B_{1,3}=-4\beta_1+2\beta_0B_{1,1}$, with $B_{1,3}$ and $B_{1,1}$ computed in \cite{Z1} and reported in eq. \eqref{relB}.
The last equality in eq. \eqref{FFCSsolpert} reproduces the perturbative result in eq. \eqref{C1totapp} to order $g^4$, as it should be. 

\subsection{Callan-Symanzik equation for $C_{0}^{(O,O)}(p)$}\label{F3}

The CS equation in dimensional regularization for the fully renormalized OPE coefficient $C_{0}^{(O,O)}(p)$ in the momentum representation\footnote{See \cite{Vector} for an analogous example.}:
\bea\label{C0OO_app}
C_{0}^{(O,O)}(p)=Z_{O}^2 C_0^{(O_0,O_0)}(p)+ p^{2\Delta_{O_0}-4}\mu^{2(1-\delta_O)\eps}Z_{0 \textrm{c.t.}} 
\eea
 with $\tilde\Delta_{O_0}=\Delta_{O_0}-\delta_O\eps$ the canonical dimension of $O$ in $\td=4-2\eps$ dimensions, follows from the renormalization-scale independence of the bare coefficient:
\be \label{CSC0momreg_mu}
\bigg(\mu \frac{\partial}{\partial \mu} +\beta(g,\epsilon) \f{\partial}{\partial g}+ 2\gamma_{O}(g)
\bigg) C_0^{(O,O)}(p,\mu,g(\mu)) = p^{2\Delta_{O_0}-4}\mu^{2(1-\delta_O)\eps}\gamma_{0 \textrm{c.t.}}(g) 
\ee
with:
\be
\mu \frac{d}{d \mu}=\mu \frac{\partial}{\partial \mu} +\beta(g,\epsilon) \f{\partial}{\partial g}
\ee
and:
\bea\label{gamma0ct}
\gamma_{0 \textrm{c.t.}}(g) &=&\mu\f{dZ_{0 \textrm{c.t.}}}{d\mu} +2\gamma_{O}Z_{0 \textrm{c.t.}}+2(1-\delta_O)\eps Z_{0 \textrm{c.t.}} \nn\\
&=&\beta(g,\epsilon)\f{\partial Z_{0 \textrm{c.t.}}}{\partial g}+2\gamma_{O}Z_{0 \textrm{c.t.}}+2(1-\delta_O)\eps Z_{0 \textrm{c.t.}} 
\eea
 where $\gamma_{0 \textrm{c.t.}}$ is finite as $\eps\to 0$ and $\eps$ independent analogously to 
$\gamma_{1 \textrm{c.t.}}$ in eq. \eqref{gamma1ct}.
For the dimensionless  object $C_{0,DL}^{(O,O)}$:
\be
C_0^{(O,O)}(p,\mu,g(\mu))=p^{2\Delta_{O_0}-4}\mu^{2(1-\delta_O)\eps}C_{0, DL}^{(O,O)}(\f{p}{\mu}, g(\mu))
\ee
that solely depends on the ratio $\f{p}{\mu}$, we also write:
\be \label{CSC0momreg_p}
\bigg(p \cdot \frac{\partial}{\partial p} -\beta(g,\epsilon) \f{\partial}{\partial g}- 2\gamma_{O}(g)
\bigg)  C_{0,DL}^{(O,O)}(\f{p}{\mu},g(\mu)) = -\gamma_{0 \textrm{c.t.}}(g) 
\ee
The CS equation in $d=4$ dimensions is obtained from eq. \eqref{CSC0momreg_mu} by removing the explicit $\eps$ dependence. Hence, for $d=4$:
\be \label{CSC0momreg_mu_4}
\bigg(\mu \frac{\partial}{\partial \mu} +\beta(g) \f{\partial}{\partial g}+ 2\gamma_{O}(g)
\bigg) C_0^{(O,O)}(p,\mu,g(\mu)) = p^{2\Delta_{O_0}-4}\gamma_{0 \textrm{c.t.}}(g) 
\ee

\subsection{General solution for $C_0^{(O,O)}(p)$ } \label{F4}

The general solution of eq. \eqref{CSC0momreg_mu} is:
\bea\label{CSC0momregsol}
C_0^{(O,O)}(p,\mu,g(\mu))&=&p^{2\Delta_{O_0}-4}\mu^{2(1-\delta_O)\eps}
{Z^{(O)}}^2(\tilde g(p),g(\mu))\Big({\cal G}_2^{(O)}(\tilde g(p))\nn\\
&&~~~+\Delta_2^{(O)}(\tilde g(p),g(\mu))\Big)
\eea
where:
\bea\label{ZOpmom}
Z^{(O)}(\tilde g(p),g(\mu))&=&\exp{\int_{g(\mu)}^{\tilde g(p)} \f{\gamma_{O}(g)}{\beta(g,\epsilon) }dg}
\eea
and:
\bea\label{DeltaOOp}
\Delta_2^{(O)}(\tilde g(p),g(\mu))&=& - \int_{g(\mu)}^{\tilde g(p)} \f{\gamma_{0 \textrm{c.t.}}(g) }{\beta(g,\epsilon) } \exp{\bigg(-2\int_{g}^{\tilde g(p)} \f{\gamma_{O}(g')}{\beta(g',\epsilon) }dg'\bigg)}\, dg\nn\\
&=& - \int_{g(\mu)}^{\tilde g(p)} \f{\gamma_{0 \textrm{c.t.}}(g) }{\beta(g,\epsilon) } {Z^{(O)}}^{-2}(\tilde g(p),g)\, dg
\eea
In $d=4$ dimensions, the solution of eq. \eqref{CSC0momreg_mu_4} reads:
\bea\label{CSC0momregsol_4}
C_0^{(O,O)}(p,\mu,g(\mu))&=&p^{2\Delta_{O_0}-4}
{Z^{(O)}}^2( g(p),g(\mu))\Big({\cal G}_2^{(O)}(g(p))\nn\\
&&~~~+\Delta_2^{(O)}( g(p),g(\mu))\Big)
\eea
where:
\bea\label{ZOpmom4}
Z^{(O)}( g(p),g(\mu))&=&\exp{\int_{g(\mu)}^{ g(p)} \f{\gamma_{O}(g)}{\beta(g) }dg}
\eea
and:
\bea\label{DeltaOOp4}
\Delta_2^{(O)}( g(p),g(\mu))&=& - \int_{g(\mu)}^{ g(p)} \f{\gamma_{0 \textrm{c.t.}}(g) }
{\beta(g) } \exp{\bigg(-2\int_{g}^{ g(p)} \f{\gamma_{O}(g')}{\beta(g') }dg'\bigg)}\, dg\nn\\
&=& - \int_{g(\mu)}^{ g(p)} \f{\gamma_{0 \textrm{c.t.}}(g) }
{\beta(g) } {Z^{(O)}}^{-2}( g(p),g)\, dg
\eea
Equivalently, eq. \eqref{CSC0momregsol_4} is obtained by taking $\eps\to 0$ in eq. \eqref{CSC0momregsol}.

\subsection{Specializing to $C_0^{(F^2,F^2)}(p)$} \label{F5}

Eq. \eqref{C0OO_app} reads for $O=F^2$: 
\bea\label{C0FF_app}
C_{0}^{(F^2,F^2)}(p)=Z_{F^2}^2 C_0^{(F^2_0,F^2_0)}(p)+ p^{4}\mu^{-2\eps}Z_{0 \textrm{c.t.}} 
\eea
where $\tilde\Delta_{F^2_0}=\Delta_{F^2_0}-\delta_{F^2}\eps=\td=4-2\eps$ is the canonical dimension of $F^2$ in $\td=4-2\eps$ dimensions.
The general solution of eq. \eqref{CSC0momreg_mu} for $O=F^2$ then reads:
\bea\label{CSC0FFmomregsol}
C_0^{(F^2,F^2)}(p,\mu,g(\mu))&=&p^{4}\mu^{-2\eps}
{Z^{(F^2)}}^2(\tilde g(p),g(\mu))\Big({\cal G}_2^{(F^2)}(\tilde g(p))
+\Delta_2^{(F^2)}(\tilde g(p),g(\mu))\Big)\nn\\
\eea
with $Z^{(F^2)}$ in eq. \eqref{ZFpmom}:
\bea
Z^{(F^2)}(\tilde g(p),g(\mu))&=&\f{\beta(\tilde g(p),\eps)}{\tilde g(p)} \f{g(\mu)}{\beta(g(\mu),\eps)}\nn\\
&\sim& \f{\tilde g^2(p)}{g^2(\mu)} \bigg(\f{\mu}{p}\bigg)^{-2\eps}\nn\\
&=& \f{1}{1-\beta_0g^2(\mu)\f{(\mu/p)^{2\eps}-1}{\eps}+\cdots}\nn\\
&=& 1-\beta_0g^2(\mu)\log\f{p^2}{\mu^2}+\cdots
\eea
where we have employed eq. \eqref{sol_eps_AF} with the replacement $z\to \f{1}{p}$ and its expansion in $g(\mu)$ as $\eps\to 0$, and:
\bea\label{DeltaFFp}
\Delta_2^{(F^2)}(\tilde g(p),g(\mu))&=& - \int_{g(\mu)}^{\tilde g(p)} \f{\gamma_{0 \textrm{c.t.}}(g) }{\beta(g,\epsilon) } \exp{\bigg(-2\int_{g}^{\tilde g(p)} \f{\gamma_{F^2}(g')}{\beta(g',\epsilon) }dg'\bigg)}\, dg\nn\\
&=& - \int_{g(\mu)}^{\tilde g(p)} \f{\gamma_{0 \textrm{c.t.}}(g) }{\beta(g,\epsilon) }
{Z^{(F^2)}}^{-2}(\tilde g(p),g)\, dg
\eea
Given $Z_{0 \textrm{c.t.}}$ to order $g^2$ according to eq. \eqref{C0S_add}:
\be\label{Z0pc1}
Z_{0 \textrm{c.t.}} =\f{N^2-1}{4\pi^2}\bigg\{-\f{1}{\eps}+\f{g^2\beta_0}{\eps^2}+\f{g^2 c_1}{\eps}\bigg\} +\cdots
\ee
we obtain for $\gamma_{0 \textrm{c.t.}}$ in eq. \eqref{gamma0ct} for $O=F^2$ to order $g^2$:
\bea\label{gamma0ctF}
\gamma_{0 \textrm{c.t.}}(g) &=&
\beta(g,\epsilon)\f{\partial Z_{0 \textrm{c.t.}}}{\partial g}+2\gamma_{F^2}Z_{0 \textrm{c.t.}}-2\eps Z_{0 \textrm{c.t.}} \nn\\
&=&\f{N^2-1}{4\pi^2}\bigg\{
-\eps g\, 2g\bigg(\f{c_1}{\eps}+\f{\beta_0}{\eps^2}\bigg) +\f{4\beta_0g^2}{\eps}
-2\eps\bigg(-\f{1}{\eps}+\f{g^2c_1}{\eps}+\f{g^2\beta_0}{\eps^2}\bigg) +\cdots\bigg\}\nn\\
&=&\f{N^2-1}{4\pi^2}\bigg\{2-4c_1g^2+\cdots\bigg\}
\eea
that is, indeed, $\eps$ independent to order $g^2$.
The solution in $d=4$ dimensions is obtained from eq. \eqref{CSC0momregsol_4} for $O=F^2$:
 \be\label{CSCFmomregsol_4}
C_0^{(F^2,F^2)}(p,\mu,g(\mu))=p^{4}
{Z^{(F^2)}}^2( g(p),g(\mu))\Big({\cal G}_2^{(F^2)}(g(p))
+\Delta_2^{(F^2)}( g(p),g(\mu))\Big)
\ee
where:
\bea\label{ZFpmom4}
Z^{(F^2)}( g(p),g(\mu))&=&\exp{\int_{g(\mu)}^{ g(p)} \f{\gamma_{F^2}(g)}{\beta(g) }dg}
\nn\\&=&\f{\beta( g(p))}{ g(p)} \f{g(\mu)}{\beta(g(\mu))}\nn\\
&=&\f{-\beta_0 g^2(p)+\cdots}{-\beta_0 g^2(\mu)+\cdots}\nn\\
&=&1-\beta_0g^2(\mu)\log\f{p^2}{\mu^2}+\cdots
\eea
 and the RG-invariant coefficient:
\bea
{\cal G}_2^{(F^2)}(g(p))&=&\bar{\cal G}_2^{(F^2)}(0)(1+\eta_2^{(F^2)}g^2(p)+\cdots)\nn\\
&=&\bar{\cal G}_2^{(F^2)}(0)(1+\eta_2^{(F^2)}g^2(\mu)+\cdots)
\eea
have been perturbatively expanded to order $g^2$, 
and:
\bea\label{DeltaFFp4}
\Delta_2^{(F^2)}( g(p),g(\mu))&=& 
 - \int_{g(\mu)}^{ g(p)} \f{\gamma_{0 \textrm{c.t.}}(g) }
{\beta(g) } {Z^{(F^2)}}^{-2}( g(p),g)\, dg
\eea
so that:
\be\label{ZDF}
{Z^{(F^2)}}^2( g(p),g(\mu))\Delta_2^{(F^2)}( g(p),g(\mu))
= -\bigg(\f{g(\mu)}{\beta(g(\mu))}\bigg)^2\int_{g(\mu)}^{ g(p)} \gamma_{0 \textrm{c.t.}}(g) \f{\beta(g)}
{g^2} \, dg
\ee
Inserting eq. \eqref{gamma0ctF}  into eq. \eqref{ZDF} we further obtain to order $g^2$:
\bea\label{ZDFg2}
&&{Z^{(F^2)}}^2( g(p),g(\mu))\Delta_2^{(F^2)}( g(p),g(\mu))\nn\\
&&= -\f{1}{\beta_0^2g^4(\mu)}\bigg(1-2\f{\beta_1}{\beta_0}g^2(\mu)+\cdots\bigg)
\f{N^2-1}{4\pi^2}\nn\\
&&~~~~~~~~~~
\int_{g(\mu)}^{ g(p)} (2-4c_1g^2+\cdots )(-\beta_0 g-\beta_1g^3+\cdots) \, dg\nn\\
&&=-\f{1}{\beta_0^2g^4(\mu)}\bigg(1-2\f{\beta_1}{\beta_0}g^2(\mu)+\cdots\bigg)\f{N^2-1}{4\pi^2}\nn\\
&&~~\bigg\{-\beta_0\Big(g^2(p)-g^2(\mu)\Big) +\Big(c_1\beta_0-\f{1}{2}\beta_1\Big)\Big(g^4(p)-g^4(\mu)\Big)+\cdots\bigg\}\nn\\
&&=-\f{1}{\beta_0^2g^4(\mu)}\bigg(1-2\f{\beta_1}{\beta_0}g^2(\mu)+\cdots\bigg)\f{N^2-1}{4\pi^2}\nn\\
&&~~\bigg\{
-\beta_0\bigg(-g^4(\mu)\beta_0\log\f{p^2}{\mu^2}+g^6(\mu)\Big(-\beta_1\log\f{p^2}{\mu^2}+
\beta_0^2\log^2\f{p^2}{\mu^2}\Big)+\cdots\bigg)
\nn\\
&&~~+\Big(c_1\beta_0-\f{1}{2}\beta_1\Big)
\Big(-2g^6(\mu)\beta_0\log\f{p^2}{\mu^2}+\cdots
\Big)+\cdots\bigg\}\nn\\
&&=\f{N^2-1}{4\pi^2}\f{1}{\beta_0^2g^4(\mu)}\bigg\{-\beta_0^2g^4(\mu)\log\f{p^2}{\mu^2}
+g^6(\mu)\bigg(\beta_0^3\log^2\f{p^2}{\mu^2}
+2c_1\beta_0^2\log\f{p^2}{\mu^2}\bigg)+\cdots\bigg\}\nn\\
&&=\f{N^2-1}{4\pi^2}\bigg\{
-\log\f{p^2}{\mu^2}+\beta_0g^2(\mu)\log^2\f{p^2}{\mu^2}+2c_1g^2(\mu)\log\f{p^2}{\mu^2}+\cdots\bigg\}
\eea
where we have employed eq. \eqref{g24mom}. 
Hence, setting $g(\mu)=g$, the complete solution in eq. \eqref{CSCFmomregsol_4} reads to order $g^2$:
\bea
&&C_0^{(F^2,F^2)}(p,\mu,g)=p^{4}
\bigg\{\bar{\cal G}_2^{(F^2)}(0)
\bigg(1-2\beta_0g^2\log\f{p^2}{\mu^2}+\cdots\bigg)\bigg(1+\eta_2^{(F^2)}g^2+\cdots\bigg) \nn\\
&&~~~~~~~~~+
\f{N^2-1}{4\pi^2}\bigg( -\log\f{p^2}{\mu^2}+\beta_0 g^2 \log^2\f{p^2}{\mu^2}+2c_1g^2 \log\f{p^2}{\mu^2} \bigg)
+\cdots\bigg\}
\eea
For:
\be
\bar{\cal G}_2^{(F^2)}(0)=\f{N^2-1}{4\pi^2}
\ee
we get:
\bea\label{C0Fpsol}
C_0^{(F^2,F^2)}(p,\mu,g)&=&
p^{4}\f{N^2-1}{4\pi^2}\bigg\{1-\log\f{p^2}{\mu^2}+g^2\beta_0\log^2\f{p^2}{\mu^2}\nn\\
&&~~+2g^2(c_1-\beta_0)\log\f{p^2}{\mu^2}
+g^2\eta_2^{(F^2)}+\cdots\bigg\}
\eea
 Hence, for a given value of the scheme-dependent coefficient $c_1$ entering the additive renormalization in eq. \eqref{Z0pc1}, the scheme-dependent coefficient of the subleading $\log\f{p^2}{\mu^2}$ to order $g^2$ that solves of the CS equation is $2(c_1-\beta_0)$.
Eq. \eqref{C0S_tot}, given eqs. \eqref{C0S_mr} and \eqref{C0S_add}, is of the form in eq. \eqref{C0Fpsol} and it is a solution of the CS equation for any value of $c_1$.

\end{document}